%% file: arxiv-preprint.tex
\documentclass{article}

\usepackage{graphicx}
\usepackage{rotating}
\usepackage{fullpage}
\usepackage{url}
\usepackage{natbib}
\usepackage{subfigure}
\usepackage{fancyvrb}
\usepackage{amsmath,amssymb}
\usepackage{multicol}
\usepackage{hyperref}
\hypersetup{
    colorlinks,
    citecolor=black,
    filecolor=black,
    linkcolor=black,
    urlcolor=black
}

\newcommand{\TableStart}[2]{
  \begin{table}[ht!]
    \scriptsize
    \centering
    \begin{tabular}{#1}
      \hline
      #2
      \hline
}
\newcommand{\TableEnd}[2]{
      \hline
    \end{tabular}
    \caption{#2}
    \label{#1}
  \end{table}
}

\begin{document}

\title{Organizing the Aggregate: Languages for Spatial Computing}
\author{Jacob Beal (Raytheon BBN Technologies, USA)\thanks{Work
    partially sponsored by DARPA; the views and conclusions contained
    in this document are those of the authors and not DARPA or the
    U.S. Government.},\\
  Stefan Dulman (Delft Univ., the Netherlands),\\
  Kyle Usbeck (Raytheon BBN Technologies, USA),\\
  Mirko Viroli (Univ. Bologna, Italy),\\
  Nikolaus Correll (Univ. Colorado Boulder, USA)}

\date{April 2nd, 2012\\ To appear as a chapter in the book ``Formal and
  Practical Aspects\\ of Domain-Specific Languages: Recent
  Developments''}

\maketitle

% Abstract should be 100-150 words
\begin{abstract}
  As the number of computing devices embedded into engineered systems
  continues to rise, there is a widening gap between the needs of the
  user to control aggregates of devices and the complex technology of
  individual devices.
  Spatial computing attempts to bridge this gap for systems with local
  communication by exploiting the connection between physical locality
  and device connectivity.
  A large number of spatial computing domain specific languages (DSLs) have
  emerged across diverse domains, from biology and reconfigurable
  computing, to sensor networks and agent-based systems.
  In this chapter, we develop a framework for analyzing and comparing
  spatial computing DSLs, survey the current state of the art, and
  provide a roadmap for future spatial computing DSL investigation.
\end{abstract}

\tableofcontents

%\todo[inline]{Check for English/american spelling, generally spellcheck}
%\todo[inline]{Make sure all references are correct.}
%\todo[inline]{Need to compact all code examples}
%\todo[inline]{For each reference example, we need to explain why we chose that DSL}

\section{Introduction}

Computation has become cheap enough and powerful enough that large
numbers of computing devices can be embedded into nearly any aspect of
our environment.  A widening gap, however, exists between the
potential users of embedded systems (biologists, architects, emergency
response teams, etc.), and the increasingly complex technology available.  
Typically, the users know what they want from the
aggregate of devices, but the programming or design interfaces that
they are presented with operate mainly at the level of individual
devices and their interactions.  Thus, biologists who want to monitor
wildlife with mesh-networked sensors end up having to debug real-time
code for parallel algorithms in order to minimize Joules-per-packet, and
architects who want to create responsive buildings end up worrying
about how to create distributed algorithms to ensure correct time-synchronization 
across the application modules.

Similar gaps are found in many other areas besides embedded systems.
For example:
\begin{itemize}
\item Robots are becoming cheaper and more reliable, such that teams
  of robots can now be used in many more applications.  For
  instance, search-and-rescue workers might use a group of unmanned
  aerial vehicles to search a wilderness area.  These
  search-and-rescue workers should be able to specify aggregate behaviors
  like sweep patterns, rather than worry about how to share
  localization and progress data in order to keep those sweep patterns
  consistent between robots.
  Modular and reconfigurable robotic systems have similar issues.

\item Reconfigurable computing devices like Field Programmable Gate Arrays (FPGAs) can solve complex
  problems extremely quickly.  Programming and code reuse are
  difficult, however, because subprograms need to be tuned for their
  layout on each particular model chip and their context of use
  relative to other subprograms.
  Similar issues are likely to emerge in multi-core systems as the
  number of cores continues to rise.

\item In synthetic biology, it will soon be desirable to program
  engineered biological organisms in terms of the structure of tissues
  or colonies, and of their interaction patterns, rather than by
  selecting particular trans-membrane proteins and adjusting their
  signal dynamics and metabolic interactions.  
  Emerging chemical and nanotechnological computing platforms are
  likely to face similar issues.
\end{itemize}

\begin{figure}[t]
\centering
\includegraphics[scale=0.5]{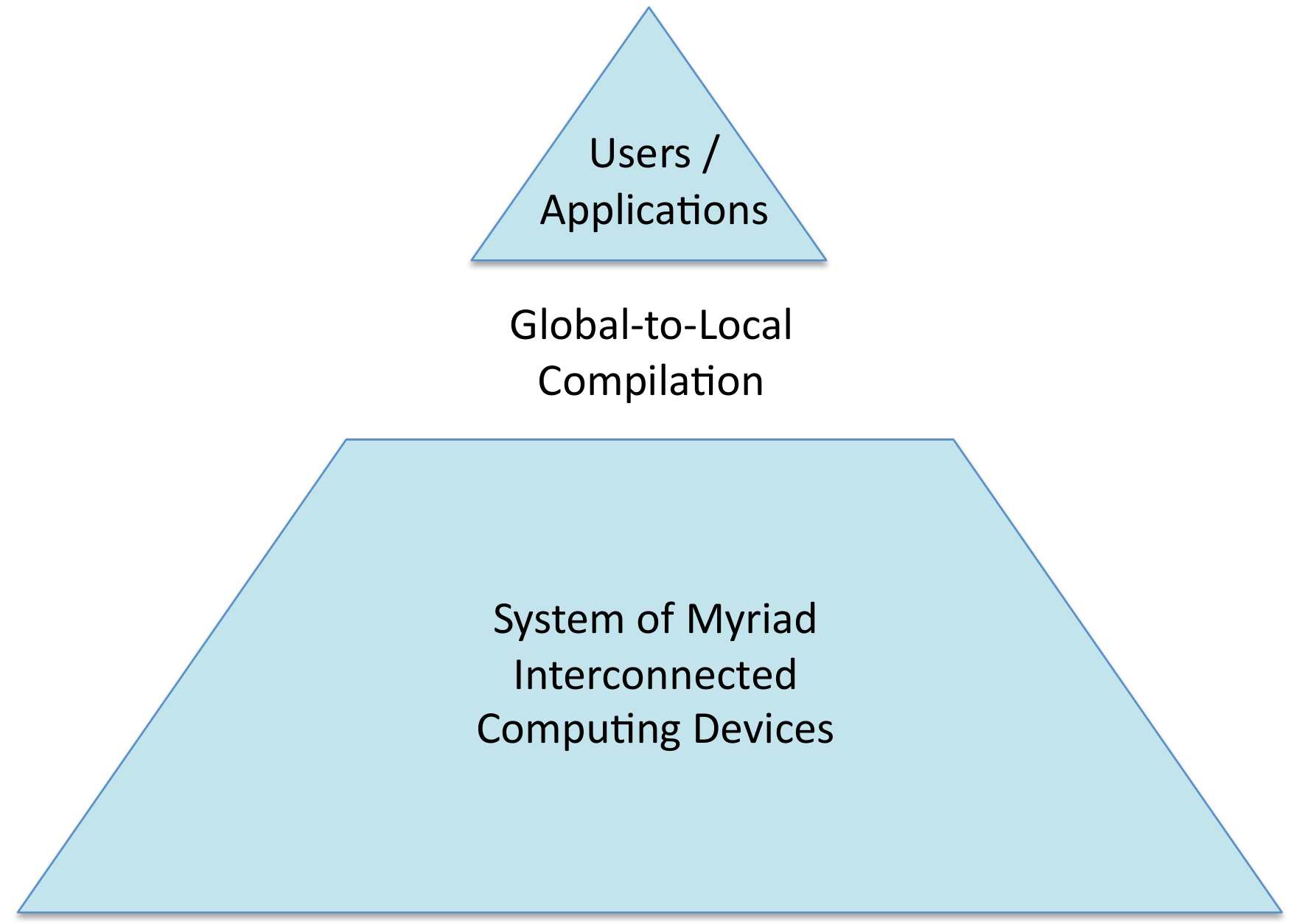}
\caption{A widening gap exists between the application needs of users
  and the implementation details of those applications on increasingly
  complex systems of interacting computing devices.  This gap must be
  crossed by {\em global-to-local compilation}: the transformation of
  specifications of aggregate behavior to actions of individual
  devices.}
\label{f:gap}
\end{figure}

Fundamentally, this gap lies between programming the aggregate and programming the individual
(Figure~\ref{f:gap}).  For the users, the application is often best
specified in terms of collective behaviors of aggregates of devices.
For example, an architect may wish to use sweeping patterns of light
on the floor and walls to guide visitors through a building.  When
implemented on a system of computing devices, however, the application
must be carried out through the actions and interactions of many
individual devices.  For example, a particular LED on the wall must
discover whether it is on a visitor's path and then synchronize with
its neighbors to ensure that it turns on and off at the correct times
to create the desired pattern.  
Thus, we see that the critical element for bridging the gap between
a user and a system of many devices is {\em global-to-local compilation}
\citep{abelson2000amorphous}, facilitated by the development of a new
programming language or languages suited to describing aggregate
behaviors.

All of the domains that we have mentioned so far share a 
property: a close relationship between the computation and the
arrangement of the computing devices in space.  These systems thus fit
the definition of {\em spatial computers}---collections of local
computational devices distributed through a physical space, in which:
\begin{itemize}
\item the difficulty of moving information between any two devices is
  strongly dependent on the distance between them, and
\item the ``functional goals'' of the system are generally defined in
  terms of the system's spatial structure.
\end{itemize}

This correlation between location and computation can be exploited to
help bridge the gap between aggregate and individual.  Geometry and
topology include many ``intermediate'' aggregate concepts, such as
regions, neighborhoods, and flows, that can be used as building blocks
for programming applications and automatically mapped to the behavior
of individual devices.

Across many different communities, researchers have built domain
specific languages (DSLs) that use spatial abstractions to simplify
the problem of programming aggregates.  These DSLs take a wide variety
of different approaches, often blending together generally applicable
spatial concepts with specific requirements of a target sub-domain or
application, and have met with varying degrees of success.  At
present, the space is quite chaotic, but the number of shared ideas
between DSLs argues that it should be possible to develop more
powerful and more generally applicable languages for spatial
computing.  Indeed, as we will see, some may be beginning to emerge.

In this chapter, our aim is to develop a clear framework for analyzing
and comparing spatial computing DSLs, to survey the current state of
the art, and to provide a roadmap for future spatial computing DSL
investigation.  We organize our investigation as follows:
\begin{itemize}
\item In Section~\ref{s:definitions}, we define a framework for
  analyzing spatial computing DSLs, separating more abstract spatial
  concerns from the pragmatic needs of particular implementations or
  application domains.

\item Following in Section~\ref{s:survey}, we survey existing spatial
  computing DSLs across a number of domains and summarize their key
  attributes.  To aid in comparison, we encode a reference example in
  representative languages.

\item We then analyze the relationship between languages in
  Section~\ref{s:analysis}, identifying common themes and gaps in the
  existing space of languages.

\item Finally, in Section~\ref{s:roadmap}, we summarize our results
  and propose a roadmap for the future development of spatial computing
  DSLs.
\end{itemize}

\section{Analytic Framework}
\label{s:definitions}

In this section, we develop a framework for analyzing spatial
computing domain specific languages.  We begin by defining the scope
of work that will be considered in this chapter.  We then
organize our analytic framework in terms of a generic aggregate
programming architecture, which separates aggregate and space-focused
aspects of a system from underlying lower level details.  Finally, for
each layer in this architecture we define the attributes that will be
examined as part of this survey.

\subsection{Chapter Scope}

For purposes of this review, we will define a spatial computing DSL as
any system construction mechanism that:
\begin{itemize}
\item is targeted at programming some class of spatial computers,
\item includes explicitly geometric or topological constructs that
  implicitly operate over aggregates of computing devices, and
\item allows an unbounded combination of systems to be specified.
\end{itemize}

Our aim is to develop both a survey and a roadmap for spatial
computing domain specific languages, however, we will also include two
classes of systems that do not fit these properties:
We will discuss some aggregate programming languages that are designed
for spatial computers, but where no constructs are explicitly
geometric or topological.  These languages typically attempt to
abolish space in some way, and comparison with these languages will
help to illuminate the benefits and costs involved in incorporating
spatial constructs in a DSL for aggregate programming.
We will also discuss some systems that are not explicitly languages,
but that still can serve as toolkits for general system specification.
These are included in the case where they include significant spatial
constructs for aggregate programming, and can therefore help define
the known design space for spatial computing DSLs or reveal thinking
relative to spatial DSLs in an application domain.

The classes of spatial computers that we consider in this chapter are
roughly: amorphous computers, natural and engineered biological
organisms, multi-agent systems, wireless sensor networks, pervasive
computing, swarm and modular robotic systems, parallel computers,
reconfigurable computers, cellular automata, and formal calculi.  The
boundaries of these fields are, of course, fuzzy and in many cases
overlapping.  We also acknowledge that, although we have tried to be
thorough, this is by no means an exhaustive list of all classes of
spatial computers, and due to the wide variety of terminology used
across fields, we may have missed some significant examples of spatial
computing DSLs.  Our belief, however, is that this survey has been
sufficiently broad to provide a clear map of the known design space
for spatial computing DSLs.

\begin{figure}
  \centering
  \includegraphics[scale=0.5]{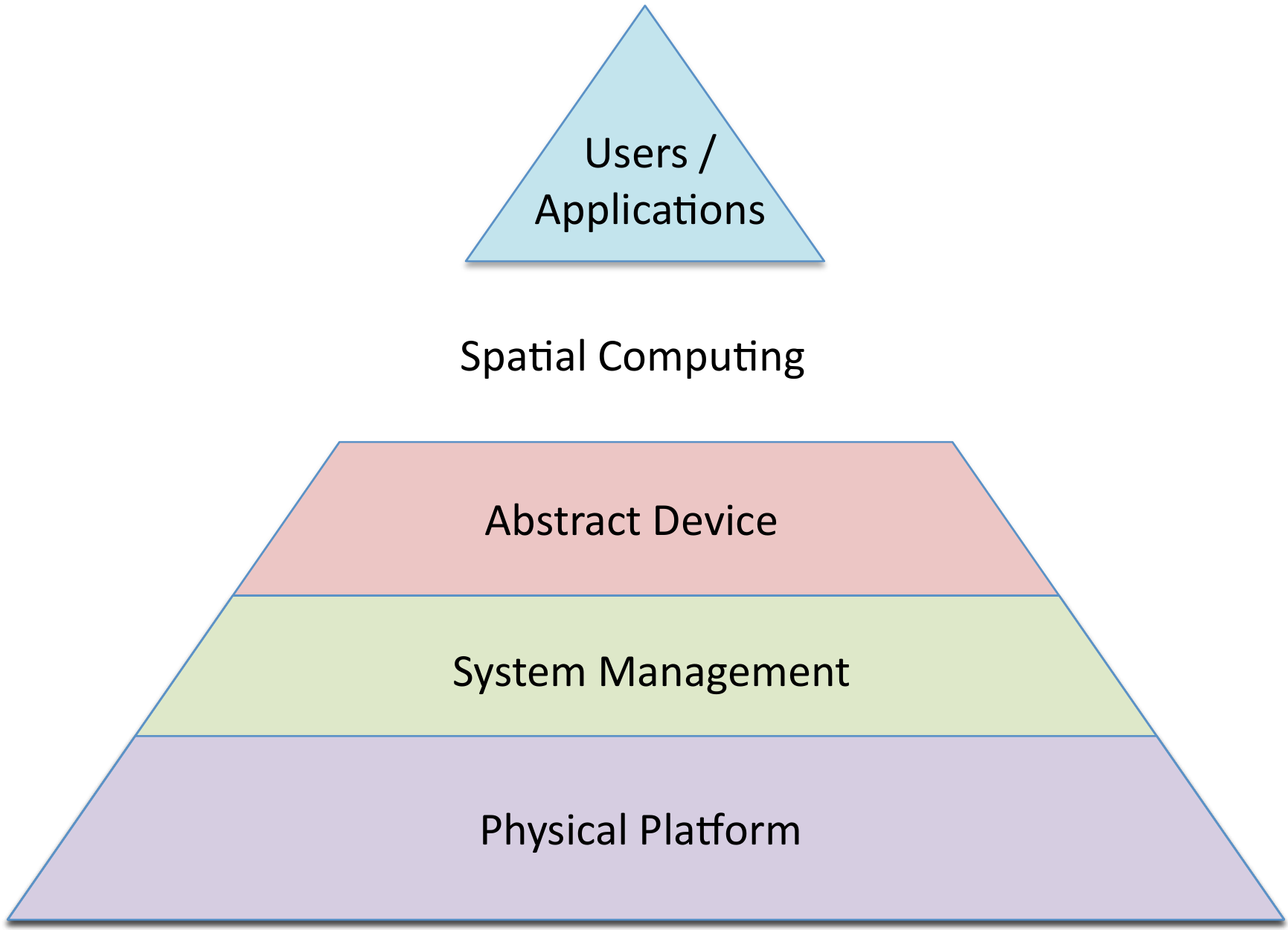}
  \caption{Spatial computing fills an architectural gap between
    individual computation devices and users wishing to control
    aggregate behavior.}
  \label{fig:layers}
\end{figure}

\subsection{Generic Aggregate Programming Architecture}

Different spatial computing DSLs operate at different levels of
abstraction and address different types of concern.  To aid us in
better understanding their relationship, we refine the simple ``gap''
diagram in Figure~\ref{f:gap} into a generic aggregate programming
architecture comprised of five layers.  We illustrate this
architecture as a pyramid in Figure~\ref{fig:layers}, the wider base
representing more detail in implementation.  The lower three layers
deal only with individual devices and their local interactions.  From
lowest to highest, these are: 
\begin{itemize}
\item The {\bf physical platform} layer at the base is the medium upon
  which computation will ultimately occur.  This could be
  anything from a smart phone or an unmanned aerial vehicle to a
  living biological cell.  It may also be a virtual device, as is the
  case for simulations.
\item The {\bf system management} layer is where the operating
  system and whatever services are built into the platform live.
  For example, on an embedded device, this layer is likely to include
  real-time process management, sensor and actuator drivers, and
  low-level networking.
\item The {\bf abstract device} layer hides details of the device,
  presenting a simplified interface to other devices with which it interacts.
  For example, a complex embedded computing platform might be abstracted
  as a graph node that periodically executes rounds of computation
  over its edges.
\end{itemize}

Above the abstract device is the gap that we have previously
discussed, between the interface for programming individual devices
and the aggregate-level needs and desires of users.  We consider the
{\bf spatial computing} abstractions and models that can connect
between individual devices and aggregates.  While there may be other
things besides spatial computing that help in bridging this gap, they
are out of scope of our discussion in this chapter.

Finally, the top of the pyramid consists of {\bf users and
  applications} that wish to deal, not with individual devices, but
with behaviors of the aggregate.

\subsection{Spatial DSL Properties}

This aggregate programming architecture forms the basis of our
analytic framework.  For each spatial computing DSL that we consider,
we will analyze that DSL in terms of which layers it focuses on and
what properties it has with respect to each layer.

As we shall see, different languages have different priorities, and no
system spans the whole range of considerations.  This is by no means a
bad thing: in a similar case, the OSI stack~\citep{zimmermann1980osi}, which
factors networking into seven different abstraction layers, has been a
powerful enabling tool for allowing the interconnection of many
different specialized tools to form the modern Internet.  
This does mean, however, that for nearly every property that we
consider, there will be at least some DSLs for which the language property is not present
or not applicable, since said property is out of scope of that
particular DSL.

\begin{figure}[t]
\centering
\includegraphics[scale=0.75]{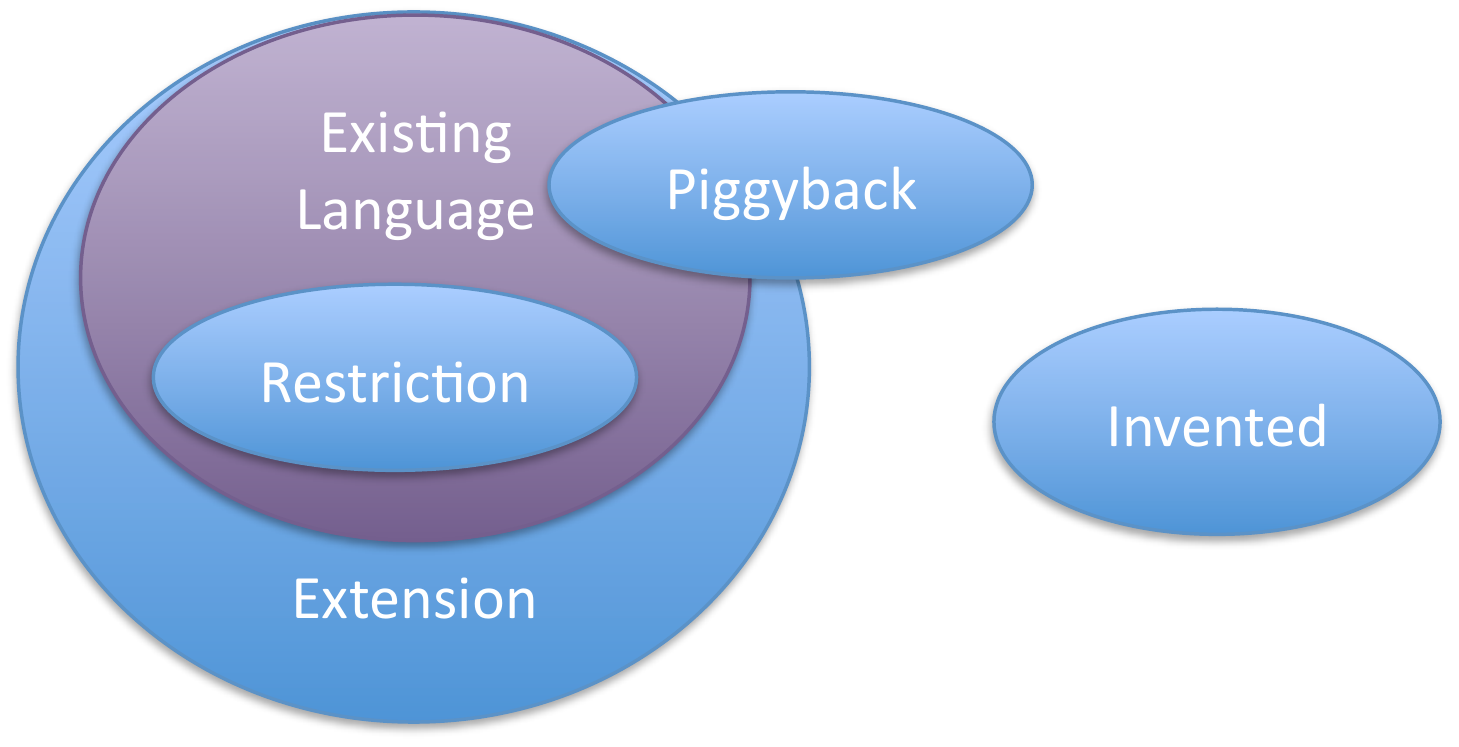}
\caption{DSL design patterns describe the relation of the new DSL to 
  existing languages: an {\em extension} contains the existing language,
  a {\em restriction} is contained within it, {\em piggyback} languages
  both extend and restrict, and {\em invented} languages are created
  from scratch.}
\label{f:dsltypes}
\end{figure}

The top-level properties we analyze for each spatial computing DSL
categorize the language itself and define the general scope that
the language covers:
\begin{itemize}
\item {\bf What type of programming language is the DSL?}  Common
  types include functional, imperative, and declarative.  
\item {\bf What is the design pattern for this DSL?}  This property
  reflects the degree to which the language is related to an existing
  conventional language, and is defined as in \citep{mernik2005and}.  DSLs may
  be based on existing languages in three ways: it may be a {\em
    restriction}, meaning that it removes some language features
  problematic for the domain, an {\em extension}, meaning that it adds
  new features, or a {\em piggyback} that adds some and removes
  others.  A DSL may also be {\em invented}, which means that a whole
  language has been created from scratch, with its own syntax and
  interpreter or compiler.

\item {\bf For what physical platforms is the language
    primarily/originally intended?}  Although spatial DSLs may
  have relevance across a number of different domains, their
  representational framework typically shows strong traces of their
  ``home'' platform. 
%  We categorize this at a high degree of
%   abstraction, into {\em wireless network}, {\em wired network}, {\em
%     mechanical structure}, {\em biological}, {\em chemical}, {\em
%     circuitry}, and {\em automata}. 
  This choice also typically
  regulates whether computing devices are assumed to be universal or
  not.
\item {\bf On which intermediate layers does the language focus?}  The
  previous questions dealt with the top and bottom layers of the
  pyramid.  Most DSLs of interest to this review focus on one or two
  of the middle layers: spatial computing, abstract device, and/or
  system management.
\end{itemize}

For the spatial computing and abstract device layers, we elaborate our
framework to include a set of key properties focused on that layer.
For the spatial computing layer, we consider what space-spanning
abstractions are supported by each DSL.  For the abstract device
layer, we consider how devices are related to space, and how
information moves through the system.  

We will not directly address the system management layer in this
document, as decisions at this layer are largely disconnected from
spatial considerations in current systems.  A useful framework for
analyzing system management properties, however, has already been
developed by the agent community.  The {\it Agent System Reference
  Model}~\citep{asrm} (ASRM) defines functional concepts that are
typical in agent frameworks, and the {\it Agent System Reference
  Architecture}~\citep{asra} (ASRA) builds on the ASRM with
architectural paradigms for these functional concepts.  As the ASRM
and ASRA address pragmatic concerns that are shared across a wide spectrum of
distributed systems, they could be applied to analyze any of the DSLs
that we survey.

\subsubsection{Types of Space-Time Operations} \label{sec:basisset}

\begin{figure}
  \centering
  \includegraphics[width=0.9\columnwidth]{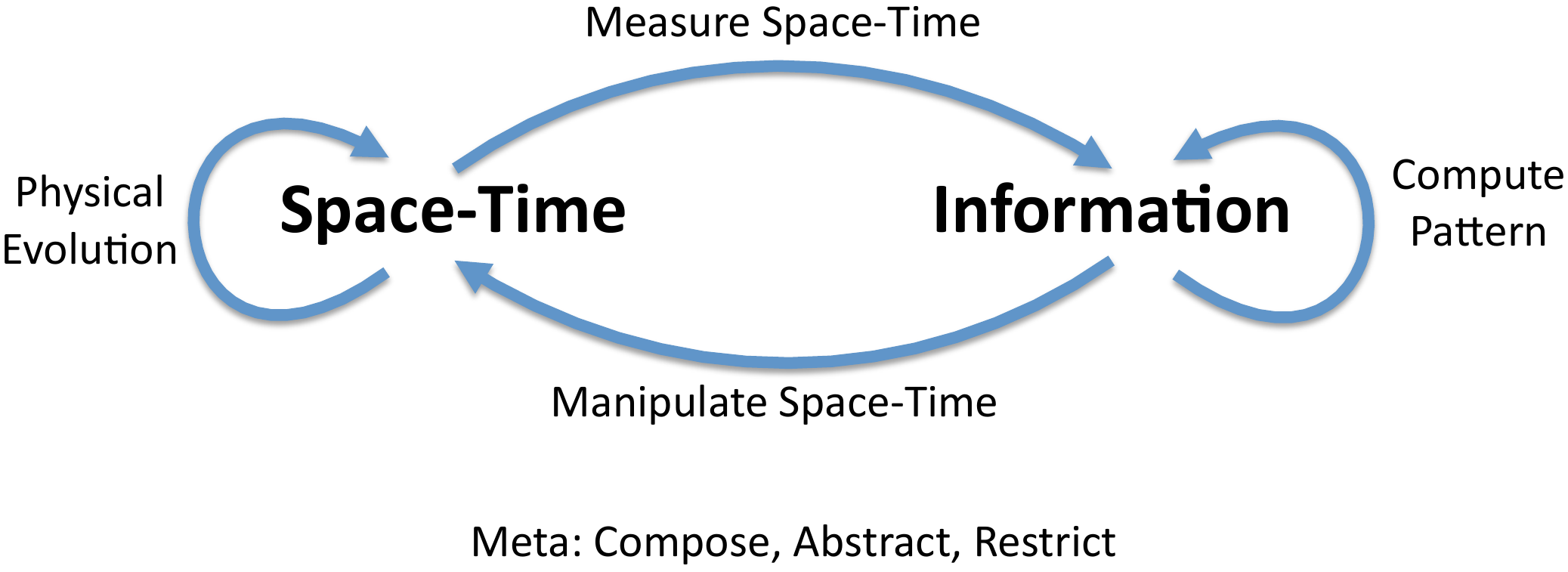}
  \caption{The basic duality of space-time and information in a spatial
    computer implies four general classes of operations, plus meta-operations
    that manipulate computations.}
  \label{fig:opclasses}
\end{figure}

Figure~\ref{fig:opclasses} shows the two base elements of a spatial
computer and the relations between them.  At a basic level, a spatial
computer is a dual entity: on the one hand, we have the volume of
space-time spanned by the computer and on the other hand, the
information that is associated with locations in that volume of
space-time.

From this duality, we may derive four classes of operations (partially
based on the universal basis set of space-time operators proposed in
\citep{bealBasisSCW10}):
\begin{itemize}
\item {\bf Measure Space-Time:} These are operators that take
  geometric properties of the space-time volume and translate them
  into information.  Examples are measures of distance, angle, time
  duration, area, density, and curvature.
\item {\bf Manipulate Space-Time:} These operators are the converse of
  measurement, being actuators that take information and modify the
  properties of the space-time volume.  Examples are moving devices,
  changing curvature (e.g., flexing a surface), expanding or
  contracting space locally (e.g., changing the size of cells in a
  tissue), or changing local physical properties such as stiffness
  that will directly affect the physical evolution of the system.
\item {\bf Compute Pattern:} Operations that stay purely in the
  informational world may be viewed at a high level of abstraction as
  computing patterns over space-time.  For example, stripes are a
  pattern over space, a timer is a pattern over time, and a
  propagating sine wave is a pattern over space-time.  This category
  includes not just computation, but most communication and any
  ``pointwise'' sensor or actuator that does not interact directly
  with geometry, such as a light or sound sensor or an LED actuator.
\item {\bf Physical Evolution:} Many physically instantiated systems
  have inherent dynamics that cause the shape of the space to change
  over time even without the use of actuators to manipulate
  space-time.  Examples include inertial motion of robots or the
  adhesive forces shaping a colony of cells.  By their nature, these
  operations are not directly part of programs, but languages may 
  assume such dynamics are in operation or have
  operations that are targeted at controlling them.
\end{itemize}

Any spatial computation can be described in terms of these four
classes of operations.  We are speaking of languages, however, so we
also need to consider the meta-operations that can be used to combine
and modulate spatial computations.  We identify two such classes of
meta-operation: {\bf Abstraction and Composition} operations hide the
implementation details of spatial computations, and allow them to be
combined together and to have multiple instances of a computation
executed.  {\bf Restriction} operations modulate a spatial computation
by selecting a particular subspace on which the computation should be
executed.

These categories of operations are of primary interest for our survey,
as the innovations of a DSL pertinent to spatial computing will
generally be closely linked to a set of space-time operations.  For
each DSL we consider, we will thus report what significant operators
are provided for each category.

\subsubsection{Abstract Device Model}
The abstract device model relates computing devices to the space that
they occupy and specifies the ways in which devices can communicate
with one another.  
The {\bf Discretization} of devices with respect to space falls into
three general types: {\em discrete} models that assume a set
of non-space-filling devices, as is typical of sensor network systems,
{\em cellular} models that assume discrete devices that do fill space,
as is typical of modular robotic systems and cellular automata, and
{\em continuous} models that assume an infinite continuum of devices that
will then be approximated by actual available hardware.

We identify three key properties of communication that describe 
the information flows between devices:
\begin{itemize}
\item {\bf Communication Region:} This is the relationship between a
  device's communication partners and space.  The most common types we will encounter
  are a distance-limited {\em neighborhood} (though not necessarily
  regular) and {\em global} communication.
\item {\bf Transmission Granularity:} Do devices {\em broadcast} the
  same thing to all neighbors, {\em unicast} a potentially
  different value to each neighbor, or {\em multicast} to groups
  of neighbors?
\item {\bf Mobility of Code:} Do devices have the same {\em uniform}
  program (which may execute differently depending on state and
  conditions), do devices have {\em heterogeneous} (but fixed)
  programs, or is code {\em mobile} and able to shift from device to
  device?
\end{itemize}

\subsection{Reference Example: ``T-Program''} \label{sec:t-prog}

As a means of evaluating and demonstrating the space-time operators which
are supported by classes of languages, we derive a {\it reference example}
and implement portions of the example in representative languages considered
during the survey.
For the example, which we refer to as the ``T-Program,'' to be
completely implemented requires the ability to perform each of the
three basic families of space-time operators described in
Section~\ref{sec:basisset}: measurement of space-time, pattern
computation, and manipulation of space-time.

\begin{figure}[t]
\centering
\subfigure[Establishing local coordinates]
{\includegraphics[scale=0.28]{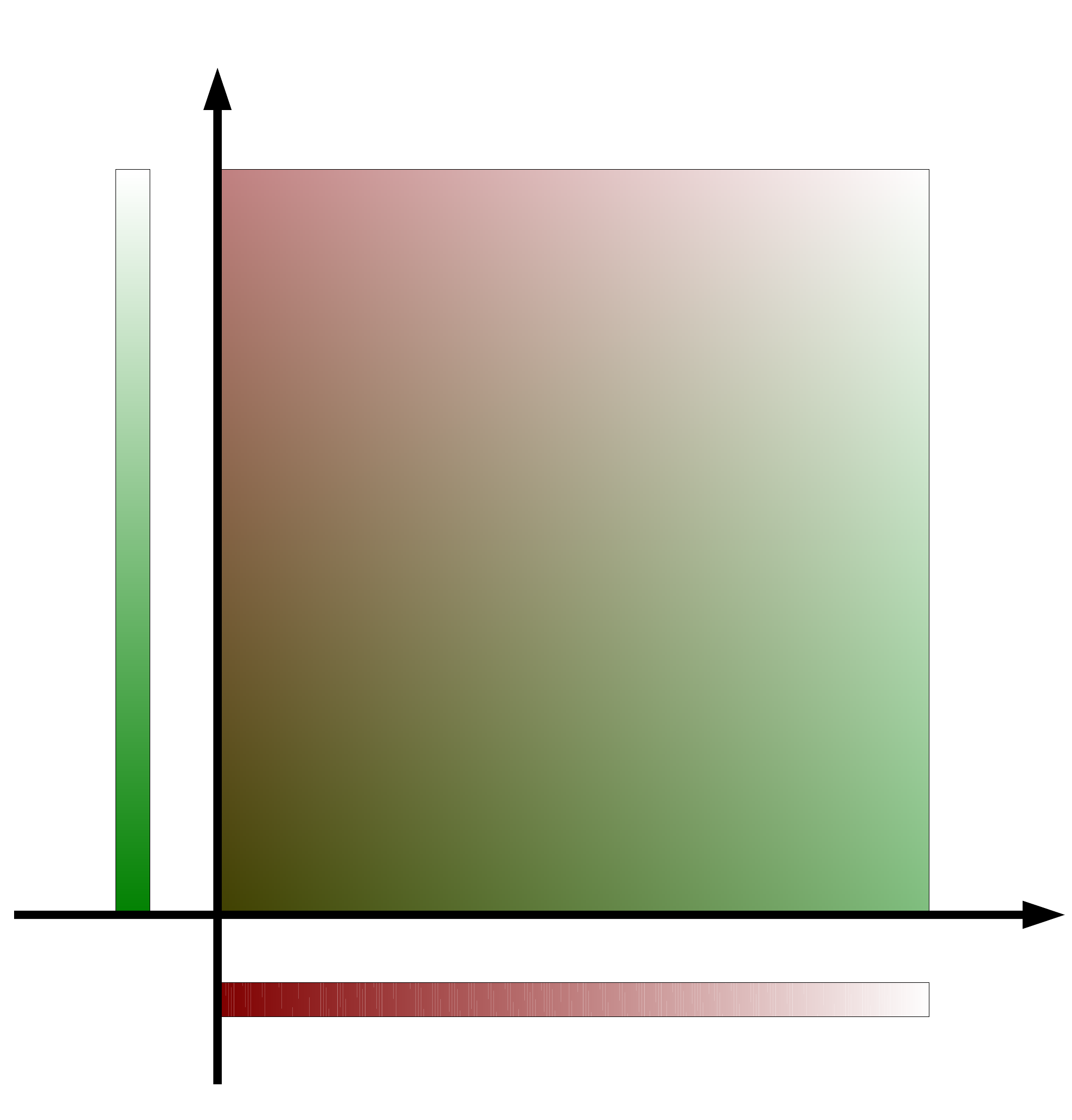}\label{fig:t-prog-ex-coord}}
\subfigure[Forming a T-shaped structure]
{\includegraphics[scale=0.28]{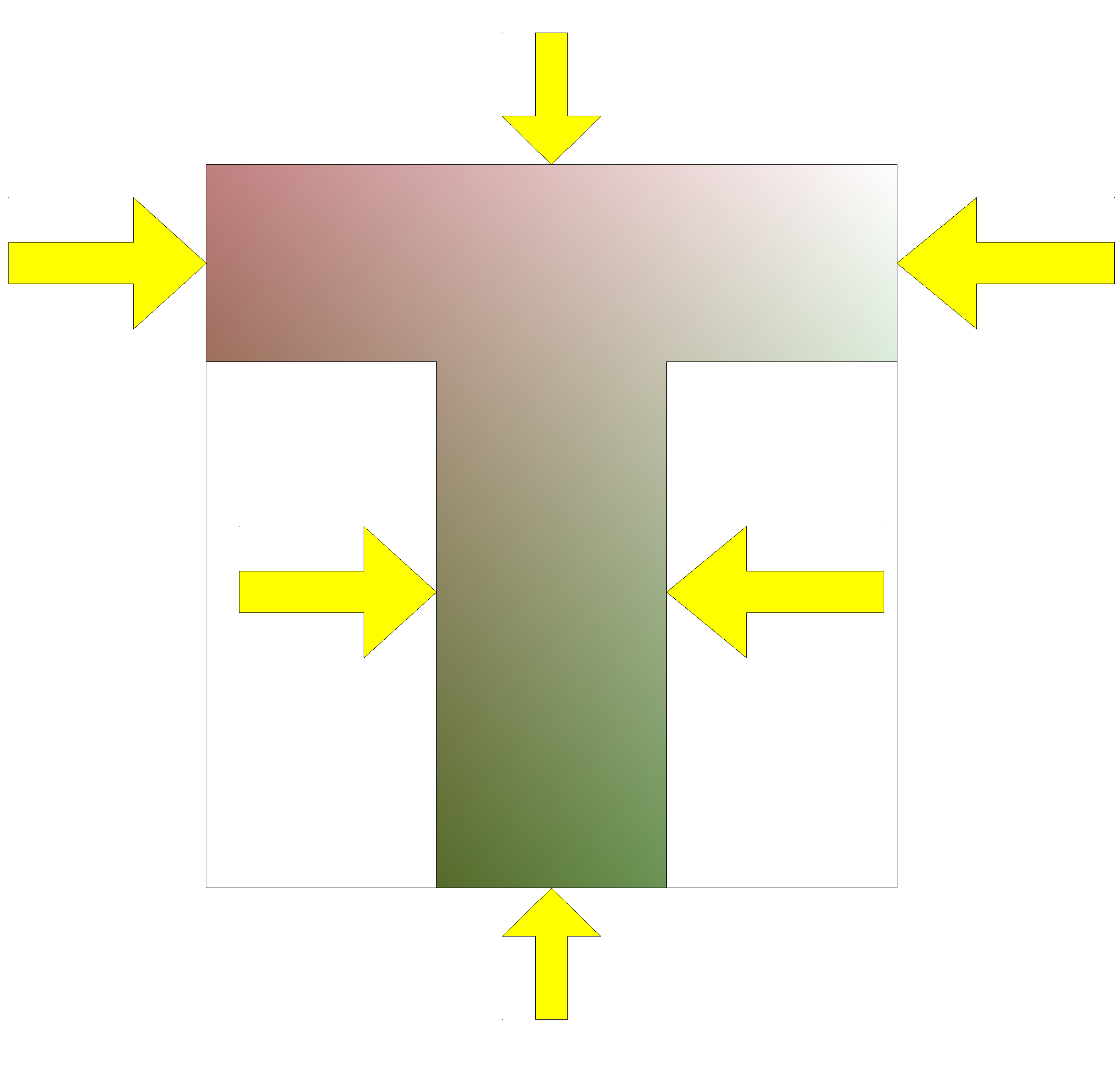}\label{fig:t-prog-ex-make-t}}
\subfigure[Ring around center of gravity]
{\hspace{0.08in}\includegraphics[scale=0.28]{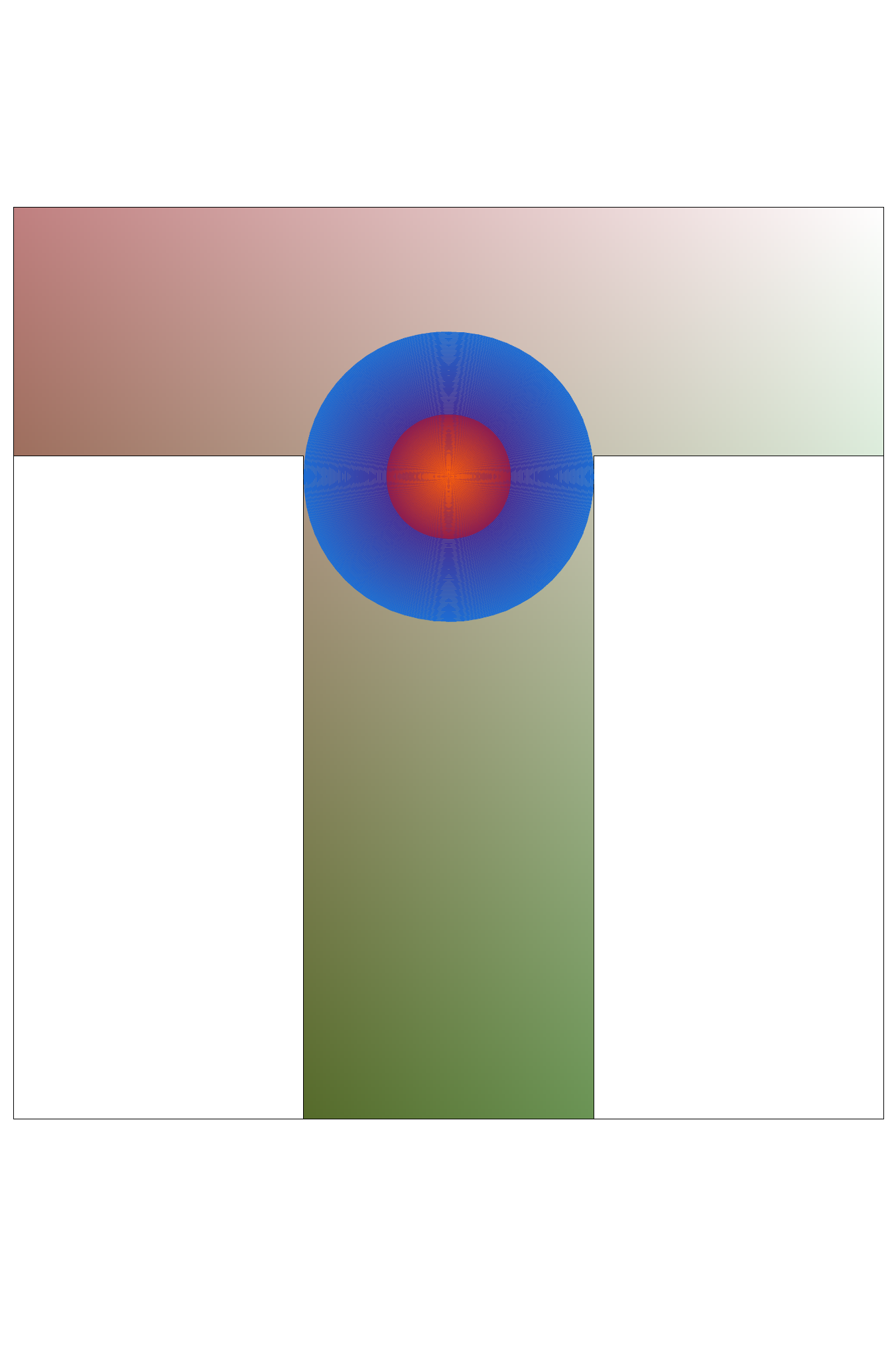}\label{fig:t-prog-ex-find-cg}\hspace{0.08in}}
\caption{The ``T program'' reference example exercises the three main
classes of space-time operations: 
measurement of space-time to organize local coordinates (a) and compute
the center of gravity (c), 
manipulation of space-time to move devices into a T-shaped structure (b), and 
pattern computation to make a ring around the center of gravity (c).}
\label{f:tprogram}
\end{figure}

The three stages of the ``T-program'' are illustrated in
Figure~\ref{f:tprogram}:
\begin{itemize}
\item Cooperatively create a local coordinate system
  (Figure~\ref{fig:t-prog-ex-coord}).  This requires measurement of
  space-time.
\item Move or grow devices to create a T-shaped structure
  (Figure~\ref{fig:t-prog-ex-make-t}).  This requires manipulation of
  space-time.
\item Compute the T's center of gravity and draw a ring pattern around
  it (Figure~\ref{fig:t-prog-ex-find-cg}).  This requires measurement
  of space-time and pattern computation.
\end{itemize}
For purposes of this example, these may happen in any order, including
concurrently.  This simple challenge will show how various exemplary
languages approach the three basic categories of space-time operations
in programs.  Meta-operations are not required, but we will either
illustrate or discuss them as well.

% \subsection{Physical Platform Properties}

% dominant constraints: energy, program size, time, bandwidth
% communication model: wireless, wired, diffusive chemical, physical interlock

\input{dsl-table}

\input{spatial-table}

\input{device-table}

\section{Survey of Existing Spatial DSLs}
\label{s:survey}

We now apply our analytic framework in a survey of spatial computing
DSLs, organizing our survey roughly by major domains.  Note that the
boundaries of domains are somewhat fuzzy, and in some cases we have
placed a language in one domain or another somewhat arbitrarily.

We begin with two domains where the goals are often explicitly
spatial: amorphous computing (Section~\ref{s:amorphous}) and
biological modeling and design (Section~\ref{s:biological}).  We then
discuss the more general area of agent-based models
(Section~\ref{s:agent}), followed by four application domains that are
being driven towards an embrace of spatiality by the nature of their
problems: wireless sensor networks (Section~\ref{s:wsn}), pervasive
systems (Section~\ref{s:pervasive}), swarm and modular robotics
(Section~\ref{s:robotics}), and parallel and reconfigurable computing
(Section~\ref{s:parallel}).  Finally, we survey a few additional
computing formalisms that deal with space explicitly
(Section~\ref{s:formal}).  A theme that we will see emerge throughout
this discussion is that DSLs throughout these domains are often torn
between addressing aggregate programming with space-time operators and
addressing other domain-specific concerns, particularly so in the four
application domains surveyed.

To better enable an overall view of the field and comparison of
languages, we have collected the characteristics of the most
significant DSLs or classes of DSLs in three tables, as derived from
our analytic framework.  Table~\ref{table:dsl} identifies the general
properties of the DSL, Table~\ref{table:spatial} identifies the
classes of space-time operations that each DSL uses to raise its
abstraction level from individual devices toward aggregates, and
Table~\ref{table:device} identifies how each DSL abstracts devices and
communication.
Note that for purposes of clarity, many of the languages discussed are
not listed in these tables, only those that we feel are necessary in
order to understand the current range and capabilities of spatial
computing DSLs.

%%%%%%%%%%%%%%%%%%%%%%%%%%%%%%%%%%%%%%%%%%%%%%%%%%%%%%%%%%%%%%%%%%%%%%%%%%%%%%%
%%%%%%%%%%%%%%%%%%%%%%%%%%%%%%%%%%%%%%%%%%%%%%%%%%%%%%%%%%%%%%%%%%%%%%%%%%%%%%%

\subsection{Amorphous Computing}
\label{s:amorphous}

%%%%%%%%%%%%%%%%%%%%%%%%%%%%%%%%%%%%%%%%%%%%%%%%%%%%%%%%%%%%%%%%%%%%%%%%%%%%%%%
%%%%%%%%%%%%%%%%%%%%%%%%%%%%%%%%%%%%%%%%%%%%%%%%%%%%%%%%%%%%%%%%%%%%%%%%%%%%%%%

%%%%%%%%%%%%%%%%%%%%%%%%%%%%%%%%%%%%%%%%%%%%%%%%%%%%%%%%%%%%%%%%%%%%%%%%%%%%%%%
%% Description of field
%%%%%%%%%%%%%%%%%%%%%%%%%%%%%%%%%%%%%%%%%%%%%%%%%%%%%%%%%%%%%%%%%%%%%%%%%%%%%%%
Amorphous computing is the study of computing systems composed of
irregular arrangements of vast numbers of unreliable, locally
communicating simple computational devices.  The aim of this research
area is to deliberately weaken many of the assumptions upon which computer
science has typically relied, and to search for engineering
principles like those exploited by natural systems.
%%%%%%%%%%%%%%%%%%%%%%%%%%%%%%%%%%%%%%%%%%%%%%%%%%%%%%%%%%%%%%%%%%%%%%%%%%%%%%%
%% Breakdown of types of languages in the field
%%%%%%%%%%%%%%%%%%%%%%%%%%%%%%%%%%%%%%%%%%%%%%%%%%%%%%%%%%%%%%%%%%%%%%%%%%%%%%%
Amorphous computing languages fall into two general categories:
pattern languages and manifold programming languages.

%%%%%%%%%%%%%%%%%%%%%%%%%%%%%%%%%%%%%%%%%%%%%%%%%%%%%%%%%%%%%%%%%%%%%%%%%%%%%%%
%% 1 sub-sub-section for each language type
%%%%%%%%%%%%%%%%%%%%%%%%%%%%%%%%%%%%%%%%%%%%%%%%%%%%%%%%%%%%%%%%%%%%%%%%%%%%%%%
\subsubsection{Pattern Languages}
The majority of the languages that have emerged from amorphous
computing have been focused on the formation of robust patterns.  The
most well known of these are Coore's Growing Point Language
(GPL)~\citep{coorethesis} and Nagpal's Origami Shape Language
(OSL)~\citep{nagpal}.  The Growing Point Language is based on a
botanical metaphor and expresses a topological structure in terms of
``growing points'' that build a pattern by incrementally passing
activity through space and ``tropisms'' that attract or repel the
motion of growing points through simulated chemical signals.  The
combination of these two primitives allows the programmer to specify a
graph and its order of development.  GPL is capable of creating
arbitrarily large and complex topological patterns~\citep{gayleCoore},
and has been used for general geometric constructions as
well~\citep{dhondtBS1,dhondtBS2}.  

The Origami Shape Language is the complement, for geometric rather
than topological patterns.  In OSL, the programmer imperatively
specifies a sequence of folds, with the catalog of possible folds
taken from Huzita's axioms for origami~\citep{huzita}.  These are then
compiled into local programs such that, given an initial
identification of edges, the local interactions will compute the
desired fold lines, eventually producing the specified shape.  Like
GPL, OSL is tolerant of changes in its conditions of execution, with
distorted initial conditions producing a similarly distorted final
pattern.
%\todo[inline]{Cite Rus programmable matter origami?}

A third similar language, the Microbial Colony Language
(MCL)~\citep{weissMCL}, is a rule-based pattern language that includes
chemical signals that diffuse over space to produce patterns: the
programmer specifies the range of propagation and the length of time
that the signal will persist.  Its level of abstraction is
significantly lower than OSL or GPL, for the reason that it was
intended to hew more closely to biological realizability.

Closely related, though more a matter of cellular automata than
spatial computing, was the pattern language established by
Yamins~\citep{yamins}.  Investigating what types of patterns could be
achieved with local communication and finite state, he was able to
establish a constructive proof system that, for any pattern, either
generates a self-stabilizing program for generating that pattern or
proves the pattern is impossible to create with finite state.  While
the pattern elements were not assembled into a full language, they
have a sufficiently broad collection of primitives, possess a means of
composition, and serve as a de facto language in \citep{yamins}.

For the languages mentioned thus far, the behavior is restricted to
the patterning of a pre-existing medium.  Kondacs~\citep{kondacs}
extended this concept with a ``bitmap language'' that takes a two-dimensional
shape and decomposes it into a covering network of overlapping
circles.  The shape can then grow from any fragment: as they grow, the
circles establish overlapping coordinate systems that link to form the
shape as a whole.

Separating the formation of the pattern from the actuation, Werfel has
created a number of systems for collective construction of two- and
three-dimensional structures~\citep{werfeletal, werfelphd}.  These
systems use mixtures of local rules and reaction to environmental
state to enable a group of robots to effectively collaborate in
construction, and have been implemented with real hardware.  While the
initial forms were all ``bitmap languages,'' based on regular grids,
some recent work has generalized to include a constraint programming
system that generates adaptive patterns on the
fly~\citep{werfelAdaptiveConstruction}.  Bitmap languages
are frequently found in modular robotics as well, and to a lesser
extent in swarm robotics (see Section~\ref{s:robotics}).

A notably different approach is taken by Butera's paintable
computing~\citep{butera}, which is also often applied to pattern
formation (e.g., self-organizing text and graphics
display~\citep{ButeraText}).  The programming model is considerably
lower-level, however, uses general computation over shared
neighbor data and viral propagation of code and data, and thus has the
potential to be used for general parallel computing.

\subsubsection{Manifold Programming Languages}

On the opposite end of the spectrum, amorphous computing has also
given birth to more general languages for spatial computing.  Chief
among these is Proto~\citep{proto2006a, mitproto}, a purely functional
language with a LISP-like syntax.  Proto uses a continuous space
abstraction called the amorphous medium~\citep{AmorphousMedium} to view
any spatial computer as an approximation of a space-time manifold with
a computing device at every point.  Information flows through this
manifold at with a bounded velocity, and each device has access to the
recent past state of other devices within a nearby neighborhood.
Proto primitives are mathematical operations on fields (functions that
associate each point in space-time with a value) and come in four
types: pointwise ``ordinary'' computations (e.g., addition),
neighborhood operations that imply communication, feedback operations
that establish state variables, and restriction operations that
modulate a computation by changing its domain.  Proto programs
interact with their environment through sensors and actuators that
measure and manipulate the space occupied by devices: for example,
Proto has been applied to swarm robotics by computing vectors fields,
which are then fed to a movement actuator that interprets them as
continuous mass flow and moves devices to
approximate~\citep{ProtoSwarm}.

Two derivatives have since forked off of the Proto project,
Gas~\citep{gasPL} and PyMorphous~\citep{PyMorphous}.  Gas is very
closely related to Proto, mostly just changing its syntax and adding
some new sensors and actuators.  PyMorphous is a relatively new and
more ambitious project, aimed at producing an imperative language
equivalent of Proto, piggybacked onto Python as a library.  Besides
these, the compilation target for Proto is itself a domain specific
language, an assembly language for a stack-based virtual machine
model~\citep{protokernel} that serves as a common reference point for
platform-specific implementations of the amorphous medium model.

%%%%%%%%%%%%%%%%%%%%%%%%%%%%%%%%%%%%%%%%%%%%%%%%%%%%%%%%%%%%%%%%%%%%%%%%%%%%%%%
%% Reference Example
%%%%%%%%%%%%%%%%%%%%%%%%%%%%%%%%%%%%%%%%%%%%%%%%%%%%%%%%%%%%%%%%%%%%%%%%%%%%%%%
\subsubsection{Reference Example: Proto}\label{sec:protoexample}

\begin{figure}[t]
\centering
\subfigure[Initial distribution showing network links.]{\includegraphics[scale=0.12]{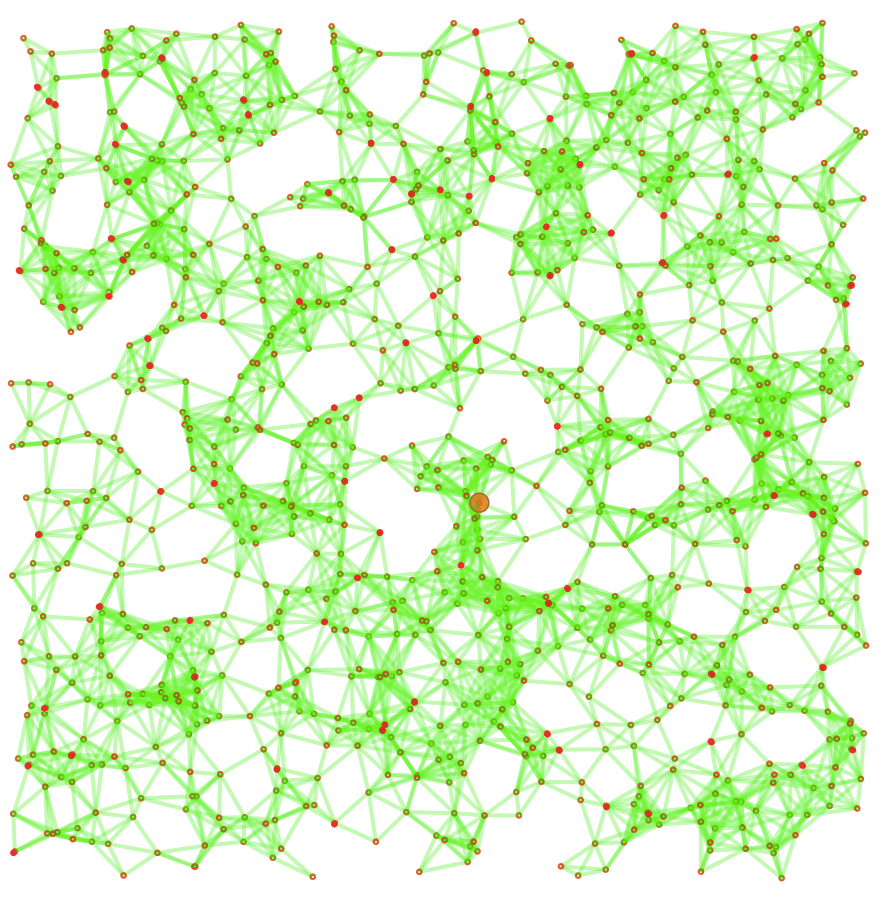}}
%\subfigure[$t = 40$s.]{\includegraphics[scale=0.10]{pt-40.png}}
\subfigure[$t = 65$s.]{\includegraphics[scale=0.12]{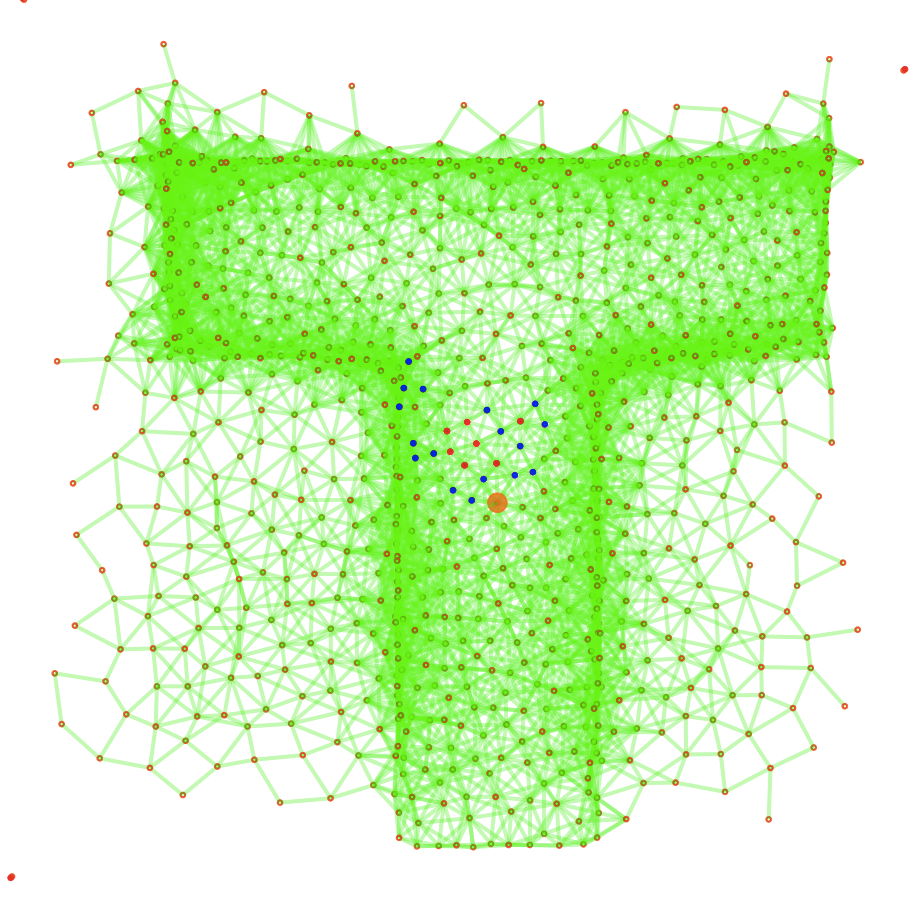}}
%\subfigure[$t = 80$s.]{\includegraphics[scale=0.10]{pt-80.png}}
%\subfigure[$t = 100$s.]{\includegraphics[width=0.3\textwidth]{pt-100.png}}
\subfigure[$t = 135$s.]{\includegraphics[scale=0.12]{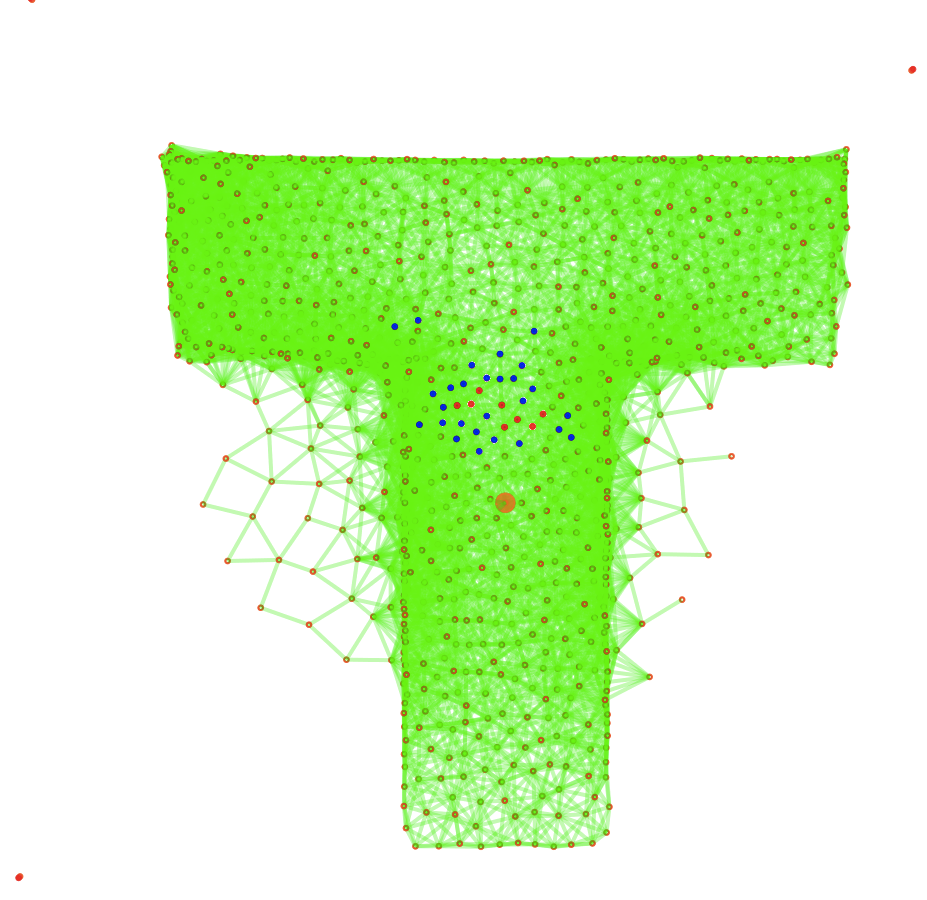}}
%\subfigure[Caption 8]{\includegraphics[width=0.3\textwidth]{proto-t-white-8.png}}
\subfigure[Stabilized center-of-gravity.]{\includegraphics[scale=0.12]{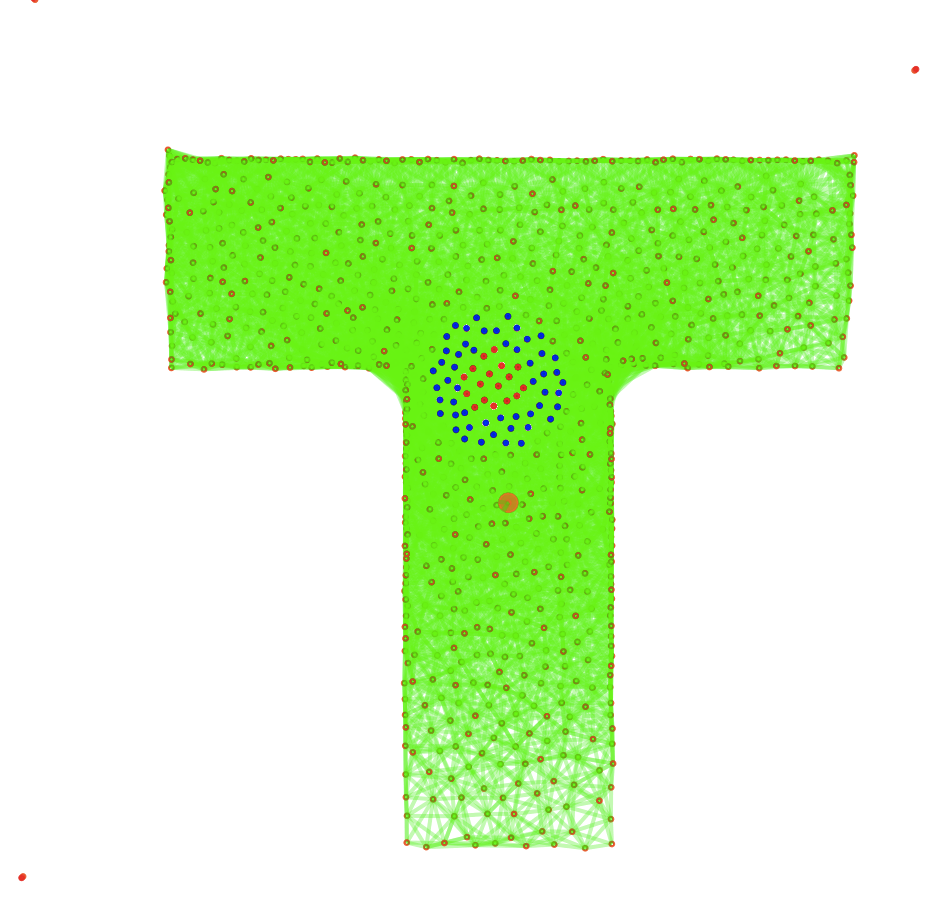}}
\caption{Creating a local coordinate system, moving into a ``T'' shape, and
finding its center of mass in Proto.  Figure~\ref{fig:proto-3d-t} shows the
same program running in three-dimensional space.}
\label{f:proto-t-program}
\end{figure}

Proto works by compiling a high-level program (written in the Proto
programming language) into a local program that is executed by every node in
the network.  The local program is executed in the Proto Virtual Machine
(VM) which runs on a variety of platforms, including a simulator (shown in
Figure~\ref{f:proto-t-program}).

The high-level Proto program for our ``T'' reference example can be
launched with the command {\tt (t-demo (sense 1))}.  This allows the user
to select, via a generic ``test'' sensor, the node to select as the origin
of the coordinate system.  The origin selection is largely inconsequential
as any node can be selected as the origin, however selecting a node close
to the center of the space requires less overall node movement.  The {\tt
t-demo} function makes use of several other functions, however their
separation is useful for code reuse and general abstraction purposes.

\DefineVerbatimEnvironment
{CodeBlock}{Verbatim}
{frame=single, fontsize=\footnotesize, samepage=true}

\begin{CodeBlock}
(def t-demo (origin)
   (let* (
          ;; establish local coordinates
          (c (abscoord origin)) 
          ;; find center-of gravity
          (cg (compute-cg origin c)) 
          ;; compute distance to center-of-gravity
          (dist-to-cg (vlen (- c cg))))
     ;; make nodes move...
     (mov (mux origin (tup 0 0 0)  ;; origin does not move
            (+ (vmul 0.2 (disperse)) ;; move away from each other
               (normalize (make-t (vmul -0.005 c)))))) ;; make "T"
     ;; turn on "bullseye" pattern
     (red (< dist-to-cg 10))
     (blue (and 
              (> dist-to-cg 10) 
              (< dist-to-cg 20)))))
\end{CodeBlock}

One useful sub-function is {\tt abscoord}, which establishes a local
coordinate system around an origin node.  This is accomplished by finding the
vector from every node to the origin node using the {\tt vec-to} function.
This implementation for finding the vector distance from one node to
another uses a self-healing distance computation (part of the core library of Proto
functions) in aggregating the vector values between all the nodes along the
path between the source (origin) and the destination (node seeking to find
its coordinates relative to the origin).

%% now talk about how this pushes information over multiple hops...  later
%the tree-aggregate function does a similar thing.
This portion of the program utilizes a common paradigm in Proto, embedding
a neighborhood operation (e.g., {\tt sum-hood}) inside a feedback operation
(e.g., {\tt rep}).  This paradigm allows data to be shared over multiple
network hops rather than just direct network neighbors (as neighborhood
operations allow).

\begin{CodeBlock}
(def vec-to (src)
     ;; establish a field of distances from the source nodes
     (let* ((d (distance-to src))
            ;; parent has the smallest distance to the source
            (parent (2nd (min-hood (nbr (tup d (mid)))))))
       ;; value is the sum of all vectors along the path 
       ;; of parents to the source node
       (rep value (tup 0 0 0)
            (mux src (tup 0 0 0)
                 (sum-hood (mux (= (nbr (mid)) parent)
                                (+ (nbr-vec) (nbr value))
                                (tup 0 0 0)))))))

(def abscoord (origin)
     (vec-to origin))
\end{CodeBlock}

The method used to compute the center of mass also makes use of the
feedback-neighborhood paradigm.  The {\tt tree-aggregate} function
constructs a tree from the node passed-in as the {\it root} argument.  It
then traverses the tree, aggregating the {\it value} parameter along the
path.  The {\tt compute-cg} function broadcasts the sum of all agents'
coordinates (in each dimension) by the total number of agents in the
network, both of which are calculated using the {\tt tree-aggregate}
function.

\begin{CodeBlock}
(def tree-aggregate (root value default-val)
     ;; establish a field of distances from the source nodes
     (let* ((d (distance-to root))
            ;; parent has the smallest distance to the source
            (parent (2nd (min-hood (nbr (tup d (mid)))))))
       ;; sumval is the sum of all values along the path 
       ;; of parents to the root node
       (rep sumval value
            (+ value (sum-hood (mux (and 
                                      (not (nbr root))
                                      (= (mid) (nbr parent)))
                                     (nbr sumval)
                                     default-val))))))

(def compute-cg (root coordinates)
     ;; center-of-gravity = sum of coordinate / number of devices
     (broadcast root (vmul
                       (/ 1 (tree-aggregate root 1 0))
                       (tree-aggregate root 
                                       coordinates
                                       (tup 0 0 0)))))
\end{CodeBlock}

The {\tt make-t} function is used to move the nodes into a ``T'' shape.  By
applying the spatial constraints, as done in {\tt x-constraints}, 
{\tt y-constraints}, and {\tt z-constraints}; the sum of the vectors
returned by each function defines the direction (and speed) in which each
node moves.

\pagebreak
\begin{multicols}{3}
\begin{CodeBlock}
(def x-constraints (x y z)
     ;; skinny part of T
     (if (< y 0.20)
       (if (< x -0.15)
         (tup 1 0 0)
         (if (> x 0.15)
           (tup -1 0 0)
           (brownian)))
       ;; top part of T
       (if (< x -0.50)
         (tup 1 0 0)
         (if (> x 0.50)
           (tup -1 0 0)
           (brownian)))))
\end{CodeBlock}

\begin{CodeBlock}
(def y-constraints (x y z)
     (if (< y -0.50)
       (tup 0 1 0)
       (if (> y 0.50)
         (tup 0 -1 0)
         (brownian))))

(def z-constraints (x y z)
     (if (< z -0.50)
       (tup 0 0 1)
       (if (> z 0.50)
         (tup 0 0 -1)
         (brownian))))
\end{CodeBlock}

\begin{CodeBlock}
(def make-t (pos)
     (let ((x (1st pos)) 
           (y (2nd pos)) 
           (z (3rd pos)))
       (+ (x-constraints x y z)
          (y-constraints x y z)
          (z-constraints x y z))))
\end{CodeBlock}
\end{multicols}

\begin{figure}[t]
\centering
\includegraphics[height=0.3\textheight]{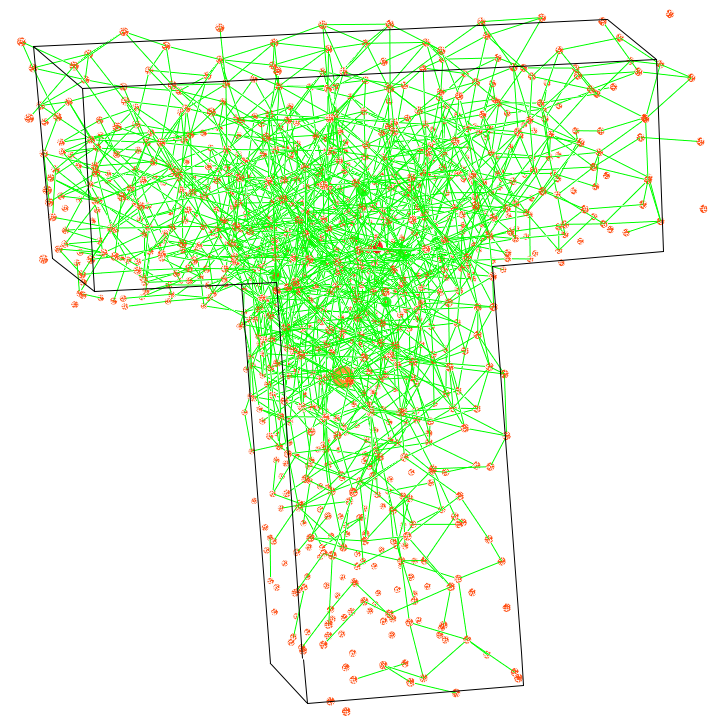}
\caption{The Proto reference example ``T'' program running in
three-dimensional space.}
\label{fig:proto-3d-t}
\end{figure}

There are a few desirable properties of this Proto program.
% 3D
Like most other Proto programs, in addition to running in two-dimensional
space, it also runs in three-dimensional space with no additional code (as
shown in Figure~\ref{fig:proto-3d-t}).
% creates a local program
Additionally, all Proto programs are first compiled to local programs that
execute in a portable Proto VM.  Thus, Proto offers a straightforward path
to execution on a real distributed network platform.
% self-adaptive/healing
Finally, there are several tools in Proto's core library that help in
designing and constructing self-healing distributed algorithms.  For
example, by simply using the naturally self-healing {\tt distance-to} function
in the implementation of the reference example, our ``T'' program inherits
this desirable property.

%%%%%%%%%%%%%%%%%%%%%%%%%%%%%%%%%%%%%%%%%%%%%%%%%%%%%%%%%%%%%%%%%%%%%%%%%%%%%%%
%% Connection to table and framework
%%%%%%%%%%%%%%%%%%%%%%%%%%%%%%%%%%%%%%%%%%%%%%%%%%%%%%%%%%%%%%%%%%%%%%%%%%%%%%%
\subsubsection{Analysis}

The characteristics of key amorphous computing DSLs are summarized in
Table~\ref{table:dsl},~\ref{table:spatial},~and~\ref{table:device},
based on the taxonomy proposed in Section~\ref{s:definitions}.

Given the goals of amorphous computing, it is unsurprising that we
find that nearly all of the languages in this domain address the
challenges of global-to-local compilation and producing predictable
aggregate behaviors from the actions of individual devices.  The
pattern formation languages use a wide variety of high-level
representations, but are ultimately limited in scope: different types
of patterns are difficult to mix together and cannot be generally be
cleanly composed.

Proto and its derivatives do not inherently provide the programmer
with quite as high a level of abstraction: the primitive operations of
Proto only deal with local neighborhoods in space-time.  Because Proto
has a functional semantics defined in terms of aggregate operations,
however, standard library functions like {\tt distance-to} and {\tt
  broadcast} can fill that gap, acting as though they themselves were
primitives.  These and other functions like them thus provide the
programmer with a toolkit of aggregate-level space-time operators.
This is a critical ingredient in the fairly general applicability of
Proto, which may be noted in its appearance below in the discussion of
biological (Section~\ref{s:biological}) and robotic
(Section~\ref{s:robotics}) domains.

%%%%%%%%%%%%%%%%%%%%%%%%%%%%%%%%%%%%%%%%%%%%%%%%%%%%%%%%%%%%%%%%%%%%%%%%%%%%%%%
%%%%%%%%%%%%%%%%%%%%%%%%%%%%%%%%%%%%%%%%%%%%%%%%%%%%%%%%%%%%%%%%%%%%%%%%%%%%%%%

\subsection{Biological Modeling and Design}
\label{s:biological}

%%%%%%%%%%%%%%%%%%%%%%%%%%%%%%%%%%%%%%%%%%%%%%%%%%%%%%%%%%%%%%%%%%%%%%%%%%%%%%%
%%%%%%%%%%%%%%%%%%%%%%%%%%%%%%%%%%%%%%%%%%%%%%%%%%%%%%%%%%%%%%%%%%%%%%%%%%%%%%%

%%%%%%%%%%%%%%%%%%%%%%%%%%%%%%%%%%%%%%%%%%%%%%%%%%%%%%%%%%%%%%%%%%%%%%%%%%%%%%%
%% Description of field
%%%%%%%%%%%%%%%%%%%%%%%%%%%%%%%%%%%%%%%%%%%%%%%%%%%%%%%%%%%%%%%%%%%%%%%%%%%%%%%
Natural biological systems often have strong locality and spatial
structure, from biofilms of single-celled organisms to the tissues of
large multicellular animals.  This spatiality is reflected in a number
of the languages that have been developed for modeling biological
systems, as well as some of the new languages that are beginning to
emerge in synthetic biology for the design of new biological
organisms.

\subsubsection{Modeling Languages}
A number of biological modeling languages have
been developed, such as Antimony~\citep{antimony},
ProMoT~\citep{ProMoT}, iBioSim~\citep{iBioSim}, and little
b~\citep{littleB}, which allow the bio-molecular reactions of cells to
be described at a somewhat higher level of abstraction.  These
generally include some spatial operations as well, in the form of
compartments through which chemicals can pass (which
can include movement from cell to cell).  These notions have been
further generalized and formalized as with P-systems and the Brane calculus
(discussed below in Section~\ref{s:formal})
The space in these cases, however, is generally extremely abstract.

More explicitly spatial are L-systems~\citep{LSystems} and
MGS~\citep{GiavittoMGS02, GiavittoMGS04}.  L-systems are a
graph-rewriting model used to model the growth and structure of
plants, and are specified in terms of rules for modification of local
geometric structures.  MGS has a somewhat similar approach, but allows
fully general rule-based computation on the much more spatially
general structure of topological complexes---it has actually been
applied much more widely than biological modeling, but biology has
been both an important inspiration and application area for MGS.  MGS
programs operate both by manipulating values locally and by
topological surgery to modify the local structures.  Coupled with a
physics model that adjusts the geometry, this has allowed MGS to
express complex models of biological phenomena with elegant
simplicity.

Another recent addition is Gro~\citep{Gro}, a Python-like language
designed for stochastic simulation of genetic regulatory networks in a
growing colony of E.~coli.  Gro includes built-in notions of chemical
reaction rates, diffusive communication, and cell growth, and allows
the programmer to construct arbitrary chemical models to control them.

\subsubsection{Synthetic Biology Languages}

More recently, the field of synthetic biology has been applying
computer science approaches to the design of engineered biological
organisms.  Of the few high-level languages that have so far emerged,
only two include spatial operations.

GEC~\citep{GEC} is a logical programming language where the programmer
describes a biological circuit in terms of design constraints.  The
spatial aspects of this language are extremely minimal: as with most
modeling languages, they only deal with motion of molecules from
compartment to compartment.

A biology-focused version of Proto~\citep{AC4GRN}, the Proto BioCompiler, has been applied in
this space as well, using a chemical diffusion model of communication
rather than local message passing.  Here, an extension to the language
associates Proto primitives with genetic regulatory network motifs,
allowing Proto programs to be compiled into genetic regulatory network
designs instead of virtual machine code.  The range of Proto constructs
that can be mapped to biological constructs at present, however, is 
fairly limited.

Other high-level biological design languages, however, such as
Eugene~\citep{Eugene} and GenoCAD~\citep{GenoCAD}, do not currently have
any spatial language constructs at all.

%Neural/ANN, and why it's not really...

%%%%%%%%%%%%%%%%%%%%%%%%%%%%%%%%%%%%%%%%%%%%%%%%%%%%%%%%%%%%%%%%%%%%%%%%%%%%%%%
%% Reference Example
%%%%%%%%%%%%%%%%%%%%%%%%%%%%%%%%%%%%%%%%%%%%%%%%%%%%%%%%%%%%%%%%%%%%%%%%%%%%%%%
\subsubsection{Reference Example: MGS}\label{s:bioexample}

MGS is a biological modeling language that allows general rule-based 
computation on topological complexes.
In this example, we illustrate MGS's spatial approach to creating a ``T''
shape for our reference example ``T-Program.''

%\begin{itemize}
%\item the first is based on cellular automata (a regular lattice), 
%\item the second is based on ``proximal'' collection that corresponds to an 
%  amorphous medium, and
%\ite the third develops the shape in an empty space, using cellular
%  complex (in 2D).
%\end{itemize}

%We do not elaborate very much on the acquisition of the "vertical" and
%the "horizontal" direction. We have made the assumption that the
%elaboration of such information of position is not central in your
%example.

%The following code segment is shared between all three models. 
The following code segment describes the local state of an entity. 
This entity will interact with the other entities in its neighborhood.
Interactions are specified using {\it transformations}, a kind of ``rewriting'' 
of the spatial structure, which is composed of the local state of all the entities.
The ``T'' shape grows starting from one (or few)
entities in two successive phases. In the first growth phase ({\tt FGP}), the growth
process follows a ``vertical'' direction. In the second growth phase ({\tt SGP}), the
two horizontal segments of the ``T'' grow in parallel.
The functions {\tt NextFGP} and {\tt NextSGP} are used to evolve the {\tt
Cell} type's counters {\tt cpt1} and {\tt cpt2}.  We start from an initial
state called {\tt seed}.

%A state is simply a data. In this case, the state is represented as a
%record that contains two counters cpt1 and cpt2. The first is used to
%control the duration of the first growth phase (the vertical part of the
%T) and cpt2 the duration of the second growth phase. 
%
%MGS is dynamically typed. However, the declaration of a type defines in
%the background:
%
%- a unary predicate (with the same name) that can be used to check that
%a avalue is of a given type
%
%- some services such as constructors, pretty-printing, etc.

\begin{CodeBlock}
//***************************************************************
// Basic growth model

record Cell = { cpt1, cpt2 } ;;
record FGP = cell + { cpt1 != 0 } ;;
record SGP = cell + { cpt1 = 0, cpt2 != 0 } ;;
record Empty = { ~cpt1, ~cpt2 } ;;
fun NextFGP (c:FGP) = c + { cpt1 = c.cpt1-1 } ;;
fun NextSGP (c:SGP) = c + { cpt2 = c.cpt2-1 } ;;

let seed = { cpt1 = 5, cpt2 = 3 } ;;
\end{CodeBlock}

The spatial structure underlying this object is a cellular complex: a space
built by aggregating elementary cells. 
We use three types of cells in this example:
\begin{itemize}
\item {\bf 0-cells} are vertices.
\item {\bf 1-cells} are edges. An edge is bound by two vertices.
\item {\bf 2-cells} are surfaces. Here the surfaces are parallelograms bound by 4
edges.  
\end{itemize}

%The boundary relationships organize the space represented by the cellular
%complex. The faces of a $p$-cell $c$ are the $q$-cell, $q < p$, which are in the
%``border'' of $c$. The co-faces of a $p$-cell $c$ are the $q$-cell $c', q > p$, such
%that $c$ is in the faces of $c'$.

The {\tt letcell} construct introduces a recursive definition of
cell relationships. 
%The definition is recursive because if a cell $c$ is the face of
%$c'$, then $c'$ is a co-face of $c$. The {\tt letcell} takes care to complete the
%information needed and for example, in the previous statement, we
%specify only the face of $f$ which implicitly adds $f$ to the co-face of the
%other cells. 
%
The collection specification builds a collection using the cells introduced
by the {\tt let} (and other cells if needed). A collection associates a
value to a cell. 
%The addition is used to amalgamate the various cells. 

\begin{CodeBlock}
//***************************************************************
// Chain implementation
// Spatial specification
let init =
  letcell v1  = new_vertex () 
  and     v2  = new_vertex () 
  and     v3  = new_vertex () 
  and     v4  = new_vertex () 
  and     e12 = new_edge v1 v2 
  and     e23 = new_edge v2 v3 
  and     e34 = new_edge v3 v4 
  and     e41 = new_edge v4 v1 
  and     f   = new_acell 2 (e12,e23,e34,e41) in
    { x = ..., y = ... } * v1 + { x = ..., y = ... } * v2 +
    { x = ..., y = ... } * v3 + { x = ..., y = ... } * v4 +
    `Basal * e12 + `Lateral * e23 + `Apical * e34 + `Lateral * e41 + seed * f
;;
\end{CodeBlock}

\begin{figure}[t]
\centering
\includegraphics[width=.3\textwidth]{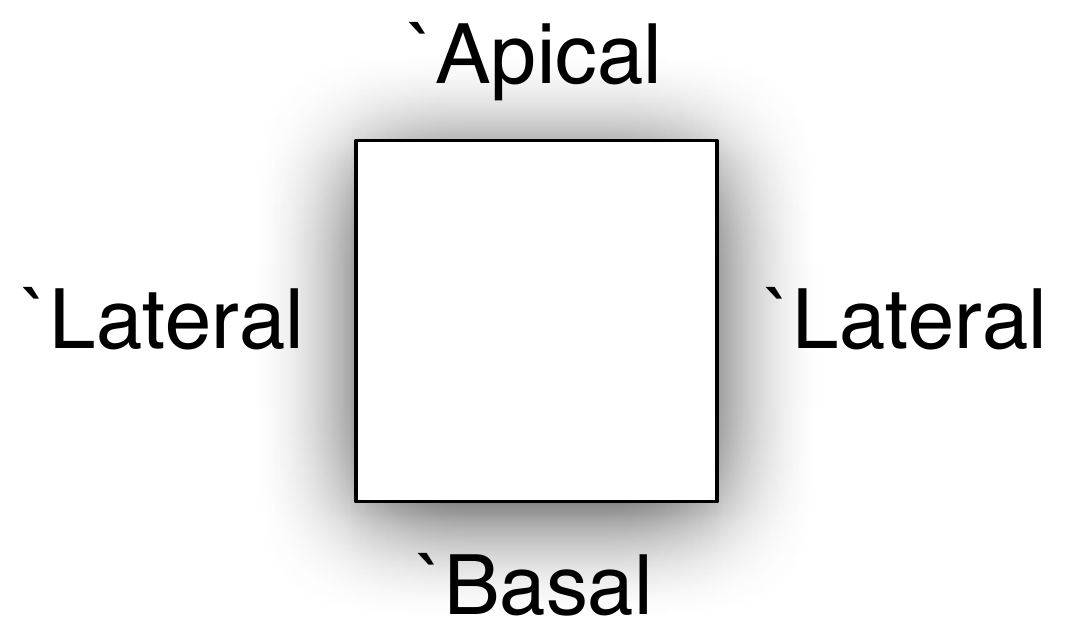}
\caption{Labeled sides of an MGS cell.  During the first growth phase,
growth occurs along the ``Apical'' side; during the second growth
phase, growth occurs along the ``Lateral'' sides.}
\label{f:mgs-sides}
\end{figure}

A record of positions $x$ and $y$ is associated to the vertices. The edges
are labeled by symbols distinguishing three kinds of edges (Apical, Basil,
and Lateral), illustrated in Figure~\ref{f:mgs-sides}.  The idea is that
the growth takes place on the ``Apical'' side during the first growth phase
(FGP) and along the ``Lateral'' sides during the second growth phase (SGP).

Next, we specify the transformation used to compute the mechanics of the
systems. For the sake of simplicity, we use a very simple
mass-spring system and Aristotelician mechanics (that is, the speed is
proportional to force, not acceleration). 
%This does not change
%the results because we are interested in the final steady state, but
%avoid a double integration. It also speed the iterations :-).  
%
%The mass spring systems is very similar to the one sketched by Radika Nagpal
%in her thesis. 
Each edge is a spring with a length of {\tt L0} (at rest) and a strength of
{\tt k}. 

The transformation 
matches only ``2-cells'' (due to the {\tt <2>} after the {\tt trans} keyword).
Then, for each cell, we compute the number of ``0-cells'' on its border.
%(expression "icellssize f 0", "i" stands for "incident"). 
The primitive {\tt icellsfold} is used to iterate over the ``0-cells'' on
each cell's border. 
%The reduction function is used to compute the barycenter of the
%corners of the parallelogram f. 

\begin{CodeBlock}
trans <2> MecaFace[k,L0,dt] = {
  f => (
    let n = icellssize f 0 in
    let g = icellsfold (fun acc v -> { x = acc.x + v.x, y = acc.y + v.y }) { x = 0.0, y = 0.0 } f 0
    in
      f + { x = g.x/n, y = g.y/n }
  )
} ;;
\end{CodeBlock}

The next transformation integrates the forces and updates 
the position of the vertexes accordingly. 
The qualifier {\tt <0,1>} means that we focus on ``0-cells'' and that the
neighborhood considered in the {\tt neighborsfold} operation meets the
following criteria: two ``0-cells'' are neighbors if they border a common
``1-cell'' (i.e., vertices are neighbors if they are linked by an edge).

\begin{CodeBlock}
trans <0,1> MecaVertex[k,L0,dt] = {
  v => (
    let Fspring = neighborsfold (fun acc v' ->
				   let d = sqrt((v'.x - v.x)*(v'.x - v.x) + (v'.y - v.y)*(v'.y - v.y)) in
				   let f = k * (d - L0) / d in
				     { x = acc.x + f*(v'.x-v.x), y = acc.y + f*(v'.y-v.y) }
				) { x = 0.0, y = 0.0 } v
    in
    let Ftot = icellsfold (fun acc g ->
			     let d = sqrt((g.x - v.x)*(g.x - v.x) + (g.y - v.y)*(g.y - v.y)) in
			     let f = k * (d - sqrt(2.0)*L0) / d in
			       { x = acc.x + f*(g.x-v.x), y = acc.y + f*(g.y-v.y) }
			  ) Fspring v 2
    in
      v + { x = v.x + dt*Ftot.x, y = v.y + dt*Ftot.y }
  )
} ;;

fun Meca(ch) = MecaVertex(MecaFace(ch)) ;;
\end{CodeBlock}

\begin{figure}[t]
\centering
\subfigure[Initial cell]{\includegraphics[scale=0.16]{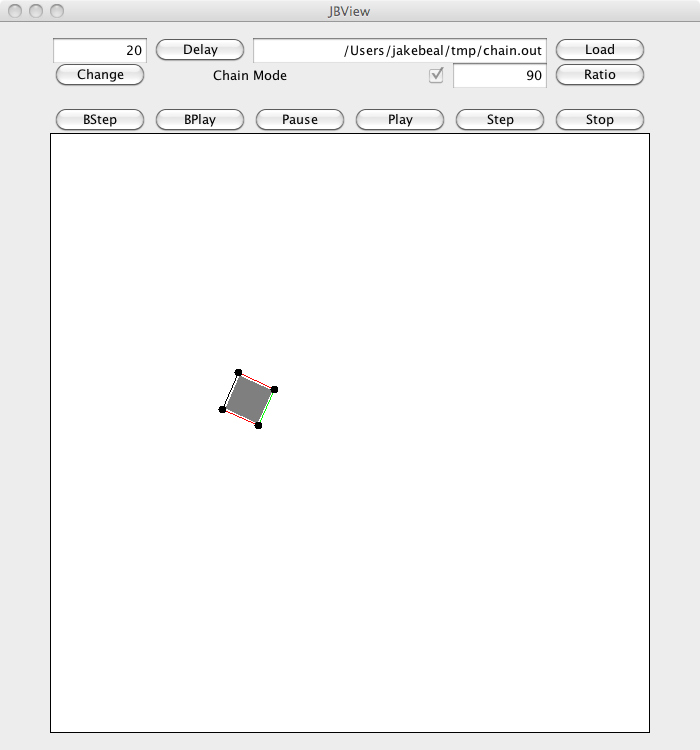}}
\subfigure[Growth of trunk]{\includegraphics[scale=0.16]{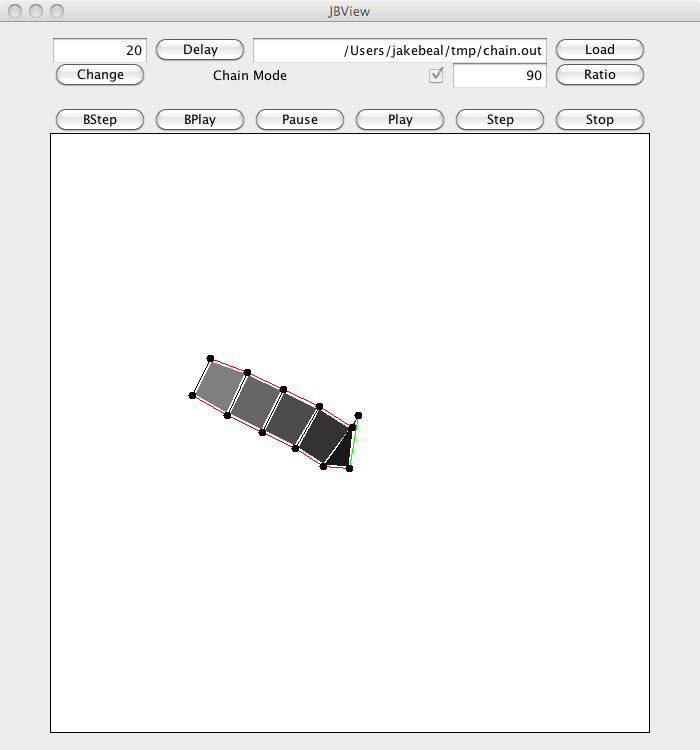}}
\subfigure[Growth of bar]{\includegraphics[scale=0.16]{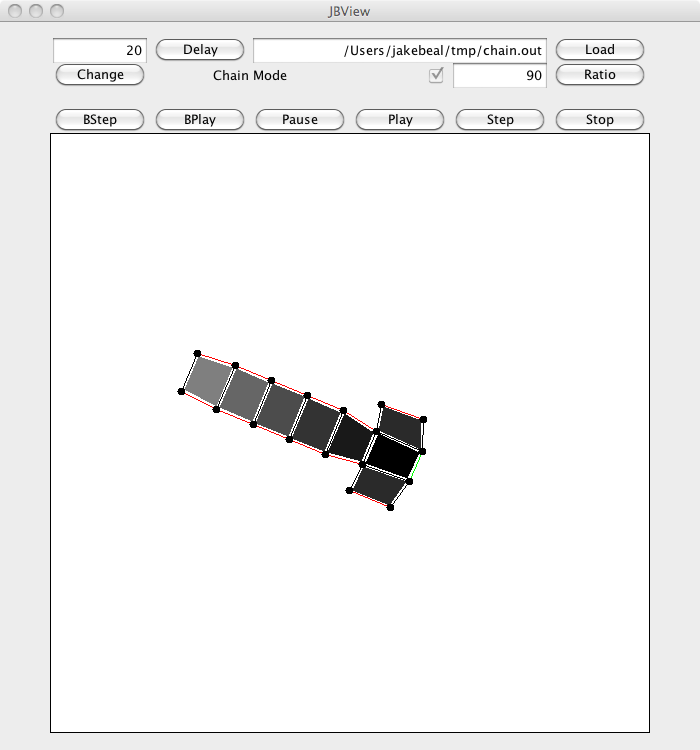}}
\subfigure[Final shape]{\includegraphics[scale=0.16]{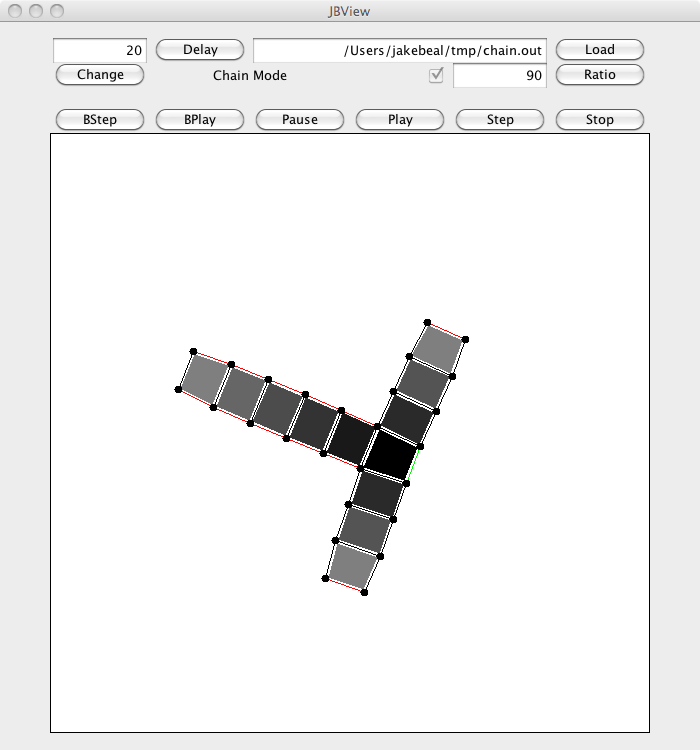}}
\caption{Construction of a ``T'' shape in MGS, beginning with an initial
  cell and using spring forces to distribute cells as they are created
  by topological surgery.}
\label{f:mgs-t-program}
\end{figure}

There is one additional transformation to compute the growth
of the ``T'' shape.
%Here we use more sophisticated pattern constructions.
%
The first evolution rule specifies the ``Apical'' growth during the first
phase.  The second evolution rule describes the growth along the
``Lateral'' edges for the second phase.
%matches one face f, on edge e12 and the two vertices bouding e12. e12
%must be labeled by the symbl `Apical. The vertices and the face remains
%in the result but the edge is removed. 
When the rule fires, it builds several new cells, replaces edge {\tt
e12} with {\tt 'e12} and updates the neighborhood relationships of the
remaining cells.  What is finally built (in both phases) is a new
parallelogram (i.e., a new face {\tt f'}). 
%The edge e12'
%replaces the edge e12 and is now lebeled by the symbol `Internal to prevent
%further development. The "opposite" edge e23 is labeled by `Apical in the
%first growth phase and `Lateral in the second growth phase.
%
Figure~\ref{f:mgs-t-program} shows the evaluation of these rules in the MGS
simulator.

\begin{CodeBlock}
// Evolution rules
patch Rules = {
  ~v1 < e12 < ~f:[dim=2, FGP(f)] > e12 > ~v2 / (e12 == `Apical) => (
    letcell v3  = new_vertex ()
    and     v4  = new_vertex ()
    and     e23 = new_edge v2 v3
    and     e34 = new_edge v3 v4
    and     e41 = new_edge v4 v1
    and     e12' = new_edge v1 v2
    and     f'  = new_acell 2 (e12,e23,e34,e41) in
      ( v2 + { x = v2.x + (v2.x-f.x) * random(0.1), y = v2.y + (v2.y-f.y) * random(0.1) } ) * v3 +
      ( v1 + { x = v1.x + (v1.x-f.x) * random(0.1), y = v1.y + (v1.y-f.y) * random(0.1) } ) * v4 +
      `Internal * e12' + `Lateral * e23 + `Apical * e34 + `Lateral * e41 + (NextFGP f) * f'    
  );

  ~v1 < e12 < ~f:[dim=2, SGP(f)] > e12 > ~v2 / (e12 == `Lateral) => (
    letcell v3  = new_vertex ()
    and     v4  = new_vertex ()
    and     e23 = new_edge v2 v3
    and     e34 = new_edge v3 v4
    and     e41 = new_edge v4 v1
    and     e12' = new_edge v1 v2
    and     f'  = new_acell 2 (e12,e23,e34,e41) in
      ( v2 + { x = v2.x + (v2.x-f.x) * random(0.1), y = v2.y + (v2.y-f.y) * random(0.1) } ) * v3 +
      ( v1 + { x = v1.x + (v1.x-f.x) * random(0.1), y = v1.y + (v1.y-f.y) * random(0.1) } ) * v4 +
      `Internal * e12' + `Basal * e23 + `Lateral * e34 + `Basal * e41 + (Next SGP f) * f'
  );
} ;;

fun Step ch = Rules(Meca* ch) ;;
\end{CodeBlock}

%% Center of Mass and Ring Pattern
The computation of a center of mass can then be implemented directly 
%on the three models 
using a {\tt fold} operator (there is a fold on any kind of
collection in MGS). This {\it fold} is very similar to the {\tt fold} operator in LISP
or other functional languages: it iterates over all the elements of a
collection and propagates an accumulator using a binary reduction function.
%Perhaps you do not like this approach: it is not very local. If you have
%some local algorithm, we will be happy to implement it.

%We do not implement the creation of a ring around the center of
%mass. 
Implementing a ring around the center-of-mass is straightforward once the 
center-of-mass is discovered. This point can diffuse some substance that
degrades over distance. The ring can be selected by the nodes that have a
level of substance in a given interval.
Implementing this kind of diffusion is straightforward. The procedure is
similar to the computation of the sum of forces that acts on a node 
%in the third model.  
\citep{mgs08a} describes a fully generic diffusion operator
(valid for all kind of spaces of all dimensionality) by implementing a
generic Lagrangian operator.

%%%%%%%%%%%%%%%%%%%%%%%%%%%%%%%%%%%%%%%%%%%%%%%%%%%%%%%%%%%%%%%%%%%%%%%%%%%%%%%
%% Connection to table and framework
%%%%%%%%%%%%%%%%%%%%%%%%%%%%%%%%%%%%%%%%%%%%%%%%%%%%%%%%%%%%%%%%%%%%%%%%%%%%%%%
\subsubsection{Analysis}

The characteristics of biological DSLs with significant spatial
operations are summarized in
Table~\ref{table:dsl},~\ref{table:spatial},~and~\ref{table:device}, based
on the taxonomy proposed in Section~\ref{s:definitions}.
Although many of the languages for this space are focused on
individual agents, those languages that do raise their abstraction
level toward the aggregate provide a rich variety of space-time
operations.  MGS in particular provides an extremely powerful
modeling language, capable of manipulating space both geometrically
and topologically, and also of succinct functional abstraction and
composition of such programs.

The design languages are much more limited at present, though this is
largely due to the current sharp limits on the ability to engineer
organisms, and will likely change as synthetic biology continues to
progress.

%%%%%%%%%%%%%%%%%%%%%%%%%%%%%%%%%%%%%%%%%%%%%%%%%%%%%%%%%%%%%%%%%%%%%%%%%%%%%%%
%%%%%%%%%%%%%%%%%%%%%%%%%%%%%%%%%%%%%%%%%%%%%%%%%%%%%%%%%%%%%%%%%%%%%%%%%%%%%%%

\subsection{Agent-Based Models}
\label{s:agent}

%%%%%%%%%%%%%%%%%%%%%%%%%%%%%%%%%%%%%%%%%%%%%%%%%%%%%%%%%%%%%%%%%%%%%%%%%%%%%%%
%%%%%%%%%%%%%%%%%%%%%%%%%%%%%%%%%%%%%%%%%%%%%%%%%%%%%%%%%%%%%%%%%%%%%%%%%%%%%%%

%%%%%%%%%%%%%%%%%%%%%%%%%%%%%%%%%%%%%%%%%%%%%%%%%%%%%%%%%%%%%%%%%%%%%%%%%%%%%%%
%% Description of field
%%%%%%%%%%%%%%%%%%%%%%%%%%%%%%%%%%%%%%%%%%%%%%%%%%%%%%%%%%%%%%%%%%%%%%%%%%%%%%%
% First, establish some basic agent terminology (e.g., agents, sensors,
% effectors (actuators), environment, etc.)
Agent-based models explained by Macal and North~\citep{macal2010tutorial}
are capable of describing any or all of three elements:
\begin{itemize}
\item A set of {\it agents}, their attributes, and their behavior(s);
\item The {\it relationships} between agents and methods of interaction; and
\item The {\it environment} in which the agents interact.
\end{itemize}
Behavioral models, such as the Belief-Desire-Intent (BDI) agent
model~\citep{rao1995bdi}, implemented in frameworks such as
Jadex~\citep{pokahr2003jadex}, describe the internals of agents. 
The agent internals can usually be reduced to Russel and Norvig's
conceptual view of agents~\citep{russellNorvig} with {\it sensors} (that
read from the environment), {\it effectors} or {\it actuators} (that change
the environment), and {\it behavioral mappings} between the sensors and
effectors.
Relationships between agents are typically encoded as topologies.  Macal
and North also explain that ``In some applications, agents interact
according to multiple topologies''~\citep{macal2010tutorial}.  For example,
a network topology may offer low-level agent communication relationships
and simultaneously a social overlay network may guide the necessity for
inter-agent messaging.
Agent environmental modeling can vary widely based on the purpose of the
modeling effort, from simple geospatial models to complex biologic models.
Often, environmental information (such as global location) is {\it sensed}
by an agent.  Likewise, {\it actuators} modify the environmental model,
which in turn serves as a blackboard for agents (i.e., stigmergy).

%%%%%%%%%%%%%%%%%%%%%%%%%%%%%%%%%%%%%%%%%%%%%%%%%%%%%%%%%%%%%%%%%%%%%%%%%%%%%%%
%% Breakdown of types of languages in the field
%%%%%%%%%%%%%%%%%%%%%%%%%%%%%%%%%%%%%%%%%%%%%%%%%%%%%%%%%%%%%%%%%%%%%%%%%%%%%%%
% Next, introduce categories
In this section, we categorize the agent-oriented DSLs as one of the
following:
\begin{itemize}
\item {\it Graphical Agent Modeling Language}.  
These languages usually extend or piggy-back some form of UML.  They focus
on modeling the internals of agents and (sometimes) their interaction
patterns using graphical tools rather than formal languages.
\item {\it Agent Framework}. 
These languages are extensions of general-purpose languages, usually
libraries, that impose common structure on agent specifications.  By
conforming to this common structure, programmers can utilize tools provided
by the framework (i.e., agent administration, logging, simulation).
\item {\it Multi-agent Modeling and Simulation Toolkit}.  
These languages focus on modeling and simulating inter-agent interactions
and environmental interactions in agent systems.
\end{itemize}
After categorizing the agent-oriented DSL, we give a brief description of
each DSL.  Then, we show and describe an implementation of the reference
example ``T'' program in a representative multi-agent-based DSL.

%%%%%%%%%%%%%%%%%%%%%%%%%%%%%%%%%%%%%%%%%%%%%%%%%%%%%%%%%%%%%%%%%%%%%%%%%%%%%%%
%% 1 sub-sub-section for each language type
%%%%%%%%%%%%%%%%%%%%%%%%%%%%%%%%%%%%%%%%%%%%%%%%%%%%%%%%%%%%%%%%%%%%%%%%%%%%%%%
% Now the breakdown of DSL's to category and a brief explanation of each.
\subsubsection{Graphical Agent Modeling Languages}

\citep{huget2005modeling} cites MAS modeling languages
Agent UML~\citep{odell1999extending, bauer2001agent}
and the Agent Modeling Language
(AML)~\citep{trencansky2005agent}.  These largely-graphical languages aim to
extend traditional UML documents with agent and multi-agent system
concepts. 
Other graphical agent modeling languages (e.g.,
Agent-DSL~\citep{kulesza2005generative, Kulesza2004AGA} and
DSML4MAS~\citep{hahn2008domain}) further enhance usability by
embedding the languages in development and simulation environments.

\subsubsection{Agent Frameworks} \label{sec:agentFrameworks}
While there are very many agent frameworks (sometimes labeled as
``architectures''), we do not intend to specify all, but simply give a list
of a few representative, popular examples.

Jade~\citep{jade} extends Java with a library for producing
FIPA-compliant~\citep{fipa} agents and managing their interactions.
AGLOBE~\citep{aglobe} also extends Java with a small, lightweight library
for implementing goal-oriented agents.
The Cognitive Agent Architecture (Cougaar)~\citep{helsinger2004cougaar,
cougaarwebsite} is another Java extension library for a
highly-configurable, QoS-adaptive intelligent agent framework.
% JESS
JESS~\citep{jesswebsite}, although written for use in Java, is an example of
a {\it declarative} rule engine and scripting environment to provide
``reasoning'' skills to agent systems.

\subsubsection{Multi-agent Modeling and Simulation Toolkits}
ASCAPE~\citep{inchiosa2002overcoming, ascapewebsite} is a Java extension for
simplification of agent model composition and agent behavior execution
using topological abstraction and rule-based execution respectively.
NetLogo~\citep{sklar2007netlogo, netlogowebsite} and
StarLogo~\citep{resnick1996starlogo, starlogowebsite} are extensions to the
Logo language for ``turtle'' sensing operations and simultaneous control of
multiple agents.
Repast~\citep{north2007declarative, repastwebsite} is a
multi-language (with extensions to Java and Logo) toolkit for modeling and
simulating MAS, with additional tools for running in high-performance
computing environments~\citep{collier2011repast}.
MASON~\citep{luke2004mason, masonwebsite} is a discrete-event multi-agent
simulation toolkit that aims to weakly-couple the MAS model from
its visualization and scale to millions of agents.

The Strictly Declarative Modeling Language
(SDML)~\citep{moss98sdml,sdmlwebsite} represents multi-agent systems using
declarative rules where each agent's beliefs are transcribed to databases.
Inter-agent communication occurs by reading and writing directly to an
agent's database or a shared container.

Swarm~\citep{swarm, swarmwebsite} is a platform and tool suite for
agent-based modeling and simulation of complex adaptive systems.
MAML~\citep{gulyas1999multi, mamlwebsite} is an Objective-C extension
language and compiler ({\tt xmc}) for helping non-programmers create Swarm
applications.
Echo~\citep{forrest1994modeling, sfiechoweb} extends genetic algorithms 
with location, resource competition, and agent interactions. Echo is
intended to capture important properties of ecological systems toward the
goal of modeling and simulating complex adaptive systems.  

Echo is also an example of a system that conducts modeling and
simulation of Cellular Automata (CA).  CA are also a topic of research
in distributed systems and parallel computing, and will be discussed
further in Section~\ref{s:parallel}

\subsubsection{Distributed Systems}

We include distributed systems DSLs within agent-based DSLs due to their
high degree of overlap.
Distributed system DSLs fall into two categories: 1) distributed system
modeling languages, and 2) information movement languages.
An example of distributed system modeling languages is the $\Psi$
Calculus~\citep{Kinny02Psi}, a formal modeling language for abstract plan
execution agents with a sense-compute-act computation cycle.
On the other hand, information movement languages aim to abstract the
process of moving information to the points in space-time where/when they
are needed.  For example, the Knowledge Query and Manipulation Language
(KQML)~\citep{Finin94kqml} focuses on agent communication for managing
distributed collaboration.

%%%%%%%%%%%%%%%%%%%%%%%%%%%%%%%%%%%%%%%%%%%%%%%%%%%%%%%%%%%%%%%%%%%%%%%%%%%%%%%
%% Reference Example
%%%%%%%%%%%%%%%%%%%%%%%%%%%%%%%%%%%%%%%%%%%%%%%%%%%%%%%%%%%%%%%%%%%%%%%%%%%%%%%
\subsubsection{Reference Example: NetLogo} \label{sec:MASexample}

This section shows and explains an implementation of the reference example
``T-Program'' (described in Section~\ref{sec:t-prog}) in
NetLogo~\citep{netlogowebsite}---a language that we have chosen as being
representative of Multi-Agent System DSLs.  The purpose of this exercise is
to demonstrate how NetLogo supports the basis set of spatial operations
described in Section~\ref{sec:basisset}.

The implementation starts with ``global'' values that are shared (and
constant) between all agents in the system.  Note that NetLogo, because of
its roots in Logo, uses the terminology {\tt turtle} to mean ``agent.''
Thus the position of the {\tt origin-turtle} (i.e., the agent elected as
the origin of the local coordinate system) is shared between all agents.

\begin{CodeBlock}
globals [ origin-turtle ]
\end{CodeBlock}

NetLogo also allows agents to track their own ``local'' values.  By
default, agents track their global position in local variables {\tt xcor}
and {\tt ycor}, however, under the constraints of our reference example, we
cannot make use of global coordinates.  Thus, we define {\tt xpos} and {\tt
ypos} as our local coordinate values.  Further, we add a state variable
{\tt is-origin} as a convenient way for an agent to determine if it has
been designated as the origin of the local coordinate system.

\begin{CodeBlock}
turtles-own [ xpos ypos is-origin ]
\end{CodeBlock}

\begin{figure}[t]
\centering
\subfigure[Initial distribution]
{\includegraphics[scale=0.18]{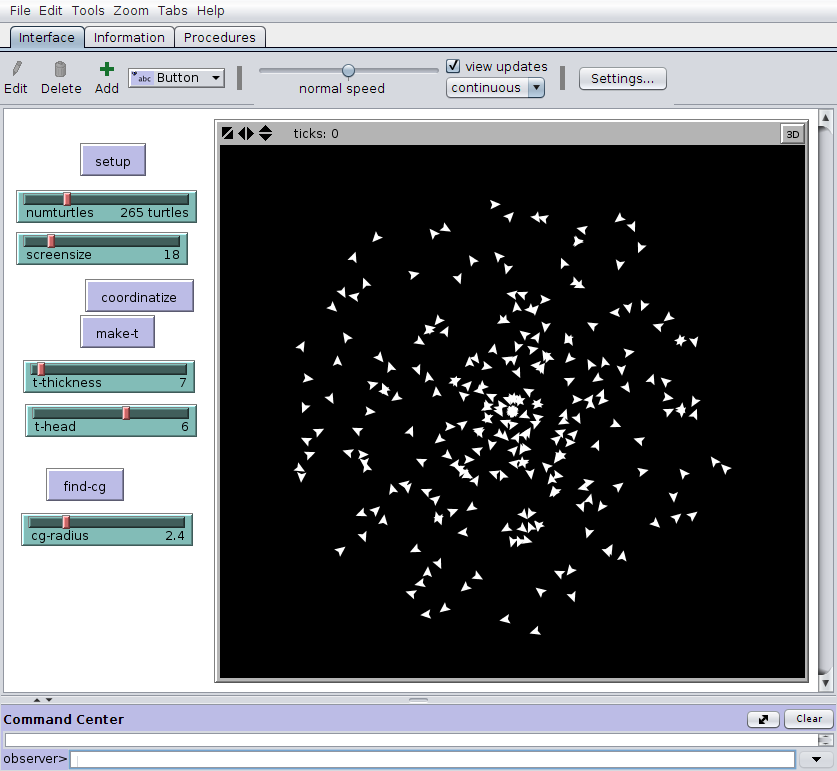}\label{f:netlogo-setup}}
%\subfigure[Creating coordinates]
%{\includegraphics[scale=0.20]{t-prog-netlogo-2.png}\label{f:netlogo-coordinatize}}
\subfigure[Creating T-shape]
{\includegraphics[scale=0.18]{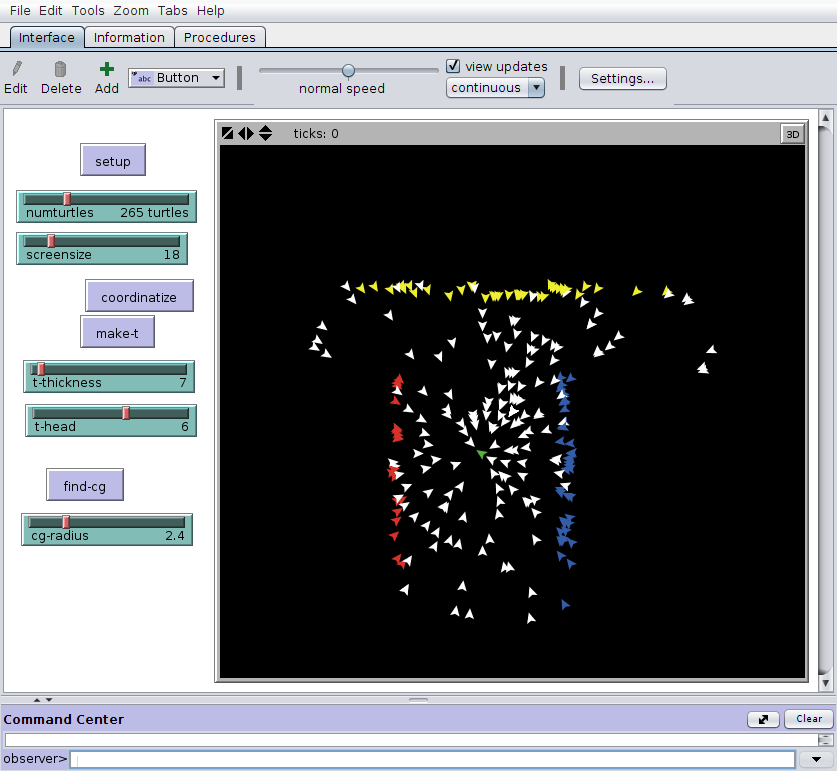}\label{f:netlogo-maket}}
\subfigure[Finding CoG]
{\includegraphics[scale=0.18]{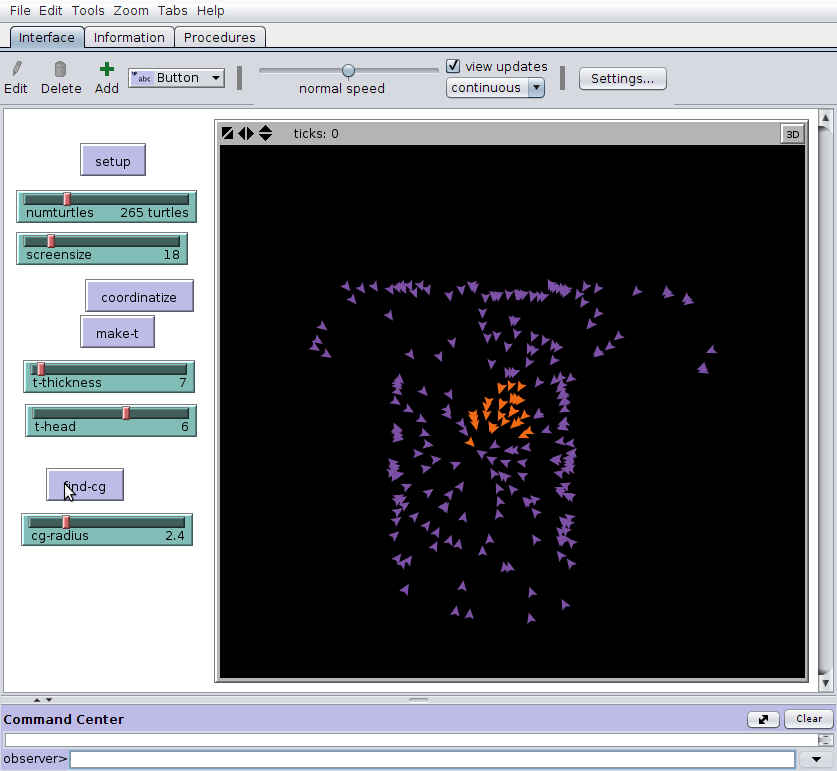}\label{f:netlogo-findcg}}
\caption{NetLogo execution of the ``T program'' reference example.}
\end{figure}

The {\tt setup} method resets the state of the world and randomly
distributes and rotates a certain number of turtles.  The variables {\tt
numturtles} and {\tt screensize} are configuration values which are set at
configuration time and have global scope.  Figure~\ref{f:netlogo-setup}
shows a simulation of the execution of the {\tt setup} method.

\begin{CodeBlock}
to setup
  clear-all        ;; clear the world
  crt numturtles   ;; make new turtles
  ask turtles
    [ 
      set color white       ;; all turtles turn white
      fd random screensize  ;; ...move forward one step
      rt random 360         ;; ...and turn a random amount
      set is-origin false   ;; default all to non-origin
    ]
end
\end{CodeBlock}

To establish a local coordinate system, we make use of each agents' ability
to determine the direction and distance to an ``origin'' node, selected in
the {\tt coordinatize} function.  The notion of the network is abstracted
and it is assumed that any agent can determine the vector (i.e., direction
and distance) to any other agent (via the function {\tt compute-vec-to}).
We then use the function {\tt polar-to-cartesian} to convert distance ({\it
r}) and direction ({\it theta}) into Cartesian coordinates.

\begin{CodeBlock}
to-report polar-to-cartesian [theta r]
  let x (r * cos(theta))
  let y (r * sin(theta))
  report list x y
end

to compute-vec-to [agent]
  ifelse is-origin = true
  [
    set xpos 0
    set ypos 0
  ]
  [
    let cartesian polar-to-cartesian towards agent distance agent
    set xpos item 1 cartesian
    set ypos (-1 * item 0 cartesian)
    face agent
  ]
end
\end{CodeBlock}

\begin{CodeBlock}
to coordinatize
  if is-turtle? origin-turtle
  [ ask origin-turtle
    [ 
      set is-origin false
      set color white
    ]
  ]
  set origin-turtle one-of turtles
  ask origin-turtle 
  [ 
    set is-origin true
    set color green 
  ]
  ask turtles
  [ compute-vec-to origin-turtle ]
end
\end{CodeBlock}

The next section of the program, the {\tt make-t} function, selects nodes
that fall within ranges of the coordinate space and moves them to create a
``T'' shape around the origin.  Note that the {\tt mov} command first
ensures that an agent is facing a given direction, then proceeds forward
for a given number of steps.

\begin{CodeBlock}
to mov [head dist]
  set heading head
  fd dist
  compute-vec-to origin-turtle
end

to make-t
  ;; right side of the lower part of the T
  ask turtles with [(xpos > t-thickness) and (ypos < t-head)]
    [ 
      set color red
      ;; move into the body of the T
      while [ xpos > t-thickness ]
        [mov 90 1]
    ]
  ;; left side of the lower part of the T
  ask turtles with [(xpos < (-1 * t-thickness)) and (ypos < t-head)]
    [ 
      set color blue
      ;; move into the body of the T
      while [ xpos < (-1 * t-thickness) ]
        [mov 270 1]
    ]
  ;; top of the T
  ask turtles with [ypos > (t-head + t-thickness)]
    [ 
      set color yellow
      ;; move into the body of the T
      while [ypos > (t-head + t-thickness)]
        [mov 180 1]
    ] 
end
\end{CodeBlock}

The center-of-gravity is computed by dividing the sum of all agents'
positions in each dimension by the number of agents.  This approach assumes
that every node is weighted equally and every node is connected.  Finally,
if agents fall within a certain radius from the center-of-gravity, they are
colored orange.

\begin{CodeBlock}
to find-cg
  ;; center-of-gravity = sum of coordinate / number of devices
  ask turtles
  [ 
    ifelse (abs(xpos - (sum [ xpos ] of turtles / numturtles)) < cg-radius) 
    [ 
      ifelse (abs(ypos - (sum [ ypos ] of turtles / numturtles)) < cg-radius) 
      [set color orange] 
      [set color violet] 
    ]
    [set color violet]
  ]
end
\end{CodeBlock}

%%%%%%%%%%%%%%%%%%%%%%%%%%%%%%%%%%%%%%%%%%%%%%%%%%%%%%%%%%%%%%%%%%%%%%%%%%%%%%%
%% Connection to table and framework
%%%%%%%%%%%%%%%%%%%%%%%%%%%%%%%%%%%%%%%%%%%%%%%%%%%%%%%%%%%%%%%%%%%%%%%%%%%%%%%
\subsubsection{Analysis}
%\todo[inline]{Need to add connection to table and framework.}

The characteristics of the DSL classes are summarized in
Table~\ref{table:dsl},~\ref{table:spatial},~and~\ref{table:device}, based
on the taxonomy proposed in Section~\ref{s:definitions}.
Table~\ref{table:dsl} shows the {\it linguistic} properties for each class
of agent-based DSL.  Table~\ref{table:spatial} similarly depicts each
class's {\it spatial} properties, and Table~\ref{table:device} summarizes
each class's {\it abstract device} properties.

% DSL Table
First we discuss the linguistic properties of the agent-based DSLs
(summarized in Table~\ref{table:dsl}).
% Graphical Agent Modeling Languages
Graphical Agent Modeling Languages typically target end-users with a
graphical UI.  Often, these languages are simply graphical representations
(or extensions) of other agent modeling languages (e.g., UML), paradigms
(e.g., Belief-Desire-Intent (BDI)~\citep{rao1995bdi}), or meta-models (e.g.,
FAML~\citep{beydoun2009faml}).  Graphical Agent
Modeling Languages function at a conceptual level---mapping actions or
behaviors to the components of the agent system.
% Agent Frameworks
Agent Frameworks, whose functions and designs are analyzed thoroughly in
the Agent System Reference Model (ASRM)~\citep{asrm} and Agent System
Reference Architecture (ASRA)~\citep{asra}, are often extensions of
general-purpose languages that provide common tools to agent system
designers and developers.  The Foundation for Intelligent Physical Agents
(FIPA)~\citep{fipa} provides specifications for which these language
extensions can be implemented and interchangeably utilized.  Most often,
these libraries extend imperative languages (as was the case for
Jade~\citep{jade}, Cougaar~\citep{helsinger2004cougaar}, and
AGLOBE~\citep{aglobe}), but there are also some declarative agent frameworks
(e.g., JESS~\citep{jesswebsite}).  The platform scope of agent frameworks,
however, varies widely.
%Multi-Agent Modeling and Simulation Toolkits
Multi-Agent Modeling and Simulation Toolkits, which focus on modeling and
simulating inter-agent and environmental interactions, have a broad scope
in terms of their DSL design.  Their types and patterns vary from
LOGO-based scripting and simulation environments (e.g.,
NetLogo~\citep{sklar2007netlogo}, StarLogo~\citep{resnick1996starlogo}) to
full tool-suites designed for use by non-programmers (e.g.,
Swarm~\citep{swarm}, MAML~\citep{gulyas1999multi}).

% Spatial Table
As shown in Table~\ref{table:spatial},
% Really only Multi-Agent Modeling and Simulation Toolkits have spatial
% properties
the only class of Agent-Based DSLs that exhibit spatial properties are the
Multi-Agent Modeling and Simulation Toolkits.  This fact is likely due to
tight integration between the language and simulation environments,
allowing the toolkit to expose language features that are unavailable in
many distributed systems (e.g., distance, movement).

% Device Table
Abstract device properties of agent-based DSLs are shown in
Table~\ref{table:device}.
% Discretization
Agent-based DSLs typically offer discrete modalities for interacting with
agents, although some (e.g., Echo~\citep{forrest1994modeling}) offer
cellular discretization.  
% Comm. Region
Similarly, most agent-based DSLs attempt to abstract the notion of the
network from the programmer, offering global communication ranges.  Notable
exceptions are in the Multi-Agent Modeling and Simulation Toolkit class,
where first-class notions of network topology and network links allow the
programmers to {\it simulate} the restriction of agent communication.
% Comm. Granularity
The granularity of communication is typically unicast (i.e., agent to
agent), however some Multi-Agent Modeling and Simulation Toolkits allow
communication to occur in a multicast style (i.e., to a set of agents).
This is ideally demonstrated by the {\it ask} feature of NetLogo in the
reference example in Section~\ref{sec:MASexample}.  NetLogo allows unicast
communication by specifying a single turtle (e.g., {\it ask origin-turtle})
or multicast communication by specifying a set of turtles (e.g., {\it ask
turtles}).
% Code Mobility
Finally, mobile code is a feature of some agent-based DSLs because it is a
useful mechanism for distributed algorithms.  Modeling and simulation
languages and toolkits typically execute a single, uniform program, whereas
agent frameworks tend to provide features for facilitating agent (code)
mobility~\citep{usbeck2011agent}.

%%%%%%%%%%%%%%%%%%%%%%%%%%%%%%%%%%%%%%%%%%%%%%%%%%%%%%%%%%%%%%%%%%%%%%%%%%%%%%%
%%%%%%%%%%%%%%%%%%%%%%%%%%%%%%%%%%%%%%%%%%%%%%%%%%%%%%%%%%%%%%%%%%%%%%%%%%%%%%%

\subsection{Wireless Sensor Networks}
\label{s:wsn}

%%%%%%%%%%%%%%%%%%%%%%%%%%%%%%%%%%%%%%%%%%%%%%%%%%%%%%%%%%%%%%%%%%%%%%%%%%%%%%%
%%%%%%%%%%%%%%%%%%%%%%%%%%%%%%%%%%%%%%%%%%%%%%%%%%%%%%%%%%%%%%%%%%%%%%%%%%%%%%%

%%%%%%%%%%%%%%%%%%%%%%%%%%%%%%%%%%%%%%%%%%%%%%%%%%%%%%%%%%%%%%%%%%%%%%%%%%%%%%%
%% Description of field
%%%%%%%%%%%%%%%%%%%%%%%%%%%%%%%%%%%%%%%%%%%%%%%%%%%%%%%%%%%%%%%%%%%%%%%%%%%%%%%
\emph{Wireless sensor networks} are a field of research concerned with the
development of large networks made up of devices performing primarily a
sensing function. The devices in the network (\emph{nodes} or \emph{motes})
are usually built using off-the-shelf-components and include a processor, a
wireless communication interface and one or more sensors. As they are
autonomous devices, the amount of energy they can use is often limited (by
the battery on board or energy scavenging device) --- hence the main design
restriction targets \emph{energy efficiency}. In order to optimize for
this, common practice includes duty-cycling (having the mote alternate
between long power-down modes and short bursts of activity), trade-offs
between ``expensive'' wireless communication and ``cheap'' local
processing, multi-hop communication and distributed algorithms.

The main goal of wireless sensor networks is efficient data collection and
delivery to a gateway linked to infrastructure (i.e., Internet). Exceptions
exist in the form of wireless sensor networks employing actuators, networks
with multiple (mobile) gateways, etc. We will focus on the ``traditional''
sensor network made up of a collection of homogeneous or heterogeneous
static devices. Collecting data at a single gateway from a large network is
a non-scalable process (limited bandwidth being the major constraint).
Thus, techniques such as data aggregation, selection of a subset of data to
be gathered, filtering of data and in-network processing are common
operations the designer faces in most of the deployments. These techniques
are usually gathered under the saying ``the network is the tool'' reflecting
the fact that a network delivering pre-processed data or synthesized events
is often desired. 

Data collection and dissemination is heavily linked to the network topology
under which it operates. For example, organizing the network as a graph (a
tree) is a common technique that allows simple algorithms to be employed on
resource-poor devices. Aspects such as communication patterns (short or
long transmission ranges, broadcast or unicast type, group based or device
based, etc.) are a key building block of the final application and are
usually dictated by the underlying hardware platform.

In order to implement data dissemination on top of the distributed
platform, each device must be capable of supporting the execution of its
local algorithm. In order to do this, operating systems such as
TinyOS~\citep{levis2005tinyos} and Contiki~\citep{dunkels2004contiki} are
employed, providing common functionality such as hardware abstraction
layers, scheduling mechanisms for tasks, execution parallelism, etc. It is
common to see virtual machines implemented on top of these systems, in
order to extend the basic functionality offered. 

This brief introduction already outlines the main concerns faced by
designers of applications for wireless sensor networks. They are basically
driving factors for developing DSLs for the wireless sensor network
platforms. Automating and abstracting some of the following mechanisms is
the logical step to take:
\begin{itemize}
  \item \emph{Hardware and software platform} is usually specific to each
  deployed application. Usual operations that can be automated are the
  control of the hardware components via a hardware abstraction layer,
  providing common programming support in the form of an operating system,
  etc. The possibility of turning on and off components (radio, sensors,
  processing routines) reacting to the event-driven programming paradigm is
  an important feature.
  \item \emph{Communication and topology control} is being performed in
  virtually all sensor networks applications. Primitives abstracting the
  communication protocols needed for discovering, creating and maintaining
  a topology need to be provided. This is the key goal of several DSLs in
  the region-based DSLs category (presented below) while completely
  abstracted in all other categories. The basic ingredients for this are
  neighborhood discovery, routing algorithms, and transport protocols.
  \item \emph{Data dissemination} being the main goal of the sensor
  networks applications must be supported with high level primitives for
  querying the network for specific items (for example, in the form of a
  SQL-like language). This is achieved by combining networking algorithms
  with transport protocol, ensuring data delivery over optimal paths. The
  maintenance of these paths is done without the involvement of the user.
  \item \emph{Energy efficiency constraints} lead to the need of
  predefining or automatically tuning the trade-off between communication and
  local processing. Estimation of quality of service metrics at runtime is
  a common mechanism through which this is achieved. The collection and
  processing of data for deriving the metrics (e.g., the radio link quality
  between two devices) is performed in the background. 
\end{itemize}

%%%%%%%%%%%%%%%%%%%%%%%%%%%%%%%%%%%%%%%%%%%%%%%%%%%%%%%%%%%%%%%%%%%%%%%%%%%%%%%
%% Breakdown of types of languages in the field
%%%%%%%%%%%%%%%%%%%%%%%%%%%%%%%%%%%%%%%%%%%%%%%%%%%%%%%%%%%%%%%%%%%%%%%%%%%%%%%
A large number of DSLs have been built already, addressing combinations of
the previous concerns
(see~\citep{sugihara2008programming,mottola2011programming} for recent
surveys). Taking note of the previous work, we would like to extend the
analysis in~\citep{mottola2011programming} in order to bring into focus the
aspect of spatial computing. We propose an extension of the original
taxonomy, looking at the basic mechanisms used in programming the network.
We have identified five classes, as follows:

%%%%%%%%%%%%%%%%%%%%%%%%%%%%%%%%%%%%%%%%%%%%%%%%%%%%%%%%%%%%%%%%%%%%%%%%%%%%%%%
%% 1 sub-sub-section for each language type
%%%%%%%%%%%%%%%%%%%%%%%%%%%%%%%%%%%%%%%%%%%%%%%%%%%%%%%%%%%%%%%%%%%%%%%%%%%%%%%
\subsubsection{Region-based DSLs} 
By far, the largest number of DSLs
   targeting wireless sensor networks is found in the category of region-based DSLs, showing the
   programmer's need for expressing operations at the level of regions
   (i.e., neighborhoods, sets, etc.) rather than individual devices. For
   example, \emph{Abstract Regions}~\citep{welsh2004regions} offers a family
   of spatial operators that allows addressing of regions of the network.
   Additional characteristics include information about the trade-off
   communication-computation resources and extension of the underlying
   TinyOS operating system with a thread-like concurrency mechanism called
   fibers.
   A similar DSL is \emph{Hood}~\citep{hood}, which offers functionality for
   defining one-hop neighborhoods and sharing data between nodes (these
   mechanisms being transparent to the programmer). 

   \emph{Logical Neighborhoods}~\citep{mottola2006logical} is somewhat more
   general in the sense that the definition of neighborhood is relaxed, not
   being restricted to physical proximity anymore. Nodes sharing the same
   characteristics can be addressed together, even if they are spread
   throughout the network. A direct extension is \emph{Virtual
   Nodes}~\citep{ciciriello2006building}, in which each neighborhood is seen as
   a single virtual sensor. The language adds further optimization to the
   compiler and network level.
   
   Several other DSLs are built upon the same concept of addressing local
   neighborhoods, \emph{Pieces}~\citep{liu2003state} and
   \emph{TeenyLime}~\citep{costa2006teenylime} being a few examples. The
   neighborhood information can be addressed in several manners:
   \emph{EnviroSuite}~\citep{abdelzaher2004envirotrack} provides a programming
   interface aimed at tracking applications which creates objects for physical
   entities. 
   
   \emph{Snlog}~\citep{chu2006entirely} is a rule-based approach,
   similar to logic programming, where rules are executed using a one-hop
   abstraction. It follows the foundation laid by NDLog~\citep{loo2006declarative} and 
   actually belongs to a larger suite of languages including Dedalus~\citep{alvaro2009dedalus} 
   and DAHL~\citep{lopes2010applying}. In this group, the closest to the spatial languages comes 
   Netlog~\citep{grumbach2010netlog} whose semantics allow moving of facts to 
   neighbors and to any routable device with a known ID.

   \emph{Regiment}~\citep{newton2007regiment} presents itself as a functional
   language, allowing easy access to data streams (called signals). Low level
   details (such as one-hop communication, parallelism, etc.) are achieved
   from successive translations of the initial program into different
   languages stacked on top of each other (four translations are necessary to
   obtain final code). It is the closest language to spatial computing in the
   wireless sensor network domain. This is due to the combination between the
   flexibility with which users can specify regions (both hop-based and
   distance-based - similar to Abstract Regions), and the functional interface
   it offers.
   
   The concept of \emph{region} comprises several representations of the
   space. The region can be defined geometrically, as the distance from a
   certain point (as done in Regiment and Abstract Regions), the set of nodes
   within a number of hop counts from a node (as done in Hood) or a set of
   nodes complying to some predicates (as done in Logical Neighborhoods).

\subsubsection{Dataflow based DSLs}
Dataflow based DSLs are one level of abstraction higher.
   Although their execution uses neighborhoods as part of the implementation,
   the user does not need to access them directly. Instead, the applications
   can be specified as a dataflow graph, in which the user specifies how the
   software components are linked by data. The location of the software, the
   communication between devices, locating and transferring the data in the
   neighborhoods, etc. are built into the languages. The simplest example in
   this category is \emph{Active Messages}~\citep{gay2003nesc} in which
   software components are expressed in the nesC programming language. Active
   Messages is a mechanism allowing asynchronous communication between
   components via interfaces that provide commands and events. The resulting
   system is similar to a socket system, having the advantage of allowing
   modularity and enabling event-driven computation.
   
   As a conceptual extension, \emph{Abstract Task Graph}
   (ATAG)~\citep{pathak2007expressing} allows the user to express data
   transformations in a distributed system independent of the architecture.
   Abstract tasks run on individual devices communicating via abstract data,
   accessible via abstract channels. The user specifies the code for the tasks
   and the way in which they interact with data. Low-level operations, such as
   task deployment and physical communication, are abstracted away.
   \emph{MiLAN}~\citep{heinzelman2004middleware} provides automatic selection
   of sensors and groups of nodes based on the quality of service of the
   collected data. The user specifies the execution as a state machine with
   quality of service requirements for each transition. Milan selects the
   appropriate sensors to collect the data. Networking layers functionality is
   provided by the network plug-in system and a service discovery is employed
   as well. 
   
   This class of DSLs presents almost no common characteristics with the
   spatial programming approaches. The focus of these languages is on
   providing functionality by linking the right software components, spatial
   features are to be implemented on top, as part of the components
   themselves.

\subsubsection{Database-like DSLs}
Database-like DSLs treat the wireless sensor network as a
   real database. They abstract low-level communication
   functionality, focusing exclusively on data collection and aggregation. As
   example, \emph{DSWare}~\citep{li2004event} is a SQL-based approach built on
   top of an extended event concept. Paths are established in the network,
   linking the subscriptions for specific data at the gateway with the nodes
   actually producing the events. The user benefits from the functionality of
   optimizations at network level, both in the network paths creation and
   maintenance, and the data aggregation in the network.
   
   \emph{TinyDB}~\citep{tinydb} is one of the first high-level DSLs
   proposed and follows the database approach, hiding the necessary
   networking mechanisms from the user (e.g., the queries emitted in
   the network define the routing tree). Users are not involved in the
   low level aspects of data aggregation. Closely related,
   \emph{Cougar}~\citep{Yao02thecougar} (not to be confused with the
   agent language Cougaar~\citep{helsinger2004cougaar}) allows the user
   to write queries over the data in the database and the system
   optimizes the dissemination of the queries and the aggregation of
   the results. A notable example of DSL in this category is
   \emph{SINA}~\citep{shen2001sensor} which differs due to a mechanism
   which allows easy addition of new data operators, extending the
   original set of SQL commands. The authors of~\citep{Duckham:2005:MDS:1097064.1097073} 
   propose a database-like approach as well, with the difference that the querying mechanism is built upon a 
   qualitative representation of \emph{dynamic fields}. The authors identify two basic entities 
   used in accessing the network, mainly the \emph{continuants} (e.g., regions that 
   endure time) and \emph{occurents} (e.g., transitory events).
   
   This class of DSLs has two characteristics in common with those of amorphous 
   computing. First, the computation pattern relies on a continuous
   representation of space, where users specify areas of interest and time
   intervals. Second, the high level description of the queries are basically
   operators over the space/time continuum.

\subsubsection{Centralized-view DSLs}
Centralized-view DSLs are basically a set of high-level
   languages that approach wireless sensor networks from a different
   perspective than the dataflow-based DSLs. The main differences lie in the
   fact that the data collection and aggregation functionality is not
   predefined. These languages allow the user to define the application
   functionality \emph{for the whole network} with a single program. For
   example, \emph{Pleiades}~\citep{kothari2007reliable} allows sequential
   execution over the whole sensor network. The network is seen as a central
   object, a container of nodes, and users can write a ``for'' loop which
   iterates through all the nodes. Parallel execution of the basic ``for''
   instruction can be specified and the compiler transforms the initial
   program into a collection of distributed algorithms that emulate
   sequentially over the distributed system.
   
   \emph{Kairos}~\citep{gummadi2005macro} is similar to Pleiades, offering a
   centralized view of the whole network. It presents itself as an
   extension of the Python language. Somewhat different,
   \emph{MacroLab}~\citep{hnat2008macrolab} is a programming approach
   similar to Matlab - the network being represented as a matrix. Each row
   in the matrix represents one sensor node, each column a data type.
   Operations such as dot product are allowed over the whole matrix and the
   language abstracts the networking part.
   
   None of the languages in this class have common points with the spatial
   computing approach. The effort is on providing the user with a discrete
   representation of the space, where data on each device can be addressed
   individually, using sequential programming.

\subsubsection{Agent-based DSLs}
Agent-based DSLs are a special subset of DSLs for
   wireless sensor networks, making the transition to more powerful system,
   such as mobile ad-hoc networks. The basic idea is that software agents
   contain the needed functionality to process data and to perform local
   aggregation while, for example, following a certain physical event
   through the network. These languages also build upon the region
   information, but they allow more complex applications than data
   collection and dissemination.

   As notable examples in this category, we mention
   \emph{Agilla}~\citep{fok2005rapid} which is an agent framework built on
   top of TinyOs. Agents travel across the nodes together with their state.
   Each node maintains a tuple space allowing interaction between the
   agents and are addressed by location rather than network address. Basic
   functionality offered to the programmer includes the list of neighbors,
   information on the tuple space at neighboring nodes, migration of
   agents. Agilla supplies an agent manager and implements memory
   management and a virtual machine.
   \emph{SensorWare}~\citep{boulis2007sensorware} is similar to Agilla, only
   that the code is expressed in Tcl. The state of the agents is not
   maintained, thus code re-initializes each time the agent moves to a
   different node. \emph{Spatial Programming}~\citep{borcea2004spatial}
   combines a light agent-based approach, in which scripts can migrate,
   with a unique addressing of space involving geographical areas. The
   addressing of space is translated automatically to the real network
   deployment. The migration of the agents is provided automatically by all
   these frameworks. Due to the unreliability of radio communication, there
   is always a chance that the transfer of agents fails, the effort of
   ensuring safe relocation of the agents being a distinguishing factor
   between these three approaches.

   With the exception of the last presented DSL, the languages in this
   class have little in common with spatial computing. As in the
   dataflow-based category, the focus is on specifying the functionality of
   the building blocks (agents) in close relationship to the available
   events/data rather than to space/time.

%%%%%%%%%%%%%%%%%%%%%%%%%%%%%%%%%%%%%%%%%%%%%%%%%%%%%%%%%%%%%%%%%%%%%%%%%%%%%%%
%% Reference Example
%%%%%%%%%%%%%%%%%%%%%%%%%%%%%%%%%%%%%%%%%%%%%%%%%%%%%%%%%%%%%%%%%%%%%%%%%%%%%%%

\subsubsection{Reference Example: Regiment} \label{sec:Regimentexample}

Most of the ``T-Program'' example cannot be implemented by wireless
sensor network DSLs, since they only gather data and do not have the
ability to actuate.  Most also do not have the ability to measure
space that is needed for computing coordinates, but depend on global
coordinates to be provided.  Computing an aggregate property like the
center of gravity, however, is a task for which these DSLs are
typically well-suited.

The following Regiment program (adapted
from~\citep{newton2007regiment}) computes the X-coordinate of the
center of gravity by averaging of the X values of all members of a
network of sensors.  The program uses the constructs \emph{rmap} to
obtain readings of an assumed global X coordinate from all devices and
\emph{rfold} to fold them into a single signal.  The aggregation is
realized via the \emph{dosum} function which uses a tuple (total
value, counter) to represent the collected data.  After computing the
average, the result is directed towards the base.  The example shows
the great flexibility offered by the in-network aggregation, achieved
in an elegant manner and with relatively small amount of code.
  
\begin{CodeBlock}
dosum :: float, (float, int) -> (float, int) 
fun dosum(X, (sumX, count)) {
  (sumX+X, count+1)
}

Xreg = rmap(fun(nd){sense("X",nd)}, world); 
sumsig = rfold(dosum, (0,0), Xreg); 
avgsig = smap( fun((sum,cnt)) {sum / cnt}, sumsig); 

BASE <- avgsig
\end{CodeBlock}

%%%%%%%%%%%%%%%%%%%%%%%%%%%%%%%%%%%%%%%%%%%%%%%%%%%%%%%%%%%%%%%%%%%%%%%%%%%%%%%
%% Connection to table and framework
%%%%%%%%%%%%%%%%%%%%%%%%%%%%%%%%%%%%%%%%%%%%%%%%%%%%%%%%%%%%%%%%%%%%%%%%%%%%%%%
\subsubsection{Analysis}
   
% generic table information
The characteristics of the five DSL classes are summarized in
Table~\ref{table:dsl},~\ref{table:spatial},~and~\ref{table:device}, based on the
taxonomy proposed in Section~\ref{s:definitions}.
% table information
Table~\ref{table:dsl} shows that only the dataflow-based DSLs fall under
the category of language inventions (according to the taxonomy proposed
in~\citep{mernik2005and}). The predominant type of programming is
imperative, the database-like DSLs being the exception. As far as layers in
Figure~\ref{fig:layers} are concerned, the results vary widely.
From the perspective of the need of a spatial description of the languages,
only the database-like DSLs come close. Unfortunately, these DSLs are very
limited in their functionality, being specifically designed for the data
dissemination application. As far as the platform is concerned, three out
of the five categories address the network as a whole, offering a balanced
alternative (taking into account also the number of DSLs in each category).
   
Regarding the spatial characteristics of the surveyed DSLs,
Table~\ref{table:spatial} gives a clear argument against the suitability of
the wireless sensor network DSLs for filling the gap between designers
and embedded systems platforms. The only two categories having entries in
this table are the region-based DSLs and database-like DSLs. Even for these
two, most of the columns contain no entries. One of the reasons for which
wireless sensor network DSLs do not meet the spatial computing requirements
is that the application which primarily drove their development (data
dissemination) is restrictive in itself regarding the needed functionality.
Additionally, we note that the large majority of setups include static topologies 
with a single data-collection point (mobile networks and multiple-gateway setups 
are seen as exceptions). 

% \todo[inline]{SD: does it hold that in swarm robotics, the explicit need
%for actuation provided a better fitting of DSLs with spatial computing?}
% - not really. wsns usually employ some form of actuation (as in lighting leds, 
% or more generally sending messages when something is happening). in robotics, 
% the nodes of a network are closer to what is called "agent" acting in a space - this may 
% be a reason... quite far fetched pobably.

As far as the abstract device taxonomy is concerned, the results are
presented in Table~\ref{table:device}. Inspecting this data, it follows that
there is a balance between the categories of DSLs from the spatial and
communication perspective. The main design constraint of static networks
for monitoring applications is reflected in the last column where with the
exception of the agent-based DSLs, the code is mainly stationary. It is
worth noticing that the agent-based DSLs are a somewhat exceptional case in
the wireless sensor networks world, requiring powerful hardware that is at
the boundary of what is called a resource-poor device. 
% The ASRM table was removed
%The same design
%constraint is reflected even better in Table~\ref{table:asrm}.  With the
%exception of the agent-based DSLs, none of the other categories has
%anything in common with the agent technologies.

As a conclusion to this section, we note that a large collection of DSLs
exists for the field of wireless sensor networks. This study surveyed only
the ones which allow writing general applications for the sensor network
platform - several other DSLs exist targeting specific sub-problems (such
as cluster formation~\citep{frank2005algorithms}, efficient code
dissemination~\citep{levis2002mate}, etc.). We noticed also that languages
are often stacked, to combine the complementary features they offer. From
the spatial computing perspective, two categories of languages share a few
characteristics (region-based DSLs and database-like DSLs), being
nonetheless extremely limited.

%%%%%%%%%%%%%%%%%%%%%%%%%%%%%%%%%%%%%%%%%%%%%%%%%%%%%%%%%%%%%%%%%%%%%%%%%%%%%%%
%%%%%%%%%%%%%%%%%%%%%%%%%%%%%%%%%%%%%%%%%%%%%%%%%%%%%%%%%%%%%%%%%%%%%%%%%%%%%%%

\subsection{Pervasive Computing}
\label{s:pervasive}

%%%%%%%%%%%%%%%%%%%%%%%%%%%%%%%%%%%%%%%%%%%%%%%%%%%%%%%%%%%%%%%%%%%%%%%%%%%%%%%
%%%%%%%%%%%%%%%%%%%%%%%%%%%%%%%%%%%%%%%%%%%%%%%%%%%%%%%%%%%%%%%%%%%%%%%%%%%%%%%

%%%%%%%%%%%%%%%%%%%%%%%%%%%%%%%%%%%%%%%%%%%%%%%%%%%%%%%%%%%%%%%%%%%%%%%%%%%%%%%
%% Description of field
%%%%%%%%%%%%%%%%%%%%%%%%%%%%%%%%%%%%%%%%%%%%%%%%%%%%%%%%%%%%%%%%%%%%%%%%%%%%%%%
Pervasive computing is the scenario in which people, immersed in their typical
environment, are able to automatically interact with sensors and actuators
spread throughout in order to consume information of interest based on their
preferences and situation, and to produce information for other people.
Due to the intrinsic complexity of such systems, metaphors inspired by
nature are typically adopted for their design and implementation
\citep{ZV-JPCC2011}.

%%%%%%%%%%%%%%%%%%%%%%%%%%%%%%%%%%%%%%%%%%%%%%%%%%%%%%%%%%%%%%%%%%%%%%%%%%%%%%%
%% Breakdown of types of languages in the field
%%%%%%%%%%%%%%%%%%%%%%%%%%%%%%%%%%%%%%%%%%%%%%%%%%%%%%%%%%%%%%%%%%%%%%%%%%%%%%%
The computational network over which pervasive computing applications
typically run, very much resembles a WSN: it hosts mobile nodes (e.g.,
smartphones, sensors on cars, etc.) and it heavily relies upon wireless
technologies because devices spread throughout the environment are rarely
networked by wires.
On the other hand, some differences from WSNs are worth noting: pervasive
computing \emph{(i)} appears to handle a wider set of networking scenarios,
which can possibly include global communications,
\emph{(ii)} is much less constrained by limitations in energy or computational power,
and \emph{(iii)} is intrinsically more ``open'' to handle a
heterogeneous set of devices and content (data, knowledge, media)
\citep{ZV-JPCC2011}.
Accordingly, in pervasive computing the interactions of devices typically
require both techniques of self-organization to make global properties
emerge, and expressive means to elaborate information (e.g., semantic
matching, or application-specific programmed matching).

%\todo[inline]{Make sure to have an explicit list of types of languages in
%pervasive computing.}
There are relatively few examples of pervasive computing DSLs with spatial
operators: most, like the LINDA coordination language~\citep{gelernter1992coordination}
and other tuple-space languages, largely seek to abstract
the network from the programmer entirely.
Thus, each of the
following sections describe a particular language or model.  {\it Tuples on
the Air (TOTA)} is a middleware for sharing data in the form of tuples
(i.e., lists) efficiently throughout a network.  Next, the {\it Chemical
Reaction Model} draws on inspiration from chemical reactions to shape how
data spreads throughout the network.
Finally, Zones-of-Influence (ZoI)
from the Peer-It system models the pervasive devices that can influence or interact with each
other.
%The following sections describe each in more detail.  Then, an
%implementation of the reference example ``T-Program'' serves as an
%illustration of the spatial properties of Pervasive Computing DSLs.

%%%%%%%%%%%%%%%%%%%%%%%%%%%%%%%%%%%%%%%%%%%%%%%%%%%%%%%%%%%%%%%%%%%%%%%%%%%%%%%
%% 1 sub-sub-section for each language type
%%%%%%%%%%%%%%%%%%%%%%%%%%%%%%%%%%%%%%%%%%%%%%%%%%%%%%%%%%%%%%%%%%%%%%%%%%%%%%%

\subsubsection{Tuples on the Air (TOTA)}
The main example of a programming framework for pervasive computing
incorporating ideas related to spatial computing is Tuples On the
Air (TOTA)~\citep{tota}
TOTA is a tuple-based middleware
supporting field-based coordination: each node hosts a tuple space,
from which tuples can diffuse in the network through neighbors, creating
spatial fields.
In TOTA each tuple, when inserted into a node of the network, is equipped
with content (the tuple data), a diffusion rule (the policy by which the
tuple is to be cloned and diffused), and a maintenance rule (the
policy whereby the tuple should evolve due to events or elapsed time).
Hence, it carries the behavior needed to identify its region of influence.
These behaviors are written in Java, making TOTA an extension DSL.

The only spatial operator directly supported by TOTA is the {\it neighbor}
concept.
Other concepts are possible like GPS absolute/relative position,
but they must be programmed on top of the TOTA API.
Using such concepts allows for more advanced spatial mechanisms, which
would not otherwise be supported natively in TOTA.

\subsubsection{Chemical Reaction Model}
Following the idea of TOTA, the work in \citep{VCMZ-TAAS2011} aims to
create a DSL for the coordination of pervasive computing
applications---as in TOTA, it addresses management of agent interaction,
not of agent behavior.
This is a biochemical-inspired language of semantic reactions: each
reaction dictates how the population of tuples, spread through the network, should
evolve and diffuse. Evolution is meant to exactly mimic chemistry, in that
a ``concentration value'' is carried in each tuple and is updated using
stochastic chemical models after the fashion of \citep{Gillespie:1977}.  Most
importantly here, diffusion is achieved by reactions producing so-called
``firing'' tuples, which schedule them for relocation to a neighboring
device. The destination is selected probabilistically, taking into
account a transfer rate characterizing each node-to-node interaction
channel.
The proposed language actually abstracts the details of semantic
matching, which is however recognized as a key ingredient of DSLs for
pervasive computing.
A preliminary DSL including semantic and spatial aspects into a
chemical-inspired framework is presented in \citep{VNCMZ-WOA2011}.
There, spatial aspects are handled as any other semantic information: the
existence (and relative position, distance, and orientation) are treated as
a tuple, and relocation of a tuple is achieved by modifying a specific
\texttt{location} property of it.

This work is an extension of the work in
\citep{MVRPD-SERENE2011-LNCS2011}, in which the syntax of chemical
reactions is directly applied to specify pattern rules for local
processing of tuples.
A pattern can be prepended by symbol $+$ meaning that the tuple is to be searched in a neighbour $r$ of the node $n$ in which the reaction is fired.
Additionally, variable $\#D$ can be used to mean the estimated distance of $r$ from $n$, and $\#T$ is the time at which the reaction is fired.
As an example, the reactions used to create a gradient structure of tuples would be:
\[\begin{array}{l}
\langle \texttt{gradient}, \texttt{Source}, \texttt{Dist} \rangle \mapsto \langle \texttt{gradient}, \texttt{Source}, \texttt{Dist} \rangle, +\langle \texttt{gradient}, \texttt{Source}, \texttt{Dist+\#D} \rangle \\ 
\langle \texttt{gradient}, \texttt{Source}, \texttt{Dist} \rangle, \langle \texttt{gradient}, \texttt{Source}, \texttt{Dist2} \rangle  \mapsto  \langle \texttt{gradient}, \texttt{Source}, min(Dist,\texttt{Dist2}) \rangle
\end{array}\]
The former propagates a clone of the gradient tuple in any neighbour, properly replacing the distance value; the latter takes two gradients tuples for the same source and retains the one with shortest distance from the source.

\subsubsection{Spatially Scoped Tuples}

An alternate approach is to explicitly encode spatial scope into
tuples, using relative or global coordinates, and then use
localization of devices to determine their distribution.

Geo-Linda~\citep{GeoLinda} is an example of such a DSL.  Derived from
the earlier SPREAD system~\citep{SPREAD03}, it combines the tuple
manipulation of LINDA with the geometric addressing concepts of
SPREAD.  In Geo-Linda, tuples are read and published over an
assortment of geometric primitives, such as boxes, spheres, cylinders,
and cones, all defined relative to a device.  The language also
introduces primitives to detect coarse movement of devices through the
appearance or disappearance of tuples.

Another example is the Peer-It system~\citep{Ferscha2008448,spacepervasive-pmc6}, in which a ``Zone-of-Influence'' (ZoI)
model is introduced to describe whether a pervasive device may or may not
influence the activity of another one depending on their relative position
in space.

Technically, a ZoI represents a spatial region (numerical position,
direction and spatial extent), but qualitative abstractions can be used
including concepts like: being in front or behind an object,
near/medium/far from an object, being in similar/opposite/left/right
direction, and covering disjoint/overlapping/equal areas.

Though this work does not form a complete programming language, it
does present a markup language to express such ZoIs which could be
used as the basis for specifications of behavior in another pervasive
DSL.

%%%%%%%%%%%%%%%%%%%%%%%%%%%%%%%%%%%%%%%%%%%%%%%%%%%%%%%%%%%%%%%%%%%%%%%%%%%%%%%
%% Reference Example
%%%%%%%%%%%%%%%%%%%%%%%%%%%%%%%%%%%%%%%%%%%%%%%%%%%%%%%%%%%%%%%%%%%%%%%%%%%%%%%
\subsubsection{Reference Example: TOTA} \label{sec:TOTAexample}
The lack of primitives for measuring or manipulating space means
that TOTA cannot implement most of the reference ``T-Program'' example.
Assuming a center of gravity has been calculated, however, the following
code shows how a ring pattern can be computed
in TOTA by means of a gradient tuple, spreading from the center
of gravity and keeping track of the estimated distance from that source.
(activation will then happen to tuples whose distances from the source
falls within a given range):
\begin{CodeBlock}
class RingTuple extends FieldTuple{
  ...
  protected boolean decideEnter() {
    RingTuple prev = (RingTuple)tota.keyrd(this);
    return prev == null ||
           prev.hop > (this.hop + 1);
  }
  protected void changeTupleContent() {
    hop++;
    if (hop <= RingTuple.RING_MAX &&
        hop >= RingTuple.RING_MIN) {
      this.ring_activation=true;
    }
  }
  protected void decidePropagate() {
    return true;
  }
}
\end{CodeBlock}
A gradient tuple always spreads in all neighbors because of the
implementation of method \texttt{decidePropagate()}.
As it is received in a neighbor, method \texttt{decideEnter()} is executed
to decide whether this tuple (\texttt{this}) is to be stored or not, which
in this case is true if no gradient tuple is already there (\texttt{prev})
or if this tuple has a lesser \texttt{hop} counter.
If the tuple is to be stored, method \texttt{changeTupleContent}
increments the hop counter by one, which sets a flag stopping the run if
the counter is in the desired range.

%%%%%%%%%%%%%%%%%%%%%%%%%%%%%%%%%%%%%%%%%%%%%%%%%%%%%%%%%%%%%%%%%%%%%%%%%%%%%%%
%% Connection to table and framework
%%%%%%%%%%%%%%%%%%%%%%%%%%%%%%%%%%%%%%%%%%%%%%%%%%%%%%%%%%%%%%%%%%%%%%%%%%%%%%%
\subsubsection{Analysis}
%\todo[inline]{Need connection to table and framework... might be in
%individual sections.}

The characteristics of the Pervasive Computing DSL classes are summarized
in Table~\ref{table:dsl},~\ref{table:spatial},~and~\ref{table:device},
based on the taxonomy proposed in Section~\ref{s:definitions}.

% DSL Table
As summarized in Table~\ref{table:dsl}, TOTA, a Java library, is an
imperative language extension designed for wired and wireless multi-hop
networks.  The chemical reaction model, although targeted at the same types
of networks as TOTA, has an invented syntax with rule-based semantics,
while the spatially-scoped tuple approaches tend to be extensions.

% Spatial Table
Table~\ref{table:spatial} shows the spatial properties of pervasive
computing DSLs.  TOTA, like the LINDA coordination language~\citep{gelernter1992coordination}
and other tuple-space languages, largely seeks to abstract spatial
properties of the network from the programmer.  Thus, TOTA has very few
spatial properties---with the exception of neighborhood propagation.
The chemical reaction model, on the other hand, makes use of a spatial
gradient for {\it diffusing} information to network neighbors.
Spatially scoped tuples, on the other hand, offer explicit patterns
over communication neighborhoods.

% Device Table
As far as the abstract device model summarized in
Table~\ref{table:device}, all pervasive computing DSLs utilize an
immobile (i.e., uniform) program targeted for discrete devices, though
the devices might be moved by external agents, such as humans
carrying or operating them.  The difference in frameworks comes from
their communication modalities.  TOTA uses a distributed
publish-subscribe mechanism for global, multicast communication and
has a neighborhood communication range.  The chemical reaction model
and spatially-scoped tuples also use a neighborhood communication
range, however, they can also direct messages to individual users
(i.e., unicast granularity).

%%%%%%%%%%%%%%%%%%%%%%%%%%%%%%%%%%%%%%%%%%%%%%%%%%%%%%%%%%%%%%%%%%%%%%%%%%%%%%%
%%%%%%%%%%%%%%%%%%%%%%%%%%%%%%%%%%%%%%%%%%%%%%%%%%%%%%%%%%%%%%%%%%%%%%%%%%%%%%%

\subsection{Swarm and Modular Robotics}
\label{s:robotics}

%%%%%%%%%%%%%%%%%%%%%%%%%%%%%%%%%%%%%%%%%%%%%%%%%%%%%%%%%%%%%%%%%%%%%%%%%%%%%%%
%%%%%%%%%%%%%%%%%%%%%%%%%%%%%%%%%%%%%%%%%%%%%%%%%%%%%%%%%%%%%%%%%%%%%%%%%%%%%%%

Multi-robot systems tend to be spatial both due to the locality of
their communication and the physical interactions of the robots.
In swarm robotics, the robots are typically not in physical contact
and may be spread fairly broadly through space: goals are typically
specified in terms of sensing and actuation interactions with the
external environment, such as mapping or search and rescue.  In
modular robotics, on the other hand, the robots are typically in
contact and working together to form a desired physical shape.  There
is a great deal of similarity in the control problems encountered in
these fields, however, and some recent projects (e.g.,
\citep{Swarmanoid}) bridge the two domains.

%%%%%%%%%%%%%%%%%%%%%%%%%%%%%%%%%%%%%%%%%%%%%%%%%%%%%%%%%%%%%%%%%%%%%%%%%%%%%%%
%% Description of field
%%%%%%%%%%%%%%%%%%%%%%%%%%%%%%%%%%%%%%%%%%%%%%%%%%%%%%%%%%%%%%%%%%%%%%%%%%%%%%%
\subsubsection{Swarm robotics} 
Swarm robotics emphasizes multi-robot systems with large number of
agents, individual simplicity, and local interactions. One of the promises
of swarm robotics is robust and scalable operation with applications such
as environmental monitoring, search and rescue using swarms of aerial
vehicles, or oil spill clean-up --- applications that take advantage of the
ability of a swarm to cover large amounts of space in parallel. Efficient
algorithms to these problems usually require the individual swarm members
to localize themselves either globally or locally with respect to each
other.  These physical capabilities also lend themselves directly to
measuring and manipulating space-time such as they are facilitated by the
Proto~\citep{ProtoSwarm} DSL, which has been demonstrated on robot swarms
with local range and bearing capabilities.  The typical approach to
swarm control with Proto has been to compute vector fields over the swarm,
computing with them and combining them to produce a commanded velocity
for each robot.

In order to control the shape of a swarm, e.g., to implement the
reference example ``T-Program,'' a robot would need the following
capabilities:
\begin{itemize}
 \item the ability to \emph{localize} to resolve a coordinate system
   (possibly indirectly, e.g., by inference from change of neighbors).
 \item the ability to \emph{communicate} either locally or globally to exchange state information, and
 \item the existence of \emph{unique identification numbers} to identify other robots' communication messages.
\end{itemize}
 With these abilities, the swarm could be either programmed using a
 multi-agent framework such as NetLogo using the code in Section
 \ref{sec:MASexample} or a manifold language such as Proto using
 the code in Section \ref{sec:protoexample}.  Here it is worth noting that
 the choice of language might heavily bias the performance of the
 implementation: for example, the NetLogo construct \emph{ask turtles} is implemented as a loop
 whose execution time scales linearly with the number of robots (and scales
 quadratically with respect to the total number of messages exchanged in
 the swarm), as each robot queries every other robot's coordinates. The
 Proto implementation's \emph{sum-hood} operator instead implies a
 broadcast operation that scales more favorably (but whose reliable
 implementation in a congested communication environment is dependent on
 its low-level implementation). 
 However, using a language such as NetLogo limits execution to
 simulation (hence the linear behavior of \emph{ask turtles}), whereas
 Proto compiles to local behavior, thus allowing the program to run on
 a real network of robots.

%%%%%%%%%%%%%%%%%%%%%%%%%%%%%%%%%%%%%%%%%%%%%%%%%%%%%%%%%%%%%%%%%%%%%%%%%%%%%%%
%% Breakdown of types of languages in the field
%%%%%%%%%%%%%%%%%%%%%%%%%%%%%%%%%%%%%%%%%%%%%%%%%%%%%%%%%%%%%%%%%%%%%%%%%%%%%%%
An additional class of swarms are those whose individual members do not
have access to global and/or local localization. This is the case for
bacteria assembling structures at the microscale \citep{martel2010} or
miniature robots imitating the capabilities of social insects whose
capabilities to localize are also limited to exploitation of gradient
fields (pheromones, e.g.) or crude global bearing estimates based on sun,
wind or magnetic field. Robotic instances of such systems include
collaborative manipulation \citep{martinoli04}, aggregation
\citep{correllijrr2011}, and clustering \citep{martinoli99a}. The behavior of
the individual units in these kind of swarms can usually be described using
a Finite State Machine  (FSM), which has both deterministic and
probabilistic transitions, where transition probabilities reflect the uncertainty of sensing and actuation on a
miniature robot. As localization is usually not
assumed, space is abstracted by assuming an average spatial
distribution of the swarm.  This distribution can be uniform and constant,
leading to constant probabilities for robots to encounter objects and each
other in the environment, or arbitrarily parameterized, leading to a time
and space-dependent encountering probability \citep{prorok10}. A DSL
targeted to this class of systems is MDL2$\epsilon$, which allows the
description of states and state transitions using an XML-based language and
compiles to bytecode for a virtual machine JaMOS \citep{szymanski07}.
JaMOS/MDL2$\epsilon$ is helpful to the programmer in abstracting the
hardware interfaces of a specific platform and facilitates programming of
the platform as only byte code needs to be transferred. It does provide
only little conceptional benefits, however. 

In contrast, the PRISM language \citep{KNP11} is a state-based language,
based on the Reactive Modules formalism \citep{alur99}  targeted at
probabilistic model checking of stochastic communication networks,
biological reaction networks and potentially robotic swarms that can be
described as probabilistic automata such as those encoded by MDL2$\epsilon$
or those hand-coded for systems like
\citep{martinoli04,correllijrr2011,martinoli99a}. After defining a
probabilistic automaton, Markov chain, or Markov Decision process, PRISM
compiles differential equations that model the average number of agents in
a certain state if possible, or uses Gillespie simulation
\citep{Gillespie:1977} otherwise. This allows the designer to quickly understand the
average stochastic dynamics that emerge from a specific set of rules.

%%%%%%%%%%%%%%%%%%%%%%%%%%%%%%%%%%%%%%%%%%%%%%%%%%%%%%%%%%%%%%%%%%%%%%%%%%%%%%%
%% 1 sub-sub-section for each language type
%%%%%%%%%%%%%%%%%%%%%%%%%%%%%%%%%%%%%%%%%%%%%%%%%%%%%%%%%%%%%%%%%%%%%%%%%%%%%%%
\label{sec:graphgrammars}
Graph grammars are another important approach to specifying swarm
formations.
Graph grammars, or graph rewriting systems, are rule sets that transform one graph into another. In a self-assembly context, e.g., for assembling the T-shape, a desired assembly can be represented as a graph. The assembly process becomes a sequence of (labeled graph) transformations of an edgeless graph into the target graph, known as Graph Rewriting Systems on Induced Subgraphs \citep{DBLP:conf/wg/LitovskyMZ92,Klavins2006}. This graph rewriting system takes one subgraph as input and has another subgraph as output (in the process removing or adding edges and changing the labels). Here, nodes represent individual robots, the rewriting represents reconfiguration of these robots, and the labels represent robot states. Rewriting rules are of the form:
\begin{equation*}
\phi_{fi}: X\quad A\Rightarrow Y - Z
\end{equation*}
For executing a rule the labels of the two modules constituting the subgraph
are compared to the LHS (left-hand side) of the graph grammar rule $\phi$$_{fi}$. If
these subgraphs match, they are replaced by the subgraph shown on
the RHS of the $\phi_{fi}$. This replacement indicates that the
states of the original modules $X$ and $A$ are replaced by $Y$ and $Z$
respectively. The actual physical connection between robots is indicated
by the existence or absence of edges `$-$' between the nodes of the subgraph
on either side of the rule $\phi_{i}$. We therefore refer to $\phi_{fi}$ as a \textbf{construction rule}. In order to break up a connection, we can define a \textbf{reversal rule}:
\begin{equation*}
\phi_{ri}: Y-Z \Rightarrow X \quad A
\end{equation*}
Thus, connections can be made
or broken between modules, depending on whether the LHS of the rule
is satisfied, which itself might depend on the existing connection
between modules represented by the subgraph. Reversal rules, can be executed by the environment
or by active decisions of the modules themselves, which then need to exchange their states and initiate a disconnect sequence when either one detects a valid LHS. In this chapter, we use the convention that atomic modules that are not part of a structure are in state $A$.

\subsubsection{Modular Robotics}
\emph{Modular robots} are reconfigurable robots that are constructed from
modules that have the ability to autonomously attach and/or detach from
each other to re-arrange themselves into different shapes \citep{yim07}.
There also exist modular robot systems that require manual assembly and
disassembly. The promise of modular robotic systems is increased
versatility, due to their ability to reconfigure into robots with different
functions, and robustness due to their potential to self-repair. As modular
robotic systems consist of tens to hundreds of actuators and distributed
computation, they pose deep challenges to DSLs for global-to-local programming.
We explicitly consider two specific problems: generating local rules for
module re-configuration from global descriptions of a desired shape, and
generating local rules for motion generation from global descriptions of a
desired spatio-temporal pattern.

\begin{itemize}
\item \textbf{Pattern formation} 
A DSL created for modular robotic systems
is Meld \citep{ashley07,ashley09}, a declarative logic programming language.
The goal of Meld is to simplify programming of modular robots, mainly with
respect to expressiveness and size of the resulting code, as well as enable
proofs of correctness based on the formal definition of the language. Meld
is indeed able to express spatial computing algorithms such as gradient
dissemination, shortest-path routing, and localization. These are important
primitives for expressing morphology changes, and Meld has been used to
implement morphological changes of large-scale distributed modular robotic
systems in simulation. Meld does not provide primitives for solving the
global-to-local programming problems for generating arbitrary patterns,
however. Such patterns are commonly defined in the form of a 3D matrix
using arbitrary graphical interfaces or directly using data structures
provided  the high-level programming language that implements the algorithm
for local rule generation. For example, a T-shape such as the one used in our running example could be defined as computer-aided design (CAD) model, which can then be used to generate motion plans and local rules.

These algorithms operate at the \emph{abstract
device} layer. A common abstraction is the assumption that modules are
\emph{unit-compressible}, that is they can contract and extend their faces
to move other modules within the structure and make room for other modules
\citep{rus01}. Other abstractions include modules that are capable of linear
motion on a plane of modules as well as convex and concave transitions
into another plane \citep{rus02,stoy04}, modular robots that have the
ability to disassemble as their sole mode of actuation \citep{gilpin08}, or simply
modules that can be created or disappear anywhere in a 3D lattice
\citep{dewey08}, among others. Finally, modular robots are dual to
robot swarms when they are able to move in free-space (e.g., fly,
swim, or roll). In this case, graph grammars as described in
Section~\ref{sec:graphgrammars} can be derived from a desired
target-graph such as the T-shape example in a 1-to-1 mapping. As graph
grammars only encode the result of two interactions, however, they are
limited to systems in which the generation of individual robot
trajectories is trivial and can be achieved by local sensing, e.g.,
random walk with mutual attraction of matching pairs.

As such, the specification of desired shapes
can be understood as a primitive form of a \emph{declarative} DSL, in which
the compiler essentially solves a path-planning problem \citep{walter04}.
These compilers then generate a sequence of (event-driven) actuations that
can be executed by the individual modules, often via a primitive virtual
machine.

Once modular robots are assembled into a static shape, they become targets
for any spatial computing programming approach, which then allows a
programmer to
implement algorithms such as the center-of-gravity example on a T-shape.
Few works, however, are concerned with spatial computing aspects that go
beyond the pattern formation problem.

\item \textbf{Motion generation} The problem of motion generation can be
expressed as a manipulation of space-time based on computed patterns. For
example, motion of an inchworm-like structure can be generated by
activating its actuators in a sinusoidal pattern. The resulting physical
evolution will lead to the equivalent of a traveling sine wave. 
Most often, this is done by means of centralized control, but a number
of spatial languages have emerged as well.

A DSL that we have already encountered that can be readily applied to
this purpose is Proto \citep{proto2006a}.  The actuations used are
different than for moving swarm robots, since the devices typically
remain fixed with respect to one another.  Instead, the programmer
manipulates shape with operations such as adjusting the curvature of
space (corresponding to joint actuation), scaling space (corresponding
to linear actuation), or adjusting its density (omnidirectional
expansion or contraction of a robot).  These continuous representation
are more indirect, but abstract the choice of specific
platform.

There are also specialized motion generation DSLs for modular
robotics.  One particularly powerful example is the DynaRole
language~\citep{DynaRole07}, which dynamically assigns behaviors to
modules using code that migrates from a seed over the graph of
connected modules, and has recently been extended to include
gossip-based synchronization and automatic reversibility of
behaviors~\citep{DynaRole11}.  This was made more spatial with the
addition of directional labels~\citep{SpatialLabels08}, and
coupled with Modular Mechatronics Modelling Language
(M3L)~\citep{M3L11}, a DSL for high-level kinematic specification of
modular robots.  Together, these allow a behavior to be specified
abstractly using labels, then to be automatically mapped onto the
spatial realization of any compatible platform's actuators, either in
automatically generated simulations or on physical robots.

A related effort (as well as one of the targets supported by DynaRole) is ASE~\citep{ASE11}, constructed
as a C library that provides a wide variety of aggregate space-time
operations, including broadcast, gossip, distance-measuring gradient,
consensus, and synch---quite similar to the aggregate operators in
Proto's library.
Another approach uses models of diffusing hormones to organize
motion: this has been investigated in both \citep{ShenHormones04}
and \citep{HormoneRobotics10}.

\end{itemize}

%%%%%%%%%%%%%%%%%%%%%%%%%%%%%%%%%%%%%%%%%%%%%%%%%%%%%%%%%%%%%%%%%%%%%%%%%%%%%%%
%% Reference Example
%%%%%%%%%%%%%%%%%%%%%%%%%%%%%%%%%%%%%%%%%%%%%%%%%%%%%%%%%%%%%%%%%%%%%%%%%%%%%%%
\subsubsection{Reference Example: Graph Grammars}

\begin{figure}
\centering
\includegraphics[width=0.4\textwidth]{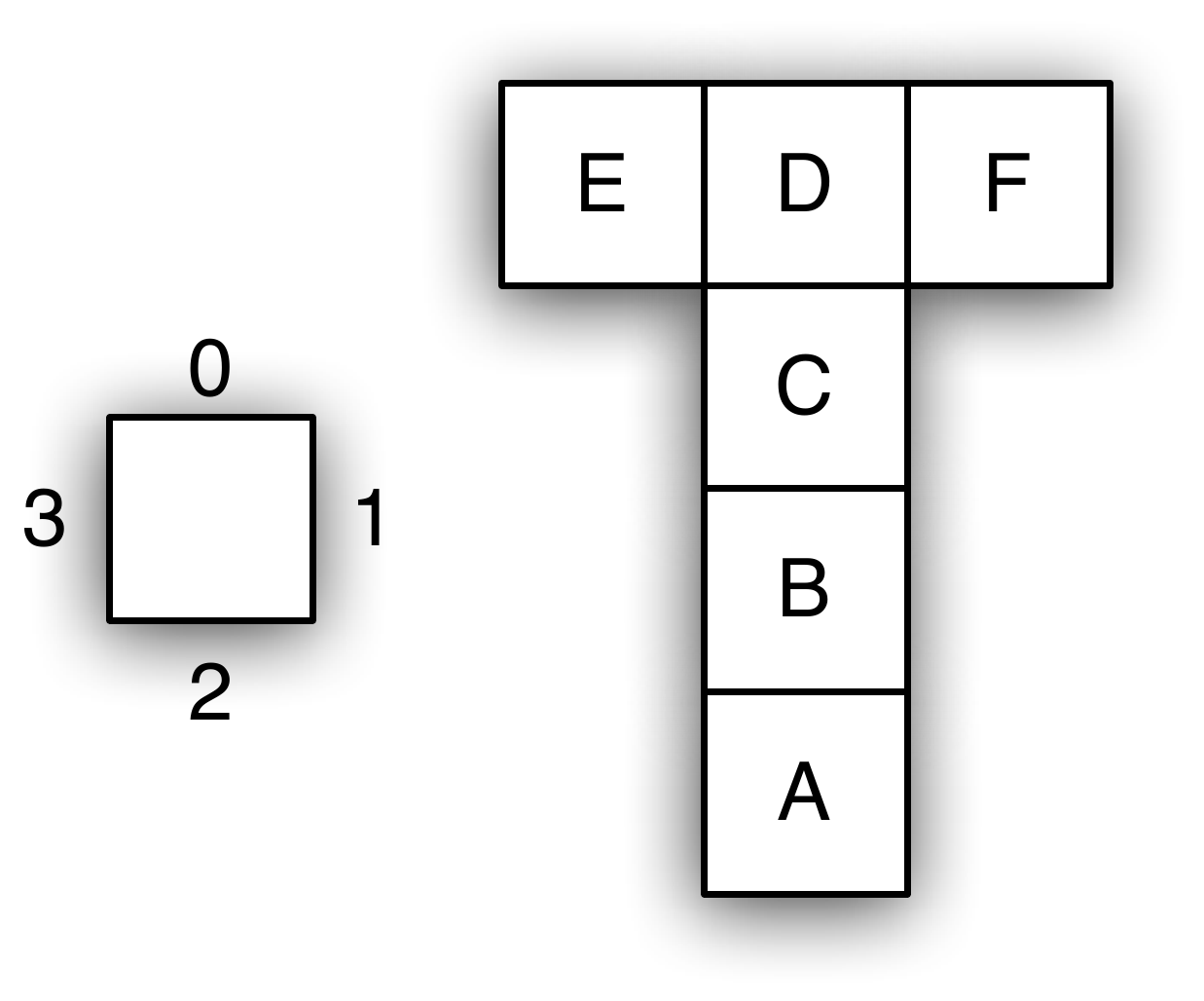} 
\caption{Construction of T-shape using a graph grammar.}
\label{f:graphex}
\end{figure}

For our reference ``T-Program'' example, in order to construct a T-Shape
consisting of 6 robots that has a width of 3 robots and a height of 4
robots, we could define the following rules that can be implemented by a
finite automata and require only local communication and orientation:
\begin{CodeBlock}
X X  -> A-B
B0 X -> B-C
C0 X -> C-D
D1 X -> D-F
D3 X -> D-E
\end{CodeBlock}
This program assumes that each robot is in state \texttt{X} initially. As
soon as two \texttt{X} meet, they begin constructing the T-shape from the
bottom, up. The notation \texttt{B0} indiciates the port of the robot at
which another robot can dock. In this example, we choose a 4-neighborhood,
labeling the ports from 0 to 3 in clockwise order, with 0 being at the
``top'' of a robot. The robots will therefore assemble a column
\texttt{ABCD} and then adding an \texttt{E} to the left and an \texttt{F}
to the right, forming a simple T-shape as shown in Figure~\ref{f:graphex}.

As soon as the shape is assembled, a coordinate system can be established
in a multi-hop fashion enabling a spatial computing approach such as Proto
to compute the center-of-gravity and produce the ring pattern.

%%%%%%%%%%%%%%%%%%%%%%%%%%%%%%%%%%%%%%%%%%%%%%%%%%%%%%%%%%%%%%%%%%%%%%%%%%%%%%%
%% Connection to table and framework
%%%%%%%%%%%%%%%%%%%%%%%%%%%%%%%%%%%%%%%%%%%%%%%%%%%%%%%%%%%%%%%%%%%%%%%%%%%%%%%
\subsubsection{Analysis}

The characteristics of robotics DSL classes are summarized in
Table~\ref{table:dsl},~\ref{table:spatial},~and~\ref{table:device}, based
on the taxonomy proposed in Section~\ref{s:definitions}.
By and large these DSLs focus on the manipulation of space.
For swarm robotics, the amorphous and agent languages already
encountered provide the most spatial approach to aggregate control,
but require that robots can obtain at least some local coordinate
information.  For modular robotics, in addition to Proto, which we
have already encountered, ASE gives a large toolkit of aggregate
space-time programming abstractions.  One concern with ASE, however,
is that to date it seems to have been tested only on small numbers of 
devices and so its scalability is not yet well established.

One of other notable things that appears in this domain are languages
with an explicit connection to formal verifiability: in particular,
Meld is both a spatial language and a predicate logic language, and
work has been done on proving correctness of Meld programs.  It is
unclear, however, how well this actually translates to modular
robotics, due to uncertainties in sensing, actuation, and communication.
Thus, despite the advantages shown by Meld and ASE, it is not yet
clear whether there is a general DSL for a large modular robotic
system that allows to efficiently maintain dynamic state and specify
the behavior of a large modular robotic system across both shape
reconfiguration and locomotion in response to external events.

%%%%%%%%%%%%%%%%%%%%%%%%%%%%%%%%%%%%%%%%%%%%%%%%%%%%%%%%%%%%%%%%%%%%%%%%%%%%%%%
%%%%%%%%%%%%%%%%%%%%%%%%%%%%%%%%%%%%%%%%%%%%%%%%%%%%%%%%%%%%%%%%%%%%%%%%%%%%%%%

\subsection{Parallel and Reconfigurable Computing}
\label{s:parallel}

%%%%%%%%%%%%%%%%%%%%%%%%%%%%%%%%%%%%%%%%%%%%%%%%%%%%%%%%%%%%%%%%%%%%%%%%%%%%%%%
%%%%%%%%%%%%%%%%%%%%%%%%%%%%%%%%%%%%%%%%%%%%%%%%%%%%%%%%%%%%%%%%%%%%%%%%%%%%%%%

%%%%%%%%%%%%%%%%%%%%%%%%%%%%%%%%%%%%%%%%%%%%%%%%%%%%%%%%%%%%%%%%%%%%%%%%%%%%%%%
%% Description of field
%%%%%%%%%%%%%%%%%%%%%%%%%%%%%%%%%%%%%%%%%%%%%%%%%%%%%%%%%%%%%%%%%%%%%%%%%%%%%%%
In {\em parallel computing} and {\em reconfigurable computing}, the
primary focus is on fast execution of complex computations.  Parallel
computing has tended to focus on architectures with many processors,
while reconfigurable computing has tended to focus on single chips
with many configurable processing elements.  In both cases, however,
the extremely high speed at which computations are executed mean that
time delays from communication between computing devices are a
dominant concern, whether those devices are individual transistors or
entire processors.  In high performance systems, the computing devices
are tightly packed together in 2D or 3D space, and thus the
communication cost is typically strongly-correlated with the distance
between devices.

Parallel and reconfigurable architectures have thus long embraced
spatiality.  Notable examples include cellular automata
machines~\citep{CAMs} such as the CAM-8~\citep{cam8}, processor grids
such as the Connection Machines~\citep{cm5}, tiled architectures such
as RAW~\citep{RAWarchitecture} and Warp~\citep{Warp}, reconfigurable
fabrics like Tartan~\citep{Tartan} and WaveScalar~\citep{WaveScalar},
and massively multicore systems like SVM/Microgrid~\citep{Microgrid}.
Indeed, DeHon's papers on the likely future evolution of such
architectures~\citep{DeHon99, DeHon02} are one of the origins of the
term ``spatial computing.''

%%%%%%%%%%%%%%%%%%%%%%%%%%%%%%%%%%%%%%%%%%%%%%%%%%%%%%%%%%%%%%%%%%%%%%%%%%%%%%%
%% Breakdown of types of languages in the field
%%%%%%%%%%%%%%%%%%%%%%%%%%%%%%%%%%%%%%%%%%%%%%%%%%%%%%%%%%%%%%%%%%%%%%%%%%%%%%%
The computation to be executed, however, often does not have a
structure with an obvious mapping to such spatial architectures.  As a
result, programming languages for parallel and reconfigurable systems
have embraced spatiality to a much lesser degree than the
architectures that they target.  With regards to spatial computing, we
classify languages for parallel and reconfigurable computing into
three categories: dataflow languages, topological languages, and field
languages.

%%%%%%%%%%%%%%%%%%%%%%%%%%%%%%%%%%%%%%%%%%%%%%%%%%%%%%%%%%%%%%%%%%%%%%%%%%%%%%%
%% 1 sub-sub-section for each language type
%%%%%%%%%%%%%%%%%%%%%%%%%%%%%%%%%%%%%%%%%%%%%%%%%%%%%%%%%%%%%%%%%%%%%%%%%%%%%%%
\subsubsection{Dataflow Languages}

Most DSLs for parallel and reconfigurable computing are what we will
categorize as ``dataflow languages:'' languages that attempt to
accelerate computations by identifying dependencies, so that
independent operations can be executed in parallel.  In effect, these
languages conceive of a computation as a partial order, yielding one
dimension of parallelism and one dimension of order or time.  The job
of the compiler is then to pack this dataflow graph as effectively as
possible into the two- or three-dimensional space of hardware
available.

The vast majority of these DSLs are either minor variants that extend
or piggyback on C (e.g. Cilk~\citep{Cilk}), FORTRAN
(e.g. HPF~\citep{HPF}), or MATLAB or else are circuit
specification languages like VHDL.  Many are implemented
simply through pre-processor macros or library calls that encourage a
programmer to code in a way that makes it easier to extract
parallelism.  A thorough discussion of such approaches for
reconfigurable computing can be found in \citep{Mucci07} and
\citep{Jozwiak10}; a similar survey for parallel computing can be found
in \citep{BarneyParallel}.

A more sophisticated approach is taken by languages for systolic
arrays or streaming.  These languages make the dataflow model
explicit.  Systolic array languages, such as SDEF~\citep{SDEF} and
ReLaCS~\citep{ReLaCS}, are the older technology and typically assume a
highly regular structure onto which the program must be decomposed.
Streaming languages such as StreamIT~\citep{StreamIT} and
SCORE~\citep{SCORE} operate conversely, allocating resources to balance
the needs of the computation.
Most recently, APIs like OpenCL~\citep{OpenCL} take advantage of modern
GPU hardware, which is often laid out in a stream-friendly manner: one
dimension of parallelism by one dimension of pipeline sequence, curled
to fit on a rectilinear chip.

Despite the strong spatial concerns in utilization of resources,
however, the languages themselves do not contain any spatial
operations.  Rather, the programmer is expressing constraints, which
imply things about spatial structure, but often fairly indirectly.
One recent and intriguing exception is Huckleberry~\citep{Huckleberry},
which provides ``split patterns'' that explicitly manipulate the
spatial relations between stages of a recursive computation, with
patterns like 1D mixing and 1D or 2D partitions.

\subsubsection{Topological Languages}

In the cluster of DSLs that we shall call ``topological languages,''
spatial locality is explicit, but abstracted from the two or
three dimensions of actual hardware on which a computation will be
executed.
Topological languages differ from dataflow languages in that the
computation is viewed in terms of information exchange amongst a
collection of processes, rather than a unidirectional flow.
These types of languages have been formalized with the $\pi$-calculus
and its relatives (discussed below in Section~\ref{s:formal}).

The classic example of the topological approach 
is MPI~\citep{MPI}, a library extension to C and Fortran in which a
computation is specified as a collection of processes that interact by
passing messages through shared memory.  MPI is widely used for
supercomputing applications, and this continues to be the case for its
descendants, OpenMP~\citep{OpenMP} and MPI-2~\citep{MPI2}.
Erlang is another widely used topological language.  Although
initially declarative, over time it has evolved into a functional
language in which processes interact by asynchronous message
passing~\citep{Erlang,HistoryOfErlang}.
More recently, major efforts have been undertaken to build concurrent
programming languages that scale well and support modern approaches to
safety and object oriented programming: X10~\citep{X10} at IBM,
Chapel~\citep{Chapel} at Cray, and Fortress~\citep{Fortress} at Sun.
All of these include explicit statements of locality (``places,''
``locales,'' and ``regions,'' respectively) that constrain interaction
and can be combined hierarchically into aggregate locations.

The GraphStep language~\citep{GraphStep} takes a different approach,
trading generality for efficiency.  In GraphStep, a computation is
expressed as a distributed graph processing algorithm, where at each
step every graph node receives messages from its neighbors, computes
over those messages, and then sends an update message along the edges
to its neighbors.  Given a computation and a dataset,
GraphStep maps the graph onto available hardware, decomposing
complex nodes and arranging nodes to balance the communication cost.
Similar ideas have been developed for embedding Markov Random
Field~\citep{ChenMundy03} and Bayes net
computations~\citep{RejimonBhanja05} in reconfigurable hardware, but
these have not been as well developed.

\subsubsection{Field Languages}

Finally, ``field languages'' are those DSLs that make an explicit
connection between the structure space-filling hardware and
computation with arrays of two or more dimensions.  These languages
thus tend to be the most spatial of the parallel and reconfigurable
computing DSLs.

The driver for field languages tends to be the recognition that many
high-performance computing applications are based on physical phenomena
that are themselves highly spatial, such as atmospheric simulation,
machine vision, VLSI design, and biological tissue simulation.

A major early example is StarLisp~\citep{StarLisp}, a functional
programming language for the Connection Machine.  In StarLisp, the
programmer manipulates ``pvars'' (parallel variables), which were
arrays with anywhere between 1 and 16 dimensions.  Each element of a
pvar was mapped to a different processor on the Connection Machine,
taking advantage of its hypercubic architecture to allow efficient
manipulation of these fields of data, including shifts along any
combination of dimensions, using either a grid (boundaries) or torus
(wrapping) topology.  

StarLisp faded along with the Connection Machines, however, and field
languages have remained a niche in high-performance computing,
primarily supported through libraries like FLIC~\citep{FLIC} or
RSL~\citep{RSL}.  The most interesting of these from a spatial
computing perspective are the Scalable Modeling System
(SMS)~\citep{SMS03}, which exposed spatial communication structure to
the programmer through the notion of manipulations on grids
partitioned into per-processor chunks, each of which interacts with
its neighbors through a cached ``data halo,'' and
PyNSol~\citep{PyNSol05}, which attempts to raise the level of
abstraction through a Python front-end where a programmer manipulates
objects with spatial types like ``torus'' and ``grid.''  There is a
growing recognition of the need for field languages in the
high-performance community, as evidenced by the recent
``rediscovery'' of spatial computing put forth in \citep{YangPNAS2011}.

The other direction from which field languages have been developed is
cellular automata.  A number languages have been developed to allow
succinct specification of cellular automata: examples include the
CAM-8 assembly language~\citep{cam8}, ALPACA~\citep{ALPACA},
CANL~\citep{CANL}, CAOS~\citep{CAOS}, CARPET~\citep{CARPET},
CELLANG~\citep{CELLANG}, JCASim~\citep{JCASim}, and
Trend/jTrend~\citep{TrendCAs}, as well as
Echo~\citep{forrest1994modeling} and NetLogo~\citep{sklar2007netlogo}
already discussed in previous sections.
Because they are all describing the same computational model, these
languages are all fairly similar: essentially declarative
specifications of the neighborhood structure and rules for the
evolution of cells.  Their differences are mostly in syntax: ALPACA
uses pseudo-english, CANL has lisp-like syntax, CARPET's syntax is
C-like, and JCASim is hosted in Java, and so on.  Some languages allow
only 2D cellular automata, while others support 1D or 3D as well, and
CAOS supports arbitrary dimensions.  An interesting generalization is
suggested by MacLennan's continuous spatial
automata~\citep{maclennanCSA}, which generalize the concept of CAs to
continuous space using differential equations, but this has not yet
been implemented in any DSL.

%%%%%%%%%%%%%%%%%%%%%%%%%%%%%%%%%%%%%%%%%%%%%%%%%%%%%%%%%%%%%%%%%%%%%%%%%%%%%%%
%% Reference Example: None
%%%%%%%%%%%%%%%%%%%%%%%%%%%%%%%%%%%%%%%%%%%%%%%%%%%%%%%%%%%%%%%%%%%%%%%%%%%%%%%

%%%%%%%%%%%%%%%%%%%%%%%%%%%%%%%%%%%%%%%%%%%%%%%%%%%%%%%%%%%%%%%%%%%%%%%%%%%%%%%
%% Connection to table and framework
%%%%%%%%%%%%%%%%%%%%%%%%%%%%%%%%%%%%%%%%%%%%%%%%%%%%%%%%%%%%%%%%%%%%%%%%%%%%%%%
\subsubsection{Analysis}

For this section, we do not provide a reference example, as it is not
particularly suitable for the languages of this domain: no parallel or
reconfigurable computing DSL supports measurement or manipulation of
space and only cellular automata (themselves at the edge the domain)
can create patterns.  

This fact reflects a long-standing assumption of the domain: that
hardware is essentially static with respect to programs.  The
programmer then lives in an idealized rectilinear environment and the
connection between computation and hardware thus becomes entirely the
job of the compiler and system management services.  This assumption
may not last, however, as the continued evolution of computing
hardware brings issues of power density and variable performance to
greater prominence.

The characteristics of the DSL classes are summarized in
Table~\ref{table:dsl},~\ref{table:spatial},~and~\ref{table:device},
based on the taxonomy proposed in Section~\ref{s:definitions}.  At
present, amongst the three classes of DSL that we have discussed, the
predominant languages of the domain are dataflow and topological
languages, which have minimal spatial operations and focus heavily on
the lower layers of our taxonomy.  The field languages are explicitly
spatial, but support only a very narrow range of operations on
rectilinear grids.  What the languages of this domain do have,
however, that is likely to be of interest for future development of
spatial DSLs, is a wide variety of models for how to specify
program control flow in a distributed environment.

%%%%%%%%%%%%%%%%%%%%%%%%%%%%%%%%%%%%%%%%%%%%%%%%%%%%%%%%%%%%%%%%%%%%%%%%%%%%%%%
%%%%%%%%%%%%%%%%%%%%%%%%%%%%%%%%%%%%%%%%%%%%%%%%%%%%%%%%%%%%%%%%%%%%%%%%%%%%%%%

\subsection{Formal calculi for concurrency and distribution}
\label{s:formal}

%%%%%%%%%%%%%%%%%%%%%%%%%%%%%%%%%%%%%%%%%%%%%%%%%%%%%%%%%%%%%%%%%%%%%%%%%%%%%%%
%%%%%%%%%%%%%%%%%%%%%%%%%%%%%%%%%%%%%%%%%%%%%%%%%%%%%%%%%%%%%%%%%%%%%%%%%%%%%%%

%%%%%%%%%%%%%%%%%%%%%%%%%%%%%%%%%%%%%%%%%%%%%%%%%%%%%%%%%%%%%%%%%%%%%%%%%%%%%%%
%% Description of field
%%%%%%%%%%%%%%%%%%%%%%%%%%%%%%%%%%%%%%%%%%%%%%%%%%%%%%%%%%%%%%%%%%%%%%%%%%%%%%%
A special form of DSL is the formal calculus, a primitive language used to
describe in an abstract way, certain features and behaviors of a system of
interest, in order to reason about its properties and possibly guide 
compliant implementations.
Examples include process algebras
\citep{PiCalculus,Priami:1995,stoklaim,ambients}, used to model distributed
system of communicating processes, membrane computing models
\citep{PSystems}, used to reason about chemical-inspired computing systems,
and core languages of programming languages \citep{FJ}, used to formally ground
application code.
In this section we review the formal calculi that are more related to
spatial computing---that is, those with first-class concepts of space.
It is worth noting that most of them are extensions of the archetype
process algebra $\pi$-calculus \citep{PiCalculus}, which intentionally
abstracts from topological issues of the computational network, 
modeling the overall system as a flat composition of processes that interact
through channels---a sort of ``space'' accessible to all processes that know
its name.

%%%%%%%%%%%%%%%%%%%%%%%%%%%%%%%%%%%%%%%%%%%%%%%%%%%%%%%%%%%%%%%%%%%%%%%%%%%%%%%
%% 1 sub-sub-section for each language type
%%%%%%%%%%%%%%%%%%%%%%%%%%%%%%%%%%%%%%%%%%%%%%%%%%%%%%%%%%%%%%%%%%%%%%%%%%%%%%%
\subsubsection{$3\pi$ Process Algebra}
$3\pi$ was developed as an extension of $\pi$-calculus with the idea of
modeling the space where processes execute as a 3-dimensional geometric
space \citep{DBLP:conf/cie/CardelliG10}.
In $3\pi$, each process has a position and an orientation in space (a
\emph{basis}), encoded in a so-called geometric data.
Other than accessing it (symbolically), a process can also send or receive
geometric data through channels and can evolve to new processes located
elsewhere (i.e., movement).
The interesting point of this approach is that geometric data are
manipulated in an abstract way, namely, only by \emph{frame shift}
operations (translation, rotation, and scaling).
For a process $p$ to move towards process $q$, $q$ must communicate its
position to $p$, $p$ should perform a subtraction operation between its
position and $q$'s obtaining a frame shift $f$, and then $p$ should evolve
to a new process to which frame shift $f$ is applied.
So, although there is indeed a unique coordinate system in the space,
processes do not know their position in it, but can just compare their
position/orientation with respect to others.
In analogous ways, one can model a force field as a process communicating
to (the processes of) mobile agents a frame shift they should apply to
themselves (e.g., a translation in a given direction), or a
developmental-like creation of ``matter'' can be modeled by processes
being spawned incrementally to form 3D structures as observed in
nature---e.g., lung development in mice as described in
\citep{DBLP:conf/cie/CardelliG10}.
The main motivation of the proposed approach is to describe and reason
about systems of developmental biology, where the evolution of biological
matter over time might be considered as a fabric for computing.

Another significant fact is that $3\pi$ has no embedded notion of time.
Processes just execute in each node and synchronize by the exchange of
messages (both sending and receiving are blocking operations).
A notion of time could be achieved by a global process sending a ``tick''
message to all others---an approach that would hardly result in a
meaningful implementation.
Similarly, processes form a flat set, with no notion of communication by
proximity.
Like in $\pi$-calculus, a process can send a message to another only if it
holds the other's unique name, and independently of its position.
Any notion of geometrically local communication must be encoded on top of the
model, resulting in specifications where both writing and reasoning are difficult.

\subsubsection{Ambient Calculus}
A different approach than $3\pi$ is taken in the Ambient calculus
\citep{ambients} and its derivatives --- like Brane Calculi
\citep{DBLP:conf/cmsb/Cardelli04} and P-systems \citep{Gheorghe2000108} --- in
which processes execute in a spatial system of hierarchically nested
compartments.
Each process is a located in a compartment, and can execute
a number of space-aware operations such as destroying the membrane of the
compartment to which it belongs, moving outside the current compartment, entering a
nested compartment, creating a new compartment, and so on.
Communication is not a primitive in the ambient calculus, but must be
implemented through interactions such as the diffusion of ``messenger
processes'' in and out of compartments.
Although based on a more primitive notion of space than the work in $3\pi$,
ambient-related approaches are interesting for their reliance upon an unconventional
notion of space. This is more often the norm, however, when considering the
computations carried-out in biochemical systems of cells and tissues of
cells.

% \subsection{Other}

% % This chunk is from Mirko
% another interesting pointer could be Membrane Computing. Take for
% instance Ambient Calculus: it is language to model a system of
% hierarchically nested compartments, with processes being localised in
% one of them and executing actions like "moving to the father
% compartment", "moving to a son", "creating a new compartment",
% "dissolving a membrane".

% This work is spatial in the sense that the effect of computation is how
% the hierarchy of compartments evolve over time -- this being a sort of
% virtual topology, not really a riemanian space, but still a space.

% There is a plethora of works like that, ranging from Ambient calculus,
% mobile ambients, KLAIM, PSystems and other "Membrane Computing"
% approaches---the latter is a whole field per se:

%http://en.wikipedia.org/wiki/Membrane_computing
%http://ppage.psystems.eu/

%%%%%%%%%%%%%%%%%%%%%%%%%%%%%%%%%%%%%%%%%%%%%%%%%%%%%%%%%%%%%%%%%%%%%%%%%%%%%%%
%% Reference Example
%%%%%%%%%%%%%%%%%%%%%%%%%%%%%%%%%%%%%%%%%%%%%%%%%%%%%%%%%%%%%%%%%%%%%%%%%%%%%%%
\subsubsection{Reference Example: $3\pi$}\label{sec:3piexample}
\begin{figure}
\centering
\includegraphics[width=0.4\textwidth]{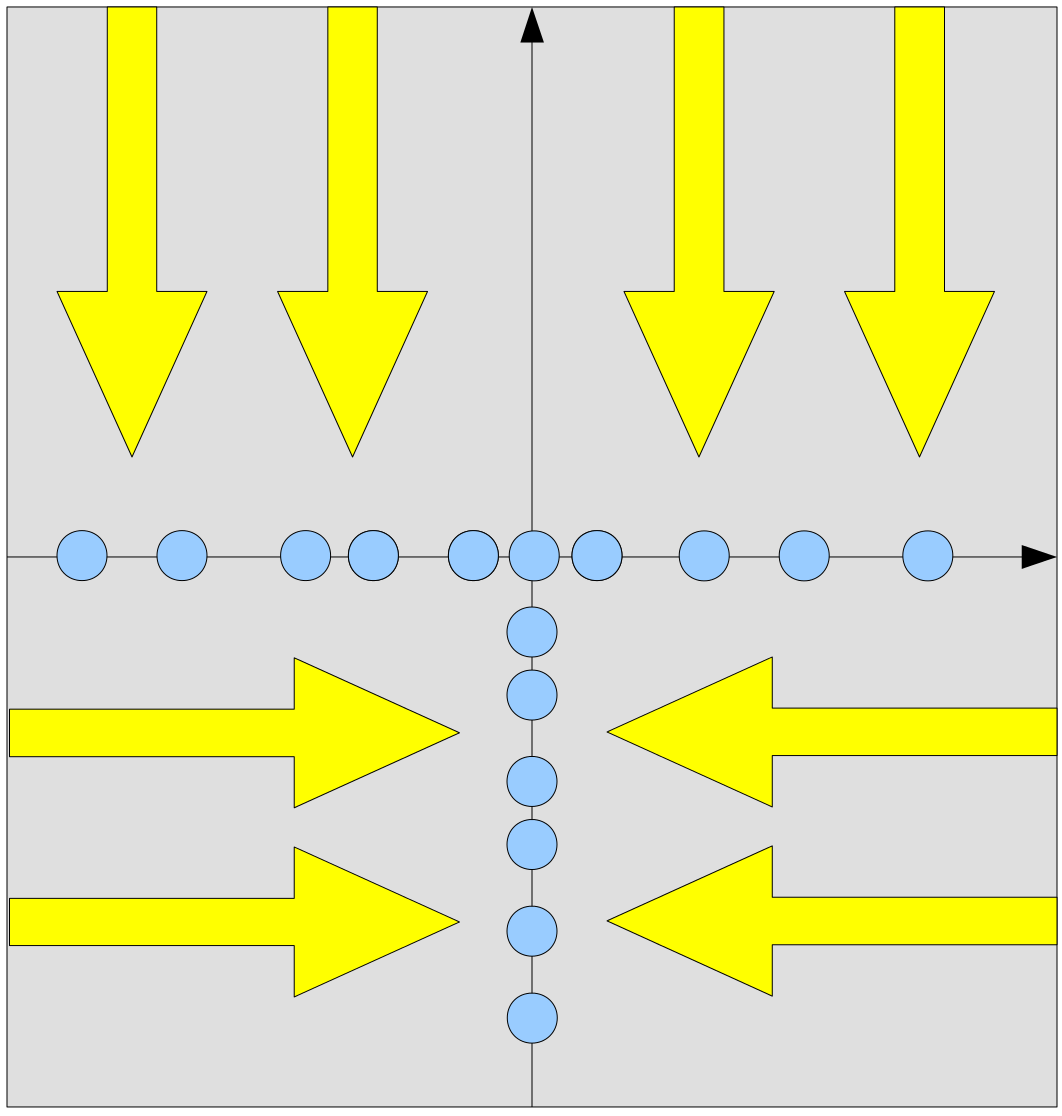} 
\caption{Creation of T-shape with $3\pi$}
\label{f:mirko}
\end{figure}

In order to illustrate programming in $3\pi$, we exploit a concrete
version of the $3\pi$ process algebra---a somewhat straightforward
variation of it which we devised for the sake of clarity, playing the same role that, e.g., PICT
\citep{DBLP:conf/birthday/PierceT00} would do for $\pi$-calculus
\citep{PiCalculus}.
Considering the reference example used in this chapter, we can create a
T-shaped structure (on the XY-plane, centered in the origin) by a force field,
namely by a process \texttt{t-force} that communicates to all interested
processes (\texttt{device}) the affine transformation they should apply to
themselves (moving to x-axis if their y-coordinate is positive, and
moving to y-axis otherwise) as shown in Figure~\ref{f:mirko}. 
This can be achieved by the specification:
\begin{CodeBlock} 
process T-force(ForceChannel) is
     ForceChannel.receive(Device-id,Device-position);      % receiving device info
     if ( scalarproduct(y-axis,Device-position) >=0)       % checking device position
        then Device-id.send( affinemap([x-axis,0,0],0) )   % moving to x-axis 
        else Device-id.send( affinemap([0,y-axis,0],0) );  % moving to y-axis
     call T-force(ForceChannel)                            % recursive call

process Device(ForceChannel) is                            
     generate Device-Id;                                   % generating a new channel
     ForceChannel.send(Device-id,myposition);              % sending device information
     Device-id.receive(Map);                               % receiving a map
     applymap Map;                                         % moving
\end{CodeBlock}

Note that an affine map is represented by a 3x3 matrix and a translation
vector, hence the map \texttt{affinemap([x-axis,0,0],0)} would, e.g.,
actually represent $\langle ((1,0,0),(0,0,0),(0,0,0)),(0,0,0) \rangle$, which
when applied to a position $(x,y,z)$ yields the new position $(x,0,0)$.

The implementation of other spatial computations, like identification of
center of gravity and drawing a ring around it, is not shown here for the
sake of brevity: the details depend very much on how the various devices
(namely, processes) are actually connected, which requires also developing
a representation of local communication.
An example approach would amount to let the process \texttt{T-force} also
\emph{(i)} initially create a new process \texttt{CG}, \emph{(ii)}
incrementally compute the average of all device positions it receives
(which as time passes tends to the center of gravity of all available
devices), and \emph{(iii)} communicate to \texttt{CG} to move towards that
average.
This would make \texttt{CG} move from the origin towards the actual
center of gravity.
Drawing a ring can then be similarly achieved by letting \texttt{CG} create a
new set of processes forming the ring.

%%%%%%%%%%%%%%%%%%%%%%%%%%%%%%%%%%%%%%%%%%%%%%%%%%%%%%%%%%%%%%%%%%%%%%%%%%%%%%%
%% Connection to table and framework
%%%%%%%%%%%%%%%%%%%%%%%%%%%%%%%%%%%%%%%%%%%%%%%%%%%%%%%%%%%%%%%%%%%%%%%%%%%%%%%
\subsubsection{Analysis}
%\todo[inline]{Need connection to table and framework.}

The characteristics of the Formal Calculi DSL classes are summarized in
Table~\ref{table:dsl},~\ref{table:spatial},~and~\ref{table:device}, based
on the taxonomy proposed in Section~\ref{s:definitions}.

% DSL
As summarized in Table~\ref{table:dsl}, both formal calculi analyzed, $3\pi$
and Mobile Ambients, are types of process algebras that extend the
$\pi$-calculus for defining independent processes that communicate through
message-passing channels.  However, whereas $3\pi$ processes are aware of
their position in space relative to other processes, ambient calculus
processes are aware of their spatial container.

% Spatial
Table~\ref{table:spatial} summarizes the spatial properties of the formal
calculi DSLs.  
$3\pi$ is capable of measuring geometric space relative to other processes,
and space is manipulated by specifying transformations on location
and orientation of processes.
The computation patterns for $3\pi$ processes are essentially
global---allowing execution of any named process.
Mobile ambients, on the other hand, can manipulate their containers (e.g.,
destroying the container's membrane) and communicate via patterns
like diffusion through container neighborhoods.

% Abstract Device
The abstract device characteristics of formal calculi DSLs are listed in
Table~\ref{table:device}.  Both process algebras focus on mobile processes
that operate on discrete devices and can address specific devices.  They
vary, however, in their communication range.  Where $3\pi$ can address
processes globally throughout the system, ambient processes communicate
with processes near their current container.

%%%%%%%%%%%%%%%%%%%%%%%%%%%%%%%%%%%%%%%%%%%%%%%%%%%%%%%%%%%%%%%%%%%%%%%%%%%%%%%
%%%%%%%%%%%%%%%%%%%%%%%%%%%%%%%%%%%%%%%%%%%%%%%%%%%%%%%%%%%%%%%%%%%%%%%%%%%%%%%
%%%%%%%%%%%%%%%%%%%%%%%%%%%%%%%%%%%%%%%%%%%%%%%%%%%%%%%%%%%%%%%%%%%%%%%%%%%%%%%
%%%%%%%%%%%%%%%%%%%%%%%%%%%%%%%%%%%%%%%%%%%%%%%%%%%%%%%%%%%%%%%%%%%%%%%%%%%%%%%
%%%%%%%%%%%%%%%%%%%%%%%%%%%%%%%%%%%%%%%%%%%%%%%%%%%%%%%%%%%%%%%%%%%%%%%%%%%%%%%

\section{Analysis}
\label{s:analysis}

%%%%%%%%%%%%%%%%%%%%%%%%%%%%%%%%%%%%%%%%%%%%%%%%%%%%%%%%%%%%%%%%%%%%%%%%%%%%%%%
%%%%%%%%%%%%%%%%%%%%%%%%%%%%%%%%%%%%%%%%%%%%%%%%%%%%%%%%%%%%%%%%%%%%%%%%%%%%%%%

As we have seen in the previous section, there are a plethora of
domain-specific languages (and DSL-like frameworks) that have been
designed to address spatial computing problems.  Although these span
many different domains, these DSLs nevertheless cluster into just a
few cross-cutting groups---a convergence that may be due to the
constraints imposed by space-time itself.  Ignoring borderline cases, we find
there to be four rough groups of spatial computing DSLs: device
abstraction languages, pattern languages, information movement
languages, and general purpose spatial languages.

\paragraph{Device Abstraction Languages}

The vast majority of DSLs for spatial computers are not particularly
spatial at all.  Instead, these {\em device abstraction languages}
attempt to simplify the aggregate programming problem by simplifying 
other complicating details from the programmer's perspective.  Therefore,
these languages tend to provide the most system management services.
They typically provide a great deal of control over how a program is implemented
locally but little (or nothing) in the way of spatial operations over
aggregates---leaving that as a problem for the programmer.  To
support distributed algorithm construction, however, these languages frequently provide strong
abstractions of local communication, which do greatly simplify the
specification of distributed algorithms.  

Device abstraction languages are found throughout all spatial
computing domains: for example, the agent languages
Repast~\citep{north2007declarative} and NetLogo~\citep{sklar2007netlogo},
neighborhood-based sensor network languages like Hood~\citep{hood}, as well as
languages like TOTA \citep{tota} in pervasive systems, MDL2$\epsilon$
\citep{szymanski07} in robotics, and MPI~\citep{MPI} in parallel computing
and GraphStep~\citep{GraphStep} in reconfigurable computing. 

\paragraph{Pattern Languages}

At a much higher level of abstraction, there are a number of {\em
  pattern languages} that focus on the distributed construction of
patterns over space.  These languages fall into three coarse
categories:
\begin{itemize}
\item {\em Bitmap} languages specify a pattern to be formed using a
  regular pattern of pixels or voxels.  These languages tend to be
  fairly rigid, and it is arguable whether they are properly languages
  at all.  These are found often in swarm and modular robotics,
  e.g., \citep{werfelphd} or \citep{stoy04}

\item {\em Geometric} languages specify the pattern as an arrangement
  of geometric constructs.  Frequently these include simple solids such as
  spheres and rectangles or Euclidean constructions such as lines and
  bisectors.  L-systems~\citep{LSystems} applied to biological modeling are an example of a geometric
  language.  When the constructions are tolerant of distortion, the
  patterns may be adaptive, as in the case of the OSL~\citep{nagpal}
  amorphous language.

\item {\em Topological} languages specify the pattern in terms of
  connectivity and hierarchical relationships between elements, which
  then need to be satisfied as best as possible by the arrangement of
  these elements in space.  Examples include the
  GPL~\citep{coorethesis} amorphous computing language and the
  ASCAPE~\citep{inchiosa2002overcoming} agent language.
\end{itemize}

In all of these cases, the pattern is specified without reference to
the process that will create it, except that the language
will often be constrained to certain classes of design.  The actual
implementation of how the pattern is computed then varies wildly from
language to language, as well as from domain to domain, including
generating a pattern on an existing surface (e.g., GPL \citep{coorethesis} in
amorphous languages and Yamins' self-stabilizing pattern language
\citep{yamins} in cellular automata), arranging swarm or modular robots
into a pattern~\citep{ashley07}, or collective construction of a
pattern by autonomous robots~\citep{werfelphd}.

\paragraph{Information Movement Languages}

The {\em information movement languages} focus on gathering
information sampled from regions of space-time and delivering it to
other regions of space-time.  As in the case of pattern languages, the
programmer typically specifies what information to gather and where it
needs to go (either in a push model or a pull model), but not how to
go about doing so or the degree of distribution~/~centralization of
collection points.  While
many of these languages do not include spatial relations, there is a
significant subclass that explicitly allow spatial constructs, which
tend to be based on space and time measuring operators.

Sensor networks include a large group of information movement
languages focused on data gathering, including languages like
TinyDB~\citep{tinydb} and Regiment~\citep{regiment}.  Although a few
languages in other domains are also focused on data, such as the
distributed system language KQML~\citep{Finin94kqml}, the information
movement languages of other domains tend to be more general, such
as the agent frameworks in Section~\ref{sec:agentFrameworks}.

\paragraph{General Purpose Spatial Languages}

Finally, we have a small but significant group of {\em General Purpose
  Spatial Languages} (GPSL).  These languages are still
domain-specific languages, in that they are specialized for spatial
computers, but are general in the sense that they have applicability
across a wide range of domains.  At their best, these languages
combine the strong spatial abstractions of pattern formation and
information movement languages with the abstract device languages'
ability to control implementation dynamics.  Unlike abstract device
languages, however, GPSLs permit a spatial aggregate view that allows
abstraction of individual devices.

The two currently most significant GPSLs that we have identified are
Proto~\citep{proto2006a} and MGS~\citep{GiavittoMGS02}.  Both of these
are relatively mature languages, have been applied successfully to a
number of different domains, and offer a wide range of spatial
operators, including meta-operators.

\subsection{Comparison of Languages}

At present, each group of spatial computing DSLs has significant
strengths and weaknesses.
\begin{itemize}
\item Device abstraction languages do little to cover the gap between
  device and aggregate, yet many of the important problems at the
  system management level are simply ignored by languages in the other
  three groups.  Device abstraction languages are currently the best
  approach for handling issues such as resource management, conflict
  management, logging, and security.

\item Pattern languages and information movement languages are
  typically excellent at carrying out a narrow set of tasks under a
  particular set of assumptions.  Their abstractions typically
  dissociate the programmer so far from the implementation, however,
  that it is impossible for the system to be reconfigured to combine
  tasks of different types or to operate under different assumptions.
  For example, a sensor network language aimed at data gathering may
  drop the rate of sampling when the network is overloaded, although
  the user may prefer to keep the rate the same but sample from fewer
  devices.

\item GPSLs combine the spatial aggregate abstractions lacking from
  device abstraction languages with the composibility and
  configurability lacking from pattern and information movement
  languages.  Their generality, however, means that the more
  specialized capabilities of these other languages are not built into
  GPSLs, and need to be implemented in libraries for the language.
  While this is not ultimately a limitation, at these languages
  present state of maturity these libraries are often not available
  and must be implemented by the programmer.
\end{itemize}

It is worth noting as well that every group contains a mixture of
different types of languages: imperative, functional, and
declarative languages exist in all groups.  Likewise,
every group of languages cuts across many different domains.  We take
this to be a confirmatory sign that, in fact, the spatial nature of a
DSL is an orthogonal attribute to the type of language.

The DSLs that we have examined can also be divided into languages that are proficient
at measuring/manipulating space-time and those that are proficient at
maintaining dynamic state. 
For example, the functional language Proto lends itself to spatial
computing with space-time operators like restriction, which can
modulate programs by changing the space-time region where they
execute.  Programs with dynamic state or state transitions, however,
are cumbersome in Proto.
On the other hand, a language such as MDL2$\epsilon$, which deals
naturally with dynamic state and state transitions, has no aggregate
space-time operators, making it cumbersome of write complex spatial
programs.
At present, there seem to be no languages that adequately address the
problem of both maintaining dynamic state and also efficiently
describing global spatio-temporal system behavior.

This lack may be due to a fundamental tension between asynchronous
parallel execution and state transitions.  In general, it is 
impossible to guarantee that these three statements always hold:
\begin{itemize}
\item Every device in the spatial computer makes the same state
  transition.
\item Decisions are made in parallel at separate devices.
\item State transitions occur more frequently than the time to send a
  message across the diameter of the spatial computer.
\end{itemize}
To understand why this is, consider the following: assume two devices,
$A$ and $B$, have a choice between two state transitions, and that the
decision will depend on the value of a piece of information that has
just appeared at $A$.  If $B$ is always to decide the same way as $A$,
then it must get $A$'s information.  Device $A$, however, might be as
far away as the whole diameter of the spatial computer.
Thus, we are always left with a choice of which property to weaken:
parallelism, speed, or coherence.

There are many reasonable ways to approach this problem, and different
languages have made different choices.  Different spatial computing
languages choose different ways to make this trade-off.  For example,
both Proto and MDL2$\epsilon$ weaken coherence, but in different ways:
Proto chooses to keep the aggregate model coherent, at the cost of
making state transitions cumbersome, while MDL2$\epsilon$ keeps state
transitions simple at the cost of having no operations that span
regions of space-time.  OSL, on the other hand, weakens speed,
executing a sequence of space-time operators by inserting
synchronization barriers where the program waits for long enough to
ensure that the last set of state transitions have completed.  There
are, however, many cases where a programmer would want to have a
mixture of strategies, such as programming a sequence of swarm
behaviors.  At present, there is no language that allows a programmer
to elegantly make trade-offs between the different options.

Another limitation of the current spatial DSLs is that almost none of
them provide any benefits beyond a concise description of the desired
global behavior.  This is in contrast to formal languages that provide
correctness guarantees and to probabilistic modeling tools such as
PRISM, which can automatically generate the expected (i.e., average)
spatio-temporal trajectories of a swarming system defined as a
probabilistic FSM.

In sum, therefore, what is currently lacking are spatial computing
DSLs that:
\begin{itemize}
\item allow general combination of aggregate programming
  and state-based programming schemes,
\item have the ability to warn about programs that use functionality
  that go beyond the physical or hardware capabilities of a specific
  platform, e.g., localization capabilities to perform space-time
  measurements or the maneuvering capabilities of a robot, and
\item both generate executables for an actual platform and also
  generate models for tool-assisted formal verification.
\end{itemize}
Whereas the first item can likely be achieved by combining existing
approaches, we expect the other two to pose deep research issues.
This is due to the fact that physical assumptions and correctness
conditions are often not explicitly defined in code, but result
implicitly from environment interaction.  One possibility is to design
a DSL so that the programmer needs to supply goal and assumption
information; another is to cross-compile a DSL to an embodied
simulator that would use transitions extracted from past experimental
data using system identification to predict the behavior of new
programs.  Likely, these challenges are areas where the connection to
particular domains will continue to play a major role.

\section{Conclusions and Future Research Directions}
\label{s:roadmap}

As we have seen, there are a large number of spatial computing DSLs
that attempt to bridge the gap between the aggregate programming needs
of users and the execution of programs on individual devices.  These
languages have emerged across a number of different application
domains, yet share broad commonalities likely due to the fact that all
are dealing with spatial constraints.

Looking forward, the increasing complexity of engineered systems is
likely to favor the increased development of GPSLs: device abstraction
languages do not offer enough leverage on the aggregate programming
problem, while pattern and information movement languages tend to be
too specialized for practical use in large-scale systems.
Besides the pragmatic questions of language and library development,
we see four key research directions that are critical to the future
development of GPSLs:
\begin{itemize}
\item Although a major focus of GPSLs is robustness to problems and
  failures, no GPSL currently exposes error handling or quality of
  service (QoS) information to the programmer.  
  Addressing error handling and QoS scalably for a distributed aggregate is
  an open research question.

\item The pragmatic issues of the system management layer are largely
  unaddressed by current GPSLs, sharply limiting their usability in
  deployed systems.  Resolving this is likely to involve building on
  top of device abstraction languages that do address issues such as
  security and logging: the key question is how to best expose these
  issues to the programmer in an aggregate abstraction.

\item No GPSL currently offers full support for first class functions.
  A key reason for this is that, although there are many ways in which
  distributed function calls have been implemented, yet no language 
  resolves some of the fundamental problems of
  identity~\citep{spatialprocesses} and scope~\citep{BealUsbeck11} in functions
  that are defined at run-time.

\item Finally, there are a number of common programming paradigms,
  such as publish/subscribe, observer/controller, and first order
  logical inference, that are difficult or impossible to implement on
  current GPSLs.  It is unclear whether the problem is due to
  limitations in existing GPSLs, or whether it is due to scalability
  problems or hidden assumptions in how these paradigms are currently
  implemented.  Future research on spatial computing DSLs needs to find
  ways to bridge this gap, so that the benefits of these programming
  paradigms can be adapted for the aggregate environment.
\end{itemize}

In addition to GPSLs, we expect that there will be a continued role
for more specialized spatial computing DSLs, as they can directly
address domain- and application-specific issues that GPSLs cannot, as
well as taking advantage of assumptions that come from a more
restricted scope.
An important research direction for future work on specialized spatial
computing DSLs, however, will be to determine whether they can be
implemented in terms of existing GPSLs, whether as libraries or as
variations of the language.  Doing so will allow a new spatial
computing DSL to bootstrap its space-time operators from the theory,
software, and system management resources of its base GPSL, rather
than needing to build up from scratch.  This will also benefit GPSLs,
by pushing their boundaries and driving them to support the needs to
the DSLs they come to host.

It is our hope that such future research directions will be aided by a
clearer understanding of the properties and relationships of spatial
computing DSLs, such as that offered by the framework, survey, and
analysis in this chapter.  

\section{Key Terms and Definitions}

\begin{itemize}
\item {\bf Spatial Computer:} a collection of local computational
  devices distributed through a physical space, in which the
  difficulty of moving information between any two devices is strongly
  dependent on the distance between them, and the ``functional goals''
  of the system are generally defined in terms of the system's spatial
  structure.

\item {\bf Global-to-Local Compilation:} transformation of a program
  for an aggregate of devices into a program that can execute on
  individual deviecs.

\item {\bf Space-Time Operations:} aggregate programming abstractions
  falling into one of five categories: measurement of space-time,
  computation of patterns over space-time, manipulation of space-time,
  physical evolution, and meta-operations.

\item {\bf Abstract Device Layer:} abstraction that hides details of
  the device where a program is executing.

\item {\bf System Management Layer:} mechanisms that provide low-level
  ``operating system'' services, such as real-time process management,
  sensor and actuator drivers, or low-level networking.

\item {\bf Physical Platform:} the actual computing device where a
  program is executed.

\item {\bf Device Abstraction Languages:} spatial computing DSLs that
  simplify aggregate programming by hiding details but without much
  power in the way of spatial abstractions.

\item {\bf Pattern Languages:} spatial computing DSLs that focus on 
  construction of patterns over space, typically at the expense of
  more general computation.

\item {\bf Information Movement Languages:} spatial computing DSLs
  that focus on gathering information in one region of space-time and
  delivering it to another region, typically at the expense of more
  general computation.

\item {\bf General Purpose Spatial Languages:} languages that provide
  a wide range of powerful spatial abstractions for aggregate
  programming, but typically require more work to apply to any
  particular domain.

\end{itemize}

% Note: this section must contain nothing but APA references
\section{Additional Reading}

%Amorphous Computing (Abelson et al 2000)
\noindent %\cite{abelson2000amorphous}
Abelson, H., Allen, D., Coore, D., Hanson, C., Homsy, G., Knight~Jr, T.,
  Nagpal, R., Rauch, E., Sussman, G., and Weiss, R. (2000).
\newblock {Amorphous computing}.
\newblock {\em Communications of the ACM}, 43(5):74--82.

%ASRM (Regli et al., 2009)
\vspace{\baselineskip} \noindent %\cite{asrm}
Regli, W.~C., Mayk, I., Dugan, C.~J., Kopena, J.~B., Lass, R.~N., Modi, P.~J.,
  Mongan, W.~M., Salvage, J.~K., and Sultanik, E.~A. (2009).
\newblock Development and specification of a reference model for agent-based
  systems.
\newblock {\em Trans. Sys. Man Cyber Part C}, 39:572--596.

%ASRA (Nguyen et al., 2010)
\vspace{\baselineskip} \noindent %\cite{asra}
Nguyen, D.~N., Usbeck, K., Mongan, W.~M., Cannon, C.~T., Lass, R.~N., Salvage,
  J., and Regli, W.~C. (2010).
\newblock A methodology for developing an agent systems reference architecture.
\newblock In {\em 11th International Workshop on Agent-oriented Software
  Engineering}, Toronto, ON.

%BioDesign Automation review chapter
\vspace{\baselineskip} \noindent %\cite{BioDesignChapter2011}
Beal, J., Phillips, A., Densmore, D., and Cai, Y. (2011b).
\newblock High-level programming languages for bio-molecular systems.

%Mottola & Picco, 2011
\vspace{\baselineskip} \noindent %\cite{mottola2011programming}
Mottola, L. and Picco, G. (2011).
\newblock Programming wireless sensor networks: Fundamental concepts and state
  of the art.
\newblock {\em ACM Computing Surveys (CSUR)}, 43(3):19.

%Mucci, Campi, Brunelli, & Nurmi, 2007
\vspace{\baselineskip} \noindent %\cite{Mucci07}
Mucci, C., Campi, F., Brunelli, C., and Nurmi, J. (2007).
\newblock Programming tools for reconfigurable processors.
\newblock In Nurmi, J., editor, {\em System-On-Chip Computing for ASICs and
  FPGAs on Processor Design}, pages 427--446. Springer.

%Jozwiak, Nedjah, & Figueroa, 2010
\vspace{\baselineskip} \noindent %\cite{Jozwiak10}
Jozwiak, L., Nedjah, N., and Figueroa, M. (2010).
\newblock Modern development methods and tools for embedded reconfigurable
  systems: A survey.
\newblock {\em Integration, the VLSI Journal}, 43(1):1--33.

%Barney, Retrieved Feb. 20, 2012
\vspace{\baselineskip} \noindent %\cite{BarneyParallel}
Barney, B. (Retrieved Feb. 20, 2012).
\newblock Introduction to parallel computing.\\
\newblock {\tt https://computing.llnl.gov/tutorials/parallel\_comp/}.

\bibliographystyle{apalike}
\bibliography{mainbib}

\end{document}

%% file: dsl-table.tex
\newcommand{\LanguageTableStart}
{\TableStart{|l|cccc|}{DSL & Type & Pattern & Platform & Layers \\}}
\newcommand{\LanguageCluster}[1]{\hline {\bf #1} & & & & \\}
\newcommand{\LanguageEntry}[5]{#1 & #2 & #3 & #4 & #5 \\}

\LanguageTableStart

\LanguageCluster{Amorphous Computing}
\LanguageEntry{Proto}{Functional}{Invention}{Any}{SC,AD}
\LanguageEntry{PyMorphous}{Imperative}{Extension}{Any Network}{SC,AD}
\LanguageEntry{ProtoVM}{Imperative}{Invention}{Any Network}{AD,SM}
\LanguageEntry{Growing Point Language}{Declarative}{Invention}{Any Network}{SC}
\LanguageEntry{Origami Shape Language}{Imperative}{Invention}{2D Mesh Network}{SC}

\LanguageCluster{Biological}
\LanguageEntry{L-systems}{Functional}{Invention}{Simulation}{SC}
\LanguageEntry{MGS}{Declarative}{Invention}{Simulation}{SC,AD}
\LanguageEntry{Gro}{Imperative}{Invention}{Simulation}{AD}
\LanguageEntry{GEC}{Functional}{Invention}{Biological cells}{AD}
\LanguageEntry{Proto BioCompiler}{Functional}{Piggyback}{Biological cells}{AD}

\LanguageCluster{Agent-Based}
\LanguageEntry{Graphical Agent Modeling Language}{Graphical}{Extension}{Conceptual}{AD}
\LanguageEntry{Agent Framework}{Imperative*}{Extension}{Any Network}{AD,SM}
\LanguageEntry{Multi-agent Modeling and Simulation Toolkit}{Any}{Any}{Any}{SC,AD,SM}
\multicolumn{5}{|l|}{* JESS, being declarative, is a notable exception in this group} \\

\LanguageCluster{Wireless Sensor Networks}
\LanguageEntry{Regions based DSLs*}{Imperative}{Extension}{Wireless Network}{AD}
\LanguageEntry{Data-flow based DSLs}{Imperative}{Invention}{Wireless Network}{AD,SM}
\LanguageEntry{Database-like DSLs}{Declarative}{Piggyback}{Wireless Network}{SC}
\LanguageEntry{Centralized-view DSLs}{Imperative}{Piggyback}{Wireless Network}{AD}
\LanguageEntry{Agent-based DSLs}{Imperative}{Extension}{Wireless Network}{AD}
\multicolumn{5}{|l|}{* Regiment, an invented functional language is a notable exception in this group} \\

\LanguageCluster{Pervasive Computing}
\LanguageEntry{TOTA}{Imperative}{Extension}{Wireless/Wired Network}{AD,SM}
\LanguageEntry{Chemical reaction model}{Declarative}{Invented}{Wireless/Wired Network}{AD,SM}
\LanguageEntry{Spatially-Scoped Tuples}{Imperative}{Extension}{Wireless/Wired Network}{AD,SM}

\LanguageCluster{Swarm \& Modular Robotics}
\LanguageEntry{Bitmap Language}{Descriptive}{Invented}{Swarms and Modular Robots}{SC}
\LanguageEntry{Graph Grammars}{Functional}{Invented}{Robot Swarms}{SC,AD}
\LanguageEntry{PRISM}{Declarative}{Invented}{Robot Swarms}{AD}
\LanguageEntry{Meld}{Declarative}{Extension}{Modular Robots}{SC,AD}
\LanguageEntry{DynaRole/M3L}{Imperative/Declarative}{Invention}{Modular Robots}{SC,AD}
\LanguageEntry{ASE}{Imperative}{Extension}{Modular Robots}{SC,AD,SM}

\LanguageCluster{Parallel \& Reconfigurable}
\LanguageEntry{Dataflow DSLs}{Any}{Any}{Parallel Hardware}{SM,AD}
\LanguageEntry{MPI}{Imperative}{Extension}{Parallel Hardware}{SC,AD,SM}
\LanguageEntry{Erlang}{Functional}{Invented}{Parallel Hardware}{SC,AD,SM}
\LanguageEntry{X10/Chapel/Fortress}{Imperative}{Invented}{Parallel Hardware}{SC,AD,SM}
\LanguageEntry{GraphStep}{Imperative}{Invented}{Parallel Hardware}{SC,AD,SM}
\LanguageEntry{StarLisp}{Functional}{Piggyback}{Parallel Hardware}{SC,AD}
\LanguageEntry{Grid Libraries}{Imperative}{Extension}{Parallel Hardware}{SC,AD,SM}
\LanguageEntry{Cellular Automata}{Declarative}{Invented}{Simulation}{SC,AD}

\LanguageCluster{Formal Calculi}
\LanguageEntry{$3\pi$}{Process Calculus}{Extension}{Abstract geometric space}{PP,AD}
\LanguageEntry{Mobile ambients}{Process Calculus}{Extension}{Abstract nested compartments}{PP,AD}

\TableEnd{table:dsl}{DSL characteristics of spatial computing languages.}

%% file: spatial-table.tex
\newcommand{\SpatialTableStart}
{\TableStart{|p{1.4in}|p{0.9in}p{0.9in}p{1.2in}p{0.5in}p{0.5in}|}{DSL & Measure & Manipulate & Pattern & Evolve & Meta \\}}
\newcommand{\SpatialCluster}[1]{\hline {\bf #1} & & & & & \\}
\newcommand{\SpatialEntry}[6]{#1 & #2 & #3 & #4 & #5 & #6 \\}

\SpatialTableStart

\SpatialCluster{Amorphous Computing}
\SpatialEntry{Proto}{Duration, Local Coordinates, Density, Curvature}{Vector Flow, Frequency, Density, Curvature}{Neighborhood, Feedback}{Modular}{Functional, Domain Restriction}
\SpatialEntry{PyMorphous}{Duration, Local Coordinates}{Vector flow}{Neighborhood}{-}{Procedural}
\SpatialEntry{ProtoVM}{Duration, Local Coordinates, Density, Curvature}{Vector Flow, Frequency, Density, Curvature}{Neighborhood, Feedback}{Modular}{Procedural}
\SpatialEntry{Growing Point Language}{-}{-}{Line growth, tropisms}{-}{-}
\SpatialEntry{Origami Shape Language}{-}{Fold}{Huzita's axioms}{-}{-}

% Butera	Imperative		Network	AD		---		---	---	Virql code	Discrete	Neighborhood		Mobile	Viral propagation	---	---	Viral propagation	Shared memory	Viral propagation	---	---	

\SpatialCluster{Biological}
\SpatialEntry{L-systems}{-}{Local Rewrite}{-}{-}{-}
\SpatialEntry{MGS}{Topological Relations, Local Coordinates}{Topological Rewrite, Geometric Location}{Neighborhood}{-}{Functional}
\SpatialEntry{Gro}{Duration, Volume}{Frequency, Growth}{Rates}{Growth, Diffusion, Reactions}{-}
\SpatialEntry{GEC}{-}{-}{Diffusion}{-}{Functional}
\SpatialEntry{Proto BioCompiler}{Duration,Density}{Frequency}{Diffusion, Feedback}{Modular}{Functional}

\SpatialCluster{Agent-Based}
\SpatialEntry{Graphical Agent Modeling Language}{-}{-}{-}{-}{-}
\SpatialEntry{Agent Framework}{-}{-}{-}{-}{-}
\SpatialEntry{Multi-agent Modeling and Simulation Toolkit}{Distance,Time}{Physical Movement}{Diffuse}{-}{-}

\SpatialCluster{Wireless Sensor Networks}
\SpatialEntry{Region-based DSLs}{Distance}{-}{Regions}{-}{- *}
\SpatialEntry{Data-flow based DSLs}{-}{-}{-}{-}{-}
\SpatialEntry{Database-like DSLs}{Distance, Time}{-}{Surfaces, Time Intervals}{-}{-}
\SpatialEntry{Centralized-view DSLs}{-}{-}{-}{-}{-}
\SpatialEntry{Agent-based DSLs}{-}{-}{-}{-}{-}
\multicolumn{6}{|l|}{* Being a functional language, Regiment offers functional composition and abstraction} \\

\SpatialCluster{Pervasive Computing}
\SpatialEntry{TOTA}{-}{-}{Neighborhood}{-}{-}
\SpatialEntry{Chemical reaction model}{Transfer rate}{-}{Neighbor diffusion}{-}{-}
\SpatialEntry{Spatially-Scoped Tuples}{Movement}{-}{Neighborhood Geometry}{-}{-}

\SpatialCluster{Swarm \& Modular Robotics}
\SpatialEntry{Bitmap Language}{-}{Physical Movement, Shape}{-}{-}{-}
\SpatialEntry{Graph Grammars}{-}{Shape}{-}{-}{-}
\SpatialEntry{PRISM}{Time}{-}{-}{-}{Grouping of states}
\SpatialEntry{Meld}{Time}{Physical Movement, Shape}{-}{-}{-}
\SpatialEntry{DynaRole/M3L}{Angles,Time}{Physical Movement, Shape, Angles}{-}{Kinematics}{-}
\SpatialEntry{ASE}{-}{Physical Movement, Shape}{Broadcast, gossip, gradient, consensus, synchronization}{-}{-}

\SpatialCluster{Parallel \& Reconfigurable}
\SpatialEntry{Dataflow Languages}{-}{-}{Array *}{-}{Procedural}
\SpatialEntry{MPI}{-}{-}{-}{-}{Procedural}
\SpatialEntry{Erlang}{-}{-}{-}{-}{Functional}
\SpatialEntry{X10/Chapel/Fortress}{-}{Locality}{Locality}{-}{Procedural, Locality}
\SpatialEntry{GraphStep}{-}{-}{Neighborhood}{-}{-}
\SpatialEntry{StarLisp}{-}{-}{Shifts}{-}{Functional}
\SpatialEntry{Grid Libraries}{-}{-}{Neighborhood}{-}{Procedural}
\SpatialEntry{Cellular Automata}{-}{-}{Neighborhood}{-}{-}
\multicolumn{6}{|l|}{* Huckleberry also offers ``split patterns''} \\

\SpatialCluster{Formal Calculi}
\SpatialEntry{$3\pi$}{Geometric position}{Translation, Rotation, Scaling}{-}{Force fields}{-}
\SpatialEntry{Mobile ambients}{-}{Compartment Change, Motion}{Neighbor diffusion}{-}{-}

\TableEnd{table:spatial}{Spatial computing operators of spatial computing languages.}

%% file: device-table.tex
\newcommand{\DeviceTableStart}
{\TableStart{|l|cccc|}{DSL & Discretization & Comm. Region & Granularity & Code Mobility \\}}
\newcommand{\DeviceCluster}[1]{\hline {\bf #1} & & & & \\}
\newcommand{\DeviceEntry}[5]{#1 & #2 & #3 & #4 & #5 \\}

\DeviceTableStart

\DeviceCluster{Amorphous Computing}
\DeviceEntry{Proto}{Continuous}{Neighborhood}{Broadcast}{Uniform}
\DeviceEntry{PyMorphous}{Discrete}{Neighborhood}{Broadcast}{Uniform}
\DeviceEntry{ProtoVM}{Discrete}{Neighborhood}{Broadcast}{Uniform}
\DeviceEntry{Growing Point Language}{Discrete}{Neighborhood}{Broadcast}{Uniform}
\DeviceEntry{Origami Shape Language}{Continuous}{Neighborhood}{Broadcast}{Uniform}

\DeviceCluster{Biological}
\DeviceEntry{L-systems}{Cellular}{Local Pattern}{N/A}{Uniform}
\DeviceEntry{MGS}{Cellular}{Local Pattern}{Multicast}{Uniform}
\DeviceEntry{Gro}{Cellular}{Chemical Diffusion}{Broadcast}{Uniform}
\DeviceEntry{GEC}{N/A}{Chemical Diffusion}{Broadcast}{Heterogeneous}
\DeviceEntry{Proto BioCompiler}{Cellular}{Chemical Diffusion}{Broadcast}{Uniform}

\DeviceCluster{Agent-Based}
\DeviceEntry{Graphical Agent Modeling Language}{Discrete}{Global}{Unicast}{-}
\DeviceEntry{Agent Framework}{Discrete}{Global}{Unicast}{Mobile}
\DeviceEntry{Multi-agent Modeling and Simulation Toolkit}{Discrete, Cellular}{Global, Neighborhood}{Unicast, Multicast}{Uniform}

\DeviceCluster{Wireless Sensor Networks}
\DeviceEntry{Region-based DSLs}{Mixed}{Region}{Multicast}{Uniform}
\DeviceEntry{Data-flow based DSLs}{Discrete}{Neighborhood}{Unicast}{Uniform}
\DeviceEntry{Database-like DSLs}{Continuous}{-}{-}{Uniform}
\DeviceEntry{Region-based DSLs}{Discrete}{-}{-}{Uniform}
\DeviceEntry{Agent-based DSLs}{Mixed}{Neighborhood}{Unicast}{Mobile}

\DeviceCluster{Pervasive Computing}
\DeviceEntry{TOTA}{Discrete}{Global, Neighborhood}{Multicast}{Uniform}
\DeviceEntry{Chemical reaction model}{Discrete}{Neighborhood}{Unicast}{Uniform}
\DeviceEntry{Spatially-Scoped Tuples}{Discrete}{Neighborhood}{Unicast}{Uniform}

\DeviceCluster{Swarm \& Modular Robotics}
\DeviceEntry{Bitmap Language}{Discrete}{-}{-}{Uniform}
\DeviceEntry{Graph Grammars}{Discrete}{Neighborhood}{Broadcast}{Uniform}
\DeviceEntry{Meld}{Discrete}{Neighborhood}{Broadcast}{Uniform}
\DeviceEntry{DynaRole/M3L}{Discrete}{Neighborhood}{Multicast}{Uniform}
\DeviceEntry{ASE}{Discrete}{Neighborhood}{Multicast}{Uniform}

\DeviceCluster{Parallel \& Reconfigurable}
\DeviceEntry{Dataflow Languages}{Discrete}{Graph}{Unicast}{Heterogeneous}
\DeviceEntry{MPI}{Discrete}{Global}{Unicast}{Heterogeneous}
\DeviceEntry{Erlang}{Discrete}{Global}{Unicast}{Heterogeneous}
\DeviceEntry{X10/Chapel/Fortress}{Discrete}{Global}{Unicast}{Heterogeneous}
\DeviceEntry{GraphStep}{Discrete}{Neighborhood}{Broadcast}{Uniform}
\DeviceEntry{StarLisp}{Cellular}{Shift}{Unicast}{Uniform}
\DeviceEntry{Grid Libraries}{Cellular}{Neighborhood}{Unicast}{Uniform}
\DeviceEntry{Cellular Automata}{Cellular}{Neighborhood}{Broadcast}{Uniform}

\DeviceCluster{Formal Calculi}
\DeviceEntry{$3\pi$}{Discrete}{Global}{Unicast}{Mobile}
\DeviceEntry{Mobile ambients}{Discrete}{Neighborhood}{Unicast}{Mobile}

\TableEnd{table:device}{Abstract device characteristics of spatial computing languages.}

%% file: arxiv-preprint.bbl
\begin{thebibliography}{}

\bibitem[Abdelzaher et~al., 2004]{abdelzaher2004envirotrack}
Abdelzaher, T., Blum, B., Cao, Q., Chen, Y., Evans, D., George, J., George, S.,
  Gu, L., He, T., Krishnamurthy, S., et~al. (2004).
\newblock Envirotrack: Towards an environmental computing paradigm for
  distributed sensor networks.
\newblock In {\em Distributed Computing Systems, 2004. Proceedings. 24th
  International Conference on}, pages 582--589. IEEE.

\bibitem[Abelson et~al., 2000]{abelson2000amorphous}
Abelson, H., Allen, D., Coore, D., Hanson, C., Homsy, G., Knight~Jr, T.,
  Nagpal, R., Rauch, E., Sussman, G., and Weiss, R. (2000).
\newblock {Amorphous computing}.
\newblock {\em Communications of the ACM}, 43(5):74--82.

\bibitem[Allen et~al., 2008]{Fortress}
Allen, E., Chase, D., Hallett, J., Luchangco, V., Maessen, J.-W., Ryu, S., Jr.,
  G. L.~S., and Tobin-Hochstadt, S. (2008).
\newblock {\em The Fortress Language Specification Version 1.0}.
\newblock Sun Microsystems.

\bibitem[Alur and Henzinger, 1999]{alur99}
Alur, R. and Henzinger, T. (1999).
\newblock Reactive modules.
\newblock {\em Formal Methods in System Design}, 15(1):7--48.

\bibitem[Alvaro, 2009]{alvaro2009dedalus}
Alvaro, P. (2009).
\newblock Dedalus: Datalog in time and space.
\newblock Technical report, DTIC Document.

\bibitem[Annaratone et~al., 1987]{Warp}
Annaratone, M., Arnould, E., Gross, T., Kung, H.~T., Lam, M., Menzilcioglu, O.,
  and Webb, J.~A. (1987).
\newblock The warp computer: Architecture, implementation, and performance.
\newblock {\em IEEE Transactions on Computers}, C-36(12):1523--1538.

\bibitem[Armstrong, 2007]{HistoryOfErlang}
Armstrong, J. (2007).
\newblock A history of erlang.
\newblock In {\em HOPL III Proceedings of the third ACM SIGPLAN conference on
  History of programming languages}, pages 6--1 -- 6--26.

\bibitem[{ASCAPE}, 2011]{ascapewebsite}
{ASCAPE} (2011).
\newblock Ascape guide.
\newblock http://ascape.sourceforge.net/.

\bibitem[Ashley-Rollman et~al., 2007]{ashley07}
Ashley-Rollman, M., Goldstein, S., Lee, P., Mowry, T., and Pillai, P. (2007).
\newblock Meld: A declarative approach to programming ensembles.
\newblock In {\em Proceedings of the {IEEE/RSJ} International Conference on
  Intelligent Robots and Systems}, pages 2794--2800.

\bibitem[Ashley-Rollman et~al., 2009]{ashley09}
Ashley-Rollman, M.~P., Lee, P., Goldstein, S.~C., Pillai, P., and Campbell,
  J.~D. (2009).
\newblock Language for large ensembles of independently executing nodes.
\newblock In {\em Proceedings of the International Conference on Logic
  Programming ({ICLP '09)}}.

\bibitem[Bachrach, 2009]{gasPL}
Bachrach, J. (2009).
\newblock Programming chained robotics in the gas programming language.
\newblock Technical report, Makani Power.

\bibitem[Bachrach and Beal, 2007]{protokernel}
Bachrach, J. and Beal, J. (2007).
\newblock Building spatial computers.
\newblock Technical Report MIT-CSAIL-TR-2007-017, MIT.

\bibitem[Bachrach et~al., 2010]{ProtoSwarm}
Bachrach, J., Beal, J., and McLurkin, J. (2010).
\newblock Composable continuous space programs for robotic swarms.
\newblock {\em Neural Computing and Applications}, 19(6):825--847.

\bibitem[Barney, 2012]{BarneyParallel}
Barney, B. (Retrieved Feb. 20, 2012).
\newblock Introduction to parallel computing.
\newblock {\tt https://computing.llnl.gov/tutorials/parallel\_comp/}.

\bibitem[Bauer et~al., 2001]{bauer2001agent}
Bauer, B., M{\"u}ller, J., and Odell, J. (2001).
\newblock Agent {{UML}}: A formalism for specifying multiagent interaction.
\newblock In {\em Agent-oriented software engineering}, volume 1957, pages
  91--103.

\bibitem[{BBN Technologies}, 2011]{cougaarwebsite}
{BBN Technologies} (2011).
\newblock {Cougaar} the cognitive agent architecture.
\newblock http://cougaar.org.

\bibitem[Beal, 2004]{AmorphousMedium}
Beal, J. (2004).
\newblock Programming an amorphous computational medium.
\newblock In {\em Unconventional Programming Paradigms International Workshop},
  volume 3566 of {\em Lecture Notes in Computer Science}, pages 121--136.
  Springer Berlin.

\bibitem[Beal, 2009]{spatialprocesses}
Beal, J. (2009).
\newblock Dynamically defined processes for spatial computers.
\newblock In {\em Spatial Computing Workshop}.

\bibitem[Beal, 2010]{bealBasisSCW10}
Beal, J. (2010).
\newblock A basis set of operators for space-time computations.
\newblock In {\em Spatial Computing Workshop}.

\bibitem[Beal and Bachrach, 2006]{proto2006a}
Beal, J. and Bachrach, J. (2006).
\newblock Infrastructure for engineered emergence on sensor/actuator networks.
\newblock {\em IEEE Intelligent Systems}.

\bibitem[Beal et~al., 2011]{AC4GRN}
Beal, J., Lu, T., and Weiss, R. (2011).
\newblock Automatic compilation from high-level languages to genetic regulatory
  networks.
\newblock {\em PLoS ONE}.

\bibitem[Beal and Usbeck, 2011]{BealUsbeck11}
Beal, J. and Usbeck, K. (2011).
\newblock On the evaluation of space-time functions.
\newblock In {\em Self-Organizing Self-Adaptive Spatial Computing Workshop}.

\bibitem[{Berkeley Software 2009 iGem Team}, 2010]{Eugene}
{Berkeley Software 2009 iGem Team} (October 2009, Retrieved May 10, 2010.).
\newblock Eugene.
\newblock http://2009.igem.org/Team:Berkeley\_Software/Eugene.

\bibitem[Beydoun et~al., 2009]{beydoun2009faml}
Beydoun, G., Low, G., Henderson-Sellers, B., Mouratidis, H., Gomez-Sanz, J.,
  Pavon, J., and Gonzalez-Perez, C. (2009).
\newblock Faml: a generic metamodel for mas development.
\newblock {\em Software Engineering, IEEE Transactions on}, 35(6):841--863.

\bibitem[Blumofe et~al., 1995]{Cilk}
Blumofe, R.~D., Joerg, C.~F., Kuszmaul, B.~C., Leiserson, C.~E., Randall,
  K.~H., and Zhou, Y. (1995).
\newblock {Cilk}: An efficient multithreaded runtime system.
\newblock In {\em Proceedings of the Fifth ACM SIGPLAN Symposium on Principles
  and Practice of Parallel Programming (PPoPP)}, pages 207--216, Santa Barbara,
  California.

\bibitem[Borcea et~al., 2004]{borcea2004spatial}
Borcea, C., Intanagonwiwat, C., Kang, P., Kremer, U., and Iftode, L. (2004).
\newblock Spatial programming using smart messages: Design and implementation.
\newblock In {\em Distributed Computing Systems, 2004. Proceedings. 24th
  International Conference on}, pages 690--699. IEEE.

\bibitem[Bordignon et~al., 2011a]{M3L11}
Bordignon, M., Stoy, K., and Schultz, U.~P. (2011a).
\newblock Generalized programming of modular robots through kinematic
  configurations.
\newblock In {\em 2011 IEEE/RSJ International Conference on Intelligent Robots
  and Systems (IROS)}, pages 3659--3666.

\bibitem[Bordignon et~al., 2011b]{DynaRole11}
Bordignon, M., Stoy, K., and Schultz, U.~P. (2011b).
\newblock Robust and reversible execution of self-reconfiguration sequences.
\newblock {\em Robotica}, 29(1):35--57.

\bibitem[Boulis et~al., 2007]{boulis2007sensorware}
Boulis, A., Han, C., Shea, R., and Srivastava, M. (2007).
\newblock Sensorware: Programming sensor networks beyond code update and
  querying.
\newblock {\em Pervasive and Mobile Computing}, 3(4):386--412.

\bibitem[Butera, 2002]{butera}
Butera, W. (2002).
\newblock {\em Programming a Paintable Computer}.
\newblock PhD thesis, MIT, Cambridge, MA, USA.

\bibitem[Butera, 2007]{ButeraText}
Butera, W. (2007).
\newblock Text display and graphics control on a paintable computer.
\newblock In {\em Int'l Conf. on Self-Adaptive and Self-Organizing Systems},
  pages 45--54.

\bibitem[Calidonna and Furnari, 2004]{CANL}
Calidonna, C. and Furnari, M. (2004).
\newblock The cellular automata network compiler system: Modules and features.
\newblock In {\em Int'l Conf. on Parallel Computing in Electrical Engineering},
  pages 271--276.

\bibitem[Cardelli, 2005]{DBLP:conf/cmsb/Cardelli04}
Cardelli, L. (2005).
\newblock Brane calculi.
\newblock In Danos, V. and Sch{\"a}chter, V., editors, {\em Computational
  Methods in Systems Biology, International Conference, CMSB 2004, Paris,
  France, May 26-28, 2004, Revised Selected Papers}, volume 3082 of {\em
  Lecture Notes in Computer Science}, pages 257--278. Springer.

\bibitem[Cardelli and Gardner, 2010]{DBLP:conf/cie/CardelliG10}
Cardelli, L. and Gardner, P. (2010).
\newblock Processes in space.
\newblock In Ferreira, F., L{\"o}we, B., Mayordomo, E., and Gomes, L.~M.,
  editors, {\em Programs, Proofs, Processes, 6th Conference on Computability in
  Europe, CiE 2010, Ponta Delgada, Azores, Portugal, June 30 - July 4, 2010.
  Proceedings}, volume 6158 of {\em Lecture Notes in Computer Science}, pages
  78--87. Springer.

\bibitem[Cardelli and Gordon, 2000]{ambients}
Cardelli, L. and Gordon, A.~D. (2000).
\newblock Mobile ambients.
\newblock {\em Theoretical Computer Science}, 240(1):177--213.

\bibitem[Caspi et~al., 2000]{SCORE}
Caspi, E., Chu, M., Huang, R., Yeh, J., Wawrzynek, J., and DeHon, A. (2000).
\newblock Stream computations organized for reconfigurable execution (score).
\newblock In {\em Conference on Field Programmable Logic and Applications
  (FPL)}.

\bibitem[{Centre for Policy Modelling}, 2011]{sdmlwebsite}
{Centre for Policy Modelling} (2011).
\newblock Strictly declarative modelling language.
\newblock http://cfpm.org/sdml/.

\bibitem[Chapel, 2011]{Chapel}
Chapel (2011).
\newblock {\em Chapel Language Specification Version 0.82}.
\newblock Cray, Inc.

\bibitem[Chen et~al., 2003]{ChenMundy03}
Chen, J., Mundy, J., Bai, Y., Chan, S.-M.~C., Petrica, P., and Bahar, R.~I.
  (2003).
\newblock A probabilistic approach to nano-computing.
\newblock In {\em Workshop on Non-Silicon Computation}.

\bibitem[Chou et~al., 2002]{TrendCAs}
Chou, H.-H., Huang, W., and Reggia, J.~A. (2002).
\newblock The trend cellular automata programming environment.
\newblock {\em Simulation}, 78(2):59--75.

\bibitem[Christensen et~al., 2007]{Swarmanoid}
Christensen, A., O'Grady, R., and Dorigo, M. (2007).
\newblock Morphology control in a multirobot system.
\newblock {\em IEEE Robotics \& Automation Magazine}, 14(4):18--25.

\bibitem[Christensen et~al., 2011]{ASE11}
Christensen, D.~J., Schultz, U.~P., and Moghadam, M. (2011).
\newblock The assemble and animate control framework for modular reconfigurable
  robots.
\newblock In {\em IROS Workshop on Reconfigurable Modular Robotics}.

\bibitem[Chu et~al., 2006]{chu2006entirely}
Chu, D., Tavakoli, A., Popa, L., and Hellerstein, J. (2006).
\newblock Entirely declarative sensor network systems.
\newblock In {\em Proceedings of the 32nd international conference on Very
  large data bases}, pages 1203--1206. VLDB Endowment.

\bibitem[Ciciriello et~al., 2006]{ciciriello2006building}
Ciciriello, P., Mottola, L., and Picco, G. (2006).
\newblock Building virtual sensors and actuators over logical neighborhoods.
\newblock In {\em Proceedings of the international workshop on Middleware for
  sensor networks}, pages 19--24. ACM.

\bibitem[Collier and North, 2011]{collier2011repast}
Collier, N. and North, M. (2011).
\newblock Repast {{SC++}}: A platform for large-scale agent-based modeling.
\newblock {\em Large-Scale Computing Techniques for Complex System
  Simulations}.

\bibitem[Collins, 2011]{Huckleberry}
Collins, R.~L. (2011).
\newblock {\em Data-Driven Programming Abstractions and Optimization for
  Multi-Core Platforms}.
\newblock PhD thesis, Columbia University.

\bibitem[Coore, 1999]{coorethesis}
Coore, D. (1999).
\newblock {\em Botanical Computing: A Developmental Approach to Generating
  Interconnect Topologies on an Amorphous Computer}.
\newblock PhD thesis, MIT.

\bibitem[Correll and Martinoli, 2011]{correllijrr2011}
Correll, N. and Martinoli, A. (2011).
\newblock Modeling self-organized aggregation in a swarm of miniature robots.
\newblock {\em The International Journal of Robotics Research. Special Issue on
  Stochasticity in Robotics and Biological Systems}, 30(5):615--626.

\bibitem[Costa et~al., 2006]{costa2006teenylime}
Costa, P., Mottola, L., Murphy, A., and Picco, G. (2006).
\newblock Teenylime: transiently shared tuple space middleware for wireless
  sensor networks.
\newblock In {\em Proceedings of the international workshop on Middleware for
  sensor networks}, pages 43--48. ACM.

\bibitem[Couderc and Banatre, 2003]{SPREAD03}
Couderc, P. and Banatre, M. (2003).
\newblock Ambient computing applications: an experience with the spread
  approach.
\newblock In {\em Hawaii International Conference on System Sciences (HICSS’
  03)}.

\bibitem[Czar et~al., 2009]{GenoCAD}
Czar, M., Cai, Y., and Peccoud, J. (2009).
\newblock Writing dna with genocad.
\newblock {\em Nucleic Acids Research}, 37(W40-7).

\bibitem[{Czech Technical Institute Agent Technology Center}, 2011]{aglobe}
{Czech Technical Institute Agent Technology Center} (2011).
\newblock Aglobe.
\newblock http://agents.felk.cvut.cz/aglobe/.

\bibitem[DeHon, 2002]{DeHon02}
DeHon, A. (2002).
\newblock Very large scale spatial computing.
\newblock In {\em Unconventional Models of Computation}, volume 2509 of {\em
  Lecture Notes in Computer Science}, pages 27--37. Springer.

\bibitem[DeHon and Wawrzynek, 1999]{DeHon99}
DeHon, A. and Wawrzynek, J. (1999).
\newblock Reconﬁgurable computing: What, why, and implications for design
  automation.
\newblock In {\em Design Automation Conference (DAC)}, pages 610--615.

\bibitem[deLorimier et~al., 2011]{GraphStep}
deLorimier, M., Kapre, N., Mehta, N., and DeHon, A. (2011).
\newblock Spatial hardware implementation for sparse graph algorithms in
  graphstep.
\newblock {\em ACM Transactions on Autonomous and Adaptive Systems (TAAS)},
  6(3).

\bibitem[Dewey et~al., 2008]{dewey08}
Dewey, D., Ashley-Rollman, M., Rosa, M.~D., Goldstein, S., Mowry, T.,
  Srinivasa, S., Pillai, P., and Campbell, J. (2008).
\newblock Generalizing metamodules to simplify planning in modular robotic
  systems.
\newblock In {\em Proceedings of the {IEEE/RSJ} International Conference on
  Intelligent Robots and Systems}, pages 1338--1345.

\bibitem[D'Hondt and D'Hondt, 2001a]{dhondtBS2}
D'Hondt, E. and D'Hondt, T. (2001a).
\newblock Amorphous geometry.
\newblock In {\em ECAL 2001}.

\bibitem[D'Hondt and D'Hondt, 2001b]{dhondtBS1}
D'Hondt, E. and D'Hondt, T. (2001b).
\newblock Experiments in amorphous geometry.
\newblock In {\em 2001 International Conference on Artificial Intelligence}.

\bibitem[Dietrich, 2011]{PyMorphous}
Dietrich, C. (2011).
\newblock Pymorphous: Python language extensions for spatial computing.
\newblock http://pymorphous.googlecode.com.

\bibitem[Duckham et~al., 2005]{Duckham:2005:MDS:1097064.1097073}
Duckham, M., Nittel, S., and Worboys, M. (2005).
\newblock Monitoring dynamic spatial fields using responsive geosensor
  networks.
\newblock In {\em Proceedings of the 13th annual ACM international workshop on
  Geographic information systems}, GIS '05, pages 51--60, New York, NY, USA.
  ACM.

\bibitem[Dunkels et~al., 2004]{dunkels2004contiki}
Dunkels, A., Gronvall, B., and Voigt, T. (2004).
\newblock Contiki-a lightweight and flexible operating system for tiny
  networked sensors.
\newblock In {\em Local Computer Networks, 2004. 29th Annual IEEE International
  Conference on}, pages 455--462. IEEE.

\bibitem[Eckart, 1997]{CELLANG}
Eckart, J.~D. (1997).
\newblock {\em Cellang: Language Reference Manual}.
\newblock Radford University.

\bibitem[Engstrom and Cappello, 1989]{SDEF}
Engstrom, B.~R. and Cappello, P.~R. (1989).
\newblock The sdef programming system.
\newblock {\em Journal of Parallel and Distributed Computing}, 7(2):201 -- 231.

\bibitem[Erlang, 2011]{Erlang}
Erlang (2011).
\newblock {\em Erlang Reference Manual User's Guide Version 5.9}.
\newblock Ericsson AB.

\bibitem[{Ernest Friedman-Hill}, 2008]{jesswebsite}
{Ernest Friedman-Hill} (2008).
\newblock {JESS}, the rule engine for the java platform.
\newblock http://herzberg.ca.sandia.gov/jess/.

\bibitem[Ferscha et~al., 2008]{Ferscha2008448}
Ferscha, A., Hechinger, M., Riener, A., dos Santos~Rocha, M., Zeidler, A.,
  Franz, M., and Mayrhofer, R. (2008).
\newblock Peer-it: Stick-on solutions for networks of things.
\newblock {\em Pervasive and Mobile Computing}, 4(3):448 -- 479.

\bibitem[Finin et~al., 1994]{Finin94kqml}
Finin, T., Fritzson, R., McKay, D., and McEntire, R. (1994).
\newblock Kqml as an agent communication language.
\newblock In {\em Proceedings of the third international conference on
  Information and knowledge management}, CIKM '94, pages 456--463, New York,
  NY, USA. ACM.

\bibitem[Fok et~al., 2005]{fok2005rapid}
Fok, C., Roman, G., and Lu, C. (2005).
\newblock Rapid development and flexible deployment of adaptive wireless sensor
  network applications.
\newblock In {\em Distributed Computing Systems, 2005. ICDCS 2005. Proceedings.
  25th IEEE International Conference on}, pages 653--662. IEEE.

\bibitem[Forrest and Jones, 1994]{forrest1994modeling}
Forrest, S. and Jones, T. (1994).
\newblock Modeling complex adaptive systems with echo.
\newblock {\em Complex systems: Mechanisms of adaptation}, pages 3--21.

\bibitem[Frank and Romer, 2005]{frank2005algorithms}
Frank, C. and Romer, K. (2005).
\newblock Algorithms for generic role assignment in wireless sensor networks.
\newblock In {\em Proceedings of the 3rd international conference on Embedded
  networked sensor systems}, pages 230--242. ACM.

\bibitem[Freiwald and Weimar, 2002]{JCASim}
Freiwald, U. and Weimar, J. (2002).
\newblock The java based cellular automata simulation system jcasim.
\newblock {\em Future Generation Computing Systems}, 18:995--1004.

\bibitem[Gay et~al., 2003]{gay2003nesc}
Gay, D., Levis, P., Von~Behren, R., Welsh, M., Brewer, E., and Culler, D.
  (2003).
\newblock The nesc language: A holistic approach to networked embedded systems.
\newblock In {\em Acm Sigplan Notices}, volume 38(5), pages 1--11. ACM.

\bibitem[Gayle and Coore, 2006]{gayleCoore}
Gayle, O. and Coore, D. (2006).
\newblock Self-organizing text in an amorphous environment.
\newblock In {\em ICCS 2006}.

\bibitem[Gelernter and Carriero, 1992]{gelernter1992coordination}
Gelernter, D. and Carriero, N. (1992).
\newblock Coordination languages and their significance.
\newblock {\em Communications of the ACM}, 35(2):97--107.

\bibitem[{George Mason University Evolutionary Computation Laboratory and
  Center for Social Complexity}, 2011]{masonwebsite}
{George Mason University Evolutionary Computation Laboratory and Center for
  Social Complexity} (2011).
\newblock {MASON} multiagent simulation.
\newblock http://cs.gmu.edu/~eclab/projects/mason/.

\bibitem[Gheorghe and Paun, 2000]{Gheorghe2000108}
Gheorghe and Paun (2000).
\newblock Computing with membranes.
\newblock {\em Journal of Computer and System Sciences}, 61(1):108 -- 143.

\bibitem[Giavitto et~al., 2002]{GiavittoMGS02}
Giavitto, J.-L., Godin, C., Michel, O., and zemyslaw Prusinkiewicz, P. (2002).
\newblock Computational models for integrative and developmental biology.
\newblock Technical Report 72-2002, Univerite d'Evry, LaMI.

\bibitem[Giavitto et~al., 2004]{GiavittoMGS04}
Giavitto, J.-L., Michel, O., Cohen, J., and Spicher, A. (2004).
\newblock Computation in space and space in computation.
\newblock Technical Report 103-2004, Univerite d'Evry, LaMI.

\bibitem[Giavitto and Spicher, 2008]{mgs08a}
Giavitto, J.-L. and Spicher, A. (2008).
\newblock Topological rewriting and the geometrization of programming.
\newblock {\em Physica D}, 237(9):1302--1314.

\bibitem[Gillespie, 1977]{Gillespie:1977}
Gillespie, D.~T. (1977).
\newblock Exact stochastic simulation of coupled chemical reactions.
\newblock {\em The Journal of Physical Chemistry}, 81(25):2340--2361.

\bibitem[Gilpin et~al., 2008]{gilpin08}
Gilpin, K., Kotay, K., Rus, D., and Vasilescu, I. (2008).
\newblock Miche: Modular shape formation by self-disassembly.
\newblock {\em I. J. Robotic Res.}, pages 345--372.

\bibitem[Govett et~al., 2003]{SMS03}
Govett, M., Middlecoff, J., Hart, L., Henderson, T., and Schaffer, D. (2003).
\newblock The scalable modeling system: directive-based code parallelization
  for distributed and shared memory computers.
\newblock {\em Parallel Computing}.

\bibitem[Grelck et~al., 2007]{CAOS}
Grelck, C., Penczek, F., and Trojahner, K. (2007).
\newblock Caos: A domain-specific language for the parallel simulation of
  cellular automata.
\newblock In {\em Parallel Computing Technologies, 9th International Conference
  (PaCT’07)}, pages 410--417. Springer-Verlag.

\bibitem[Gropp et~al., 1994]{MPI}
Gropp, W., Lusk, E., and Skjellum, A. (1994).
\newblock {\em Using MPI: Portable Parallel Programming with the Message
  Passing Interface}.
\newblock MIT Press, Cambridge, Massachusetts.

\bibitem[Grumbach and Wang, 2010]{grumbach2010netlog}
Grumbach, S. and Wang, F. (2010).
\newblock Netlog, a rule-based language for distributed programming.
\newblock {\em Practical Aspects of Declarative Languages}, pages 88--103.

\bibitem[Guly{\'a}s et~al., 1999]{gulyas1999multi}
Guly{\'a}s, L., Kozsik, T., and Corliss, J. (1999).
\newblock The multi-agent modelling language and the model design interface.
\newblock {\em Journal of Artificial Societies and Social Simulation}, 2(3):8.

\bibitem[Guly\'{a}s et~al., 2011]{mamlwebsite}
Guly\'{a}s, L., Kozsik, T., and Fazekas, S. (2011).
\newblock Multi-agent modeling language {{MAML}}.
\newblock http://www.maml.hu/.

\bibitem[Gummadi et~al., 2005]{gummadi2005macro}
Gummadi, R., Gnawali, O., and Govindan, R. (2005).
\newblock Macro-programming wireless sensor networks using kairos.
\newblock {\em Distributed Computing in Sensor Systems}, pages 466--466.

\bibitem[Hahn, 2008]{hahn2008domain}
Hahn, C. (2008).
\newblock A domain specific modeling language for multiagent systems.
\newblock In {\em Proceedings of the 7th international joint conference on
  Autonomous agents and multiagent systems-Volume 1}, pages 233--240.
  International Foundation for Autonomous Agents and Multiagent Systems.

\bibitem[Hamann et~al., 2010]{HormoneRobotics10}
Hamann, H., Stradner, J., Schmickl, T., and Crailsheim, K. (2010).
\newblock A hormone-based controller for evolutionary multi-modular robotics:
  from single modules to gait learning.
\newblock In {\em IEEE Congress on Evolutionary Computation (CEC'10)}, pages
  244--251.

\bibitem[Heinzelman et~al., 2004]{heinzelman2004middleware}
Heinzelman, W., Murphy, A., Carvalho, H., and Perillo, M. (2004).
\newblock Middleware to support sensor network applications.
\newblock {\em Network, IEEE}, 18(1):6--14.

\bibitem[Helsinger et~al., 2004]{helsinger2004cougaar}
Helsinger, A., Thome, M., and Wright, T. (2004).
\newblock Cougaar: a scalable, distributed multi-agent architecture.
\newblock In {\em IEEE Systems, Man and Cybernetics, 2004}, volume~2, pages
  1910--1917. IEEE.

\bibitem[Hnat et~al., 2008]{hnat2008macrolab}
Hnat, T., Sookoor, T., Hooimeijer, P., Weimer, W., and Whitehouse, K. (2008).
\newblock Macrolab: a vector-based macroprogramming framework for
  cyber-physical systems.
\newblock In {\em Proceedings of the 6th ACM conference on Embedded network
  sensor systems}, pages 225--238. ACM.

\bibitem[Holzmann and Ferscha, 2010]{spacepervasive-pmc6}
Holzmann, C. and Ferscha, A. (2010).
\newblock A framework for utilizing qualitative spatial relations between
  networked embedded systems.
\newblock {\em Pervasive and Mobile Computing}, 6(3):362--381.

\bibitem[HPF, 1997]{HPF}
HPF (1997).
\newblock {\em High Performance Fortran Language Specification, Version 2.0}.
\newblock High Performance Fortran Forum.

\bibitem[Huget, 2005]{huget2005modeling}
Huget, M. (2005).
\newblock Modeling languages for multiagent systems.
\newblock {\em Agent-Oriented Software Engineering (AOSE-2005)}.

\bibitem[Huzita and Scimemi, 1989]{huzita}
Huzita, H. and Scimemi, B. (1989).
\newblock The algebra of paper-folding.
\newblock In {\em First International Meeting of Origami Science and
  Technology}.

\bibitem[{IEEE Computer Society}, 2011]{fipa}
{IEEE Computer Society} (2011).
\newblock Foundation for intelligent physical agents.
\newblock http://www.fipa.org/.

\bibitem[Igarashi et~al., 2001]{FJ}
Igarashi, A., Pierce, B.~C., and Wadler, P. (2001).
\newblock Featherweight java: a minimal core calculus for java and gj.
\newblock {\em ACM Trans. Program. Lang. Syst.}, 23(3):396--450.

\bibitem[Inchiosa and Parker, 2002]{inchiosa2002overcoming}
Inchiosa, M. and Parker, M. (2002).
\newblock Overcoming design and development challenges in agent-based modeling
  using ascape.
\newblock {\em Proceedings of the National Academy of Sciences of the United
  States of America}, 99(Suppl 3):7304.

\bibitem[Jesshope et~al., 2009]{Microgrid}
Jesshope, C., Lankamp, M., and Zhang, L. (2009).
\newblock Evaluating cmps and their memory architecture.
\newblock {\em Proc Architecture of Computing Systems ARCS 2009}, LNCS
  5455:246--257.

\bibitem[Jones and Forrest, 2011]{sfiechoweb}
Jones, T. and Forrest, S. (2011).
\newblock An introduction to sfi echo.
\newblock http://tuvalu.santafe.edu/~pth/echo/how-to/how-to.html.

\bibitem[Jozwiak et~al., 2010]{Jozwiak10}
Jozwiak, L., Nedjah, N., and Figueroa, M. (2010).
\newblock Modern development methods and tools for embedded reconfigurable
  systems: A survey.
\newblock {\em Integration, the VLSI Journal}, 43(1):1--33.

\bibitem[Kinny, 2002]{Kinny02Psi}
Kinny, D. (2002).
\newblock The $\psi$ calculus: An algebraic agent language.
\newblock In Meyer, J.-J. and Tambe, M., editors, {\em Intelligent Agents
  VIII}, volume 2333 of {\em Lecture Notes in Computer Science}, pages 32--50.
  Springer Berlin / Heidelberg.

\bibitem[Klavins et~al., 2006]{Klavins2006}
Klavins, E., Ghrist, R., and Lipsky, D. (2006).
\newblock A grammatical approach to self-organizing robotic systems.
\newblock {\em Automatic Control, IEEE Transactions on}, 51(6):949--962.

\bibitem[Kondacs, 2003]{kondacs}
Kondacs, A. (2003).
\newblock Biologically-inspired self-assembly of 2d shapes, using
  global-to-local compilation.
\newblock In {\em International Joint Conference on Artificial Intelligence
  (IJCAI)}.

\bibitem[Kothari et~al., 2007]{kothari2007reliable}
Kothari, N., Gummadi, R., Millstein, T., and Govindan, R. (2007).
\newblock Reliable and efficient programming abstractions for wireless sensor
  networks.
\newblock In {\em Proceedings of the 2007 ACM SIGPLAN conference on Programming
  language design and implementation}, pages 200--210. ACM.

\bibitem[Kulesza et~al., 2004]{Kulesza2004AGA}
Kulesza, U., Garcia, A., and Lucena, C. (2004).
\newblock An aspect-oriented generative approach.
\newblock In {\em Companion to the 19th annual ACM SIGPLAN conference on
  Object-oriented programming systems, languages, and applications}, OOPSLA
  '04, pages 166--167, New York, NY, USA. ACM.

\bibitem[Kulesza et~al., 2005]{kulesza2005generative}
Kulesza, U., Garcia, A., Lucena, C., and Alencar, P. (2005).
\newblock A generative approach for multi-agent system development.
\newblock {\em Software Engineering for Multi-Agent Systems III}, pages 52--69.

\bibitem[Kwiatkowska et~al., 2011]{KNP11}
Kwiatkowska, M., Norman, G., and Parker, D. (2011).
\newblock {PRISM} 4.0: Verification of probabilistic real-time systems.
\newblock In Gopalakrishnan, G. and Qadeer, S., editors, {\em Proc. 23rd
  International Conference on Computer Aided Verification (CAV'11)}, volume
  6806 of {\em LNCS}, pages 585--591. Springer.

\bibitem[Lasser et~al., 1988]{StarLisp}
Lasser, C., Massar, J., Miney, J., and Dayton, L. (1988).
\newblock {\em Starlisp Reference Manual}.
\newblock Thinking Machines Corporation.

\bibitem[Levis and Culler, 2002]{levis2002mate}
Levis, P. and Culler, D. (2002).
\newblock Mate: a tiny virtual machine for sensor networks.
\newblock In {\em ACM Sigplan Notices}, volume 37(10), pages 85--95. ACM.

\bibitem[Levis et~al., 2005]{levis2005tinyos}
Levis, P., Madden, S., Polastre, J., Szewczyk, R., Whitehouse, K., Woo, A.,
  Gay, D., Hill, J., Welsh, M., Brewer, E., et~al. (2005).
\newblock Tinyos: An operating system for sensor networks.
\newblock {\em Ambient intelligence}, 35.

\bibitem[Li et~al., 2004]{li2004event}
Li, S., Lin, Y., Son, S., Stankovic, J., and Wei, Y. (2004).
\newblock Event detection services using data service middleware in distributed
  sensor networks.
\newblock {\em Telecommunication Systems}, 26(2):351--368.

\bibitem[Litovsky et~al., 1992]{DBLP:conf/wg/LitovskyMZ92}
Litovsky, I., M{\'e}tivier, Y., and Zielonka, W. (1992).
\newblock The power and the limitations of local computations on graphs.
\newblock In {\em WG}, pages 333--345.

\bibitem[Liu et~al., 2003]{liu2003state}
Liu, J., Chu, M., Liu, J., Reich, J., and Zhao, F. (2003).
\newblock State-centric programming for sensor-actuator network systems.
\newblock {\em IEEE Pervasive Computing}, pages 50--62.

\bibitem[Loo et~al., 2006]{loo2006declarative}
Loo, B., Condie, T., Garofalakis, M., Gay, D., Hellerstein, J., Maniatis, P.,
  Ramakrishnan, R., Roscoe, T., and Stoica, I. (2006).
\newblock Declarative networking: language, execution and optimization.
\newblock In {\em Proceedings of the 2006 ACM SIGMOD international conference
  on Management of data}, pages 97--108. ACM.

\bibitem[Lopes et~al., 2010]{lopes2010applying}
Lopes, N., Navarro, J., Rybalchenko, A., and Singh, A. (2010).
\newblock Applying prolog to develop distributed systems.
\newblock {\em Theory and Practice of Logic Programming}, 10(4-6):691--707.

\bibitem[Luke et~al., 2004]{luke2004mason}
Luke, S., Cioffi-Revilla, C., Panait, L., and Sullivan, K. (2004).
\newblock Mason: A new multi-agent simulation toolkit.
\newblock In {\em Proceedings of the 2004 SwarmFest Workshop}, volume~8.
  Citeseer.

\bibitem[Macal and North, 2010]{macal2010tutorial}
Macal, C. and North, M. (2010).
\newblock Tutorial on agent-based modelling and simulation.
\newblock {\em Journal of Simulation}, 4(3):151--162.

\bibitem[MacLennan, 1990]{maclennanCSA}
MacLennan, B. (1990).
\newblock Continuous spatial automata.
\newblock Technical Report Department of Computer Science Technical Report
  CS-90-121, University of Tennessee, Knoxville.

\bibitem[Madden et~al., 2002]{tinydb}
Madden, S.~R., Szewczyk, R., Franklin, M.~J., and Culler, D. (2002).
\newblock Supporting aggregate queries over ad-hoc wireless sensor networks.
\newblock In {\em Workshop on Mobile Computing and Systems Applications}.

\bibitem[Mallavarapu et~al., 2009]{littleB}
Mallavarapu, A., Thomson, M., Ullian, B., and Gunawardena, J. (2009).
\newblock Programming with models: modularity and abstraction provide powerful
  capabilities for systems biology.
\newblock {\em Journal of The Royal Society Interface}, 6(32):257--270.

\bibitem[Mamei and Zambonelli, 2008]{tota}
Mamei, M. and Zambonelli, F. (2008).
\newblock Programming pervasive and mobile computing applications: the {TOTA}
  approach.
\newblock {\em ACM Transactions on Software Engineering and Methodology}.

\bibitem[Margolus, 1993]{cam8}
Margolus, N. (1993).
\newblock {CAM-8}: A computer architecture based on cellular automata.
\newblock In {\em Pattern Formation and Lattice-Gas Automata}.

\bibitem[Martel and Mohammadi, 2010]{martel2010}
Martel, S. and Mohammadi, M. (2010).
\newblock Using a swarm of self-propelled natural microrobots in the form of
  flagellated bacteria to perform complex micro-assembly tasks.
\newblock In {\em In Proceedings of the Int.\ Conference on Robotics and
  Automation ({ICRA})}, pages 500--505.

\bibitem[Martinoli et~al., 2004]{martinoli04}
Martinoli, A., Easton, K., and Agassounon, W. (2004).
\newblock Modeling of swarm robotic systems: A case study in collaborative
  distributed manipulation.
\newblock {\em International Journal of Robotics Research}, 23(4):415--436.

\bibitem[Martinoli et~al., 1999]{martinoli99a}
Martinoli, A., Ijspeert, A.~J., and Mondada, F. (1999).
\newblock Understanding collective aggregation mechanisms: From probabilistic
  modelling to experiments with real robots.
\newblock {\em Robotics \& Autonomous Systems}, 29:51--63.
\newblock Special Issue on Distributed Autonomous Robotic Systems.

\bibitem[Mernik et~al., 2005]{mernik2005and}
Mernik, M., Heering, J., and Sloane, A. (2005).
\newblock When and how to develop domain-specific languages.
\newblock {\em ACM Computing Surveys (CSUR)}, 37(4):316--344.

\bibitem[Michalakes, 1994]{RSL}
Michalakes, J. (1994).
\newblock Rsl: A parallel runtime system library for regular grid finite
  difference models using multiple nests.
\newblock Technical Report ANL/MCS-TM-197, Argonne National Laboratory.

\bibitem[Michalakes, 1997]{FLIC}
Michalakes, J. (1997).
\newblock Flic: A translator for same-source parallel implementation of regular
  grid applications.
\newblock Technical Report ANL/MCS-TM-223, Argonne National Laboratory.

\bibitem[Milner, 1999]{PiCalculus}
Milner, R. (1999).
\newblock {\em Communicating and Mobile Systems: The Pi-Calculus}.
\newblock Cambridge University Press.

\bibitem[Minar et~al., 1996]{swarm}
Minar, N., Burkhart, R., Langton, C., and Askenazi, M. (1996).
\newblock The swarm simulation system, a toolkit for building multi-agent
  simulations.
\newblock Technical Report Working Paper 96-06-042, Santa Fe Institute.

\bibitem[Mirschel et~al., 2009]{ProMoT}
Mirschel, S., Steinmetz, K., Rempel, M., Ginkel, M., and Gilles, E.~D. (2009).
\newblock Promot: Modular modeling for systems biology.
\newblock {\em Bioinformatics}, 25(5):687--689.

\bibitem[Mishra et~al., 2006]{Tartan}
Mishra, M., Callahan, T., Chelcea, T., Venkataramani, G., Budiu, M., and
  Goldstein, S. (2006).
\newblock Tartan: evaluating spatial computation for whole program execution.
\newblock In {\em ASPLOS 2006}.

\bibitem[{MIT Media Lab and Schellar Teacher Education Program},
  2011]{starlogowebsite}
{MIT Media Lab and Schellar Teacher Education Program} (2011).
\newblock Starlogo.
\newblock http://education.mit.edu/starlogo/.

\bibitem[MIT Proto, 2010]{mitproto}
MIT Proto (Retrieved November 22, 2010).
\newblock {MIT Proto}.
\newblock software available at {\tt http://proto.bbn.com/}.

\bibitem[Montagna et~al., 2011]{MVRPD-SERENE2011-LNCS2011}
Montagna, S., Viroli, M., Risoldi, M., Pianini, D., and Di~Marzo~Serugendo, G.
  (2011).
\newblock Self-organising pervasive ecosystems: A crowd evacuation example.
\newblock In {\em 3rd International Workshop on Software Engineering for
  Resilient Systems}, volume 6968 of {\em Lecture Notes in Computer Science},
  pages 115--129. Springer, Geneva, Switzerland.

\bibitem[Moss et~al., 1998]{moss98sdml}
Moss, S., Gaylard, H., Wallis, S., and Edmonds, B. (1998).
\newblock Sdml: A multi-agent language for organizational modelling.
\newblock {\em Computational and Mathematical Organization Theory}, 4:43--69.
\newblock 10.1023/A:1009600530279.

\bibitem[Mottola and Picco, 2006]{mottola2006logical}
Mottola, L. and Picco, G. (2006).
\newblock Logical neighborhoods: A programming abstraction for wireless sensor
  networks.
\newblock {\em Distributed Computing in Sensor Systems}, pages 150--168.

\bibitem[Mottola and Picco, 2011]{mottola2011programming}
Mottola, L. and Picco, G. (2011).
\newblock Programming wireless sensor networks: Fundamental concepts and state
  of the art.
\newblock {\em ACM Computing Surveys (CSUR)}, 43(3):19.

\bibitem[MPI2, 2009]{MPI2}
MPI2 (2009).
\newblock {\em MPI: A Message-Passing Interface Standard Version 2.2}.
\newblock Message Passing Interface Forum.

\bibitem[Mucci et~al., 2007]{Mucci07}
Mucci, C., Campi, F., Brunelli, C., and Nurmi, J. (2007).
\newblock Programming tools for reconfigurable processors.
\newblock In Nurmi, J., editor, {\em System-On-Chip Computing for ASICs and
  FPGAs on Processor Design}, pages 427--446. Springer.

\bibitem[Myers et~al., 2009]{iBioSim}
Myers, C., Barker, N., Jones, K., Kuwahara, H., Madsen, C., and Nguyen, N.
  (2009).
\newblock ibiosim: a tool for the analysis and design of genetic circuits.
\newblock {\em Bioinformatics}, 25:2848--9.

\bibitem[Nagpal, 2001]{nagpal}
Nagpal, R. (2001).
\newblock {\em Programmable Self-Assembly: Constructing Global Shape using
  Biologically-inspired Local Interactions and Origami Mathematics}.
\newblock PhD thesis, MIT.

\bibitem[Newton et~al., 2007]{newton2007regiment}
Newton, R., Morrisett, G., and Welsh, M. (2007).
\newblock The regiment macroprogramming system.
\newblock In {\em Proceedings of the 6th international conference on
  Information processing in sensor networks}, pages 489--498. ACM.

\bibitem[Newton and Welsh, 2004]{regiment}
Newton, R. and Welsh, M. (2004).
\newblock Region streams: Functional macroprogramming for sensor networks.
\newblock In {\em First International Workshop on Data Management for Sensor
  Networks (DMSN)}.

\bibitem[Nguyen et~al., 2010]{asra}
Nguyen, D.~N., Usbeck, K., Mongan, W.~M., Cannon, C.~T., Lass, R.~N., Salvage,
  J., and Regli, W.~C. (2010).
\newblock A methodology for developing an agent systems reference architecture.
\newblock In {\em 11th International Workshop on Agent-oriented Software
  Engineering}, Toronto, ON.

\bibitem[North et~al., 2007]{north2007declarative}
North, M., Howe, T., Collier, N., and Vos, J. (2007).
\newblock A declarative model assembly infrastructure for verification and
  validation.
\newblock In {\em Advancing Social Simulation: The First World Congress}, pages
  129--140. Springer Japan.

\bibitem[Odell et~al., 1999]{odell1999extending}
Odell, J., Parunak, H., and Bauer, B. (1999).
\newblock Extending uml for agents.
\newblock {\em Ann Arbor}, 1001:48--103.

\bibitem[OpenCL, 2011]{OpenCL}
OpenCL (2011).
\newblock {\em The OpenCL Specification, Version 1.2}.
\newblock Khronos OpenCL Working Group.

\bibitem[OpenMP, 2011]{OpenMP}
OpenMP (2011).
\newblock {\em OpenMP Application Program Interface Version 3.1}.
\newblock OpenMP Architecture Review Board.

\bibitem[Palmer and G.L.~Steele, 1992]{cm5}
Palmer, J. and G.L.~Steele, J. (1992).
\newblock Connection machine model cm-5 system overview.
\newblock In {\em Fourth Symposium on the Frontiers of Massively Parallel
  Computation}, pages 474--483. IEEE Press.

\bibitem[Pathak et~al., 2007]{pathak2007expressing}
Pathak, A., Mottola, L., Bakshi, A., Prasanna, V., and Picco, G. (2007).
\newblock Expressing sensor network interaction patterns using data-driven
  macroprogramming.
\newblock In {\em Proceedings of the Fifth IEEE International Conference on
  Pervasive Computing and Communications Workshops}, pages 255--260. IEEE
  Computer Society.

\bibitem[Paun, 2002]{PSystems}
Paun, G. (2002).
\newblock {\em Membrane computing: An introduction}.
\newblock Springer.

\bibitem[Pauty et~al., 2007]{GeoLinda}
Pauty, J., Couderc, P., Banatre, M., and Berbers, Y. (2007).
\newblock Geo-linda: a geometry aware distributed tuple space.
\newblock In {\em IEEE 21st International Conference on Advanced Networking and
  Applications (AINA '07)}, pages 370--377.

\bibitem[Pedersen and Phillips, 2009]{GEC}
Pedersen, M. and Phillips, A. (2009).
\newblock Towards programming languages for genetic engineering of living
  cells.
\newblock {\em Journal of the Royal Society Interface}.

\bibitem[Pierce and Turner, 2000]{DBLP:conf/birthday/PierceT00}
Pierce, B.~C. and Turner, D.~N. (2000).
\newblock Pict: a programming language based on the pi-calculus.
\newblock In Plotkin, G.~D., Stirling, C., and Tofte, M., editors, {\em Proof,
  Language, and Interaction, Essays in Honour of Robin Milner}, pages 455--494.
  The MIT Press.

\bibitem[Pierro et~al., 2005]{stoklaim}
Pierro, A.~D., Hankin, C., and Wiklicky, H. (2005).
\newblock Continuous-time probabilistic klaim.
\newblock {\em Electr. Notes Theor. Comput. Sci.}, 128(5):27--38.

\bibitem[Pokahr et~al., 2003]{pokahr2003jadex}
Pokahr, A., Braubach, L., and Lamersdorf, W. (2003).
\newblock Jadex: Implementing a bdi-infrastructure for jade agents.
\newblock {\em EXP--in search of innovation}, 3(3):76--85.

\bibitem[Pressey, 2012]{ALPACA}
Pressey, C. (Retrieved Feb 20, 2012).
\newblock The alpaca meta-language.
\newblock {\tt http://catseye.tc/projects/alpaca/}.

\bibitem[Priami, 1995]{Priami:1995}
Priami, C. (1995).
\newblock Stochastic pi-calculus.
\newblock {\em The Computer Journal}, 38(7):578--589.

\bibitem[Prorok et~al., 2011]{prorok10}
Prorok, A., Correll, N., and Martinoli, A. (2011).
\newblock Multi-level spatial models for swarm-robotic systems.
\newblock {\em The International Journal of Robotics Research. Special Issue on
  Stochasticity in Robotics and Biological Systems.}, 30(5):574--589.

\bibitem[Prusinkiewicz and Lindenmayer, 1990]{LSystems}
Prusinkiewicz, P. and Lindenmayer, A. (1990).
\newblock {\em The Algorithmic Beauty of Plants}.
\newblock Springer-Verlag, New York, NY, USA.

\bibitem[Raimbault and Lavenier, 1993]{ReLaCS}
Raimbault, F. and Lavenier, D. (1993).
\newblock Relacs for systolic programming.
\newblock In {\em Int'l Conf. on Application-Specific Array Processors}, pages
  132--135.

\bibitem[Rao and Georgeff, 1995]{rao1995bdi}
Rao, A. and Georgeff, M. (1995).
\newblock Bdi agents: From theory to practice.
\newblock In {\em Proceedings of the first international conference on
  multi-agent systems (ICMAS-95)}, pages 312--319. San Francisco.

\bibitem[Regli et~al., 2009]{asrm}
Regli, W.~C., Mayk, I., Dugan, C.~J., Kopena, J.~B., Lass, R.~N., Modi, P.~J.,
  Mongan, W.~M., Salvage, J.~K., and Sultanik, E.~A. (2009).
\newblock Development and specification of a reference model for agent-based
  systems.
\newblock {\em Trans. Sys. Man Cyber Part C}, 39:572--596.

\bibitem[Rejimon and Bhanja, 2005]{RejimonBhanja05}
Rejimon, T. and Bhanja, S. (2005).
\newblock Scalable probabilistic computing models using bayesian networks.
\newblock In {\em 48th Midwest Symposium on Circuits and Systems}, pages
  712--715.

\bibitem[{Repast Team}, 2011]{repastwebsite}
{Repast Team} (2011).
\newblock The repast suite.
\newblock http://repast.sourceforge.net/index.html.

\bibitem[Resnick, 1996]{resnick1996starlogo}
Resnick, M. (1996).
\newblock Starlogo: an environment for decentralized modeling and decentralized
  thinking.
\newblock In {\em Conference companion on Human factors in computing systems:
  common ground}, pages 11--12. ACM.

\bibitem[Rus et~al., 2002]{rus02}
Rus, D., Butler, Z.~J., Kotay, K., and Vona, M. (2002).
\newblock Self-reconfiguring robots.
\newblock {\em Commun. ACM}, 45(3):39--45.

\bibitem[Rus and Vona, 2001]{rus01}
Rus, D. and Vona, M. (2001).
\newblock Crystalline robots: Self-reconfiguration with compressible unit
  modules.
\newblock {\em Auton. Robots}, 10(1):107--124.

\bibitem[Russell et~al., 1995]{russellNorvig}
Russell, S., Norvig, P., Canny, J., Malik, J., and Edwards, D. (1995).
\newblock {\em Artificial intelligence: a modern approach}.
\newblock Prentice hall Englewood Cliffs, NJ.

\bibitem[Saraswat et~al., 2012]{X10}
Saraswat, V., Bloom, B., Peshansky, I., Tardieu, O., and Grove, D. (2012).
\newblock {\em X10 Language Specification Version 2.2}.
\newblock IBM, Yorktown Heights, New York.

\bibitem[Schultz et~al., 2008]{SpatialLabels08}
Schultz, U., Bordignon, M., Christensen, D., and Stoy, K. (2008).
\newblock Spatial computing with labels.
\newblock In {\em Spatial Computing Workshop at IEEE SASO}.

\bibitem[Schultz et~al., 2007]{DynaRole07}
Schultz, U.~P., Christensen, D.~J., and Stoy, K. (2007).
\newblock A domain-specific language for programming self-reconfigurable
  robots.
\newblock In {\em Workshop on Automatic Program Generation for Embedded Systems
  (APGES)}, pages 28--36.

\bibitem[Shen et~al., 2001]{shen2001sensor}
Shen, C., Srisathapornphat, C., and Jaikaeo, C. (2001).
\newblock Sensor information networking architecture and applications.
\newblock {\em Personal communications, IEEE}, 8(4):52--59.

\bibitem[Shen et~al., 2004]{ShenHormones04}
Shen, W.-M., Will, P., Galstyan, A., and Chuong, C. (2004).
\newblock Hormone-inspired self-organization and distributed control of robotic
  swarms.
\newblock {\em Autonomous Robotics}, 17(1):93--105.

\bibitem[Sklar, 2007]{sklar2007netlogo}
Sklar, E. (2007).
\newblock Netlogo, a multi-agent simulation environment.
\newblock {\em Artificial life}, 13(3):303--311.

\bibitem[Smith et~al., 2009]{antimony}
Smith, L.~P., Bergmann, F.~T., Chandran, D., and Sauro, H.~M. (2009).
\newblock Antimony: a modular model definition language.
\newblock {\em Bioinformatics}, 25(18):2452--54.

\bibitem[Spezzano and Talia, 1997]{CARPET}
Spezzano, G. and Talia, D. (1997).
\newblock A high-level cellular programming model for massively parallel
  processing.
\newblock In {\em 2nd Int'l Workshop on High-Level Programming Models and
  Supportive Environments (HIPS’97)}, pages 55--63.

\bibitem[Stoy and Nagpal, 2004]{stoy04}
Stoy, K. and Nagpal, R. (2004).
\newblock Self-repair through scale independent self-reconfiguration.
\newblock In {\em Proceedings of the {IEEE/RSJ} International Conference on
  Intelligent Robots and Systems}, pages 2062--2067.

\bibitem[Sugihara and Gupta, 2008]{sugihara2008programming}
Sugihara, R. and Gupta, R. (2008).
\newblock Programming models for sensor networks: A survey.
\newblock {\em ACM Transactions on Sensor Networks (TOSN)}, 4(2):8.

\bibitem[Swanson et~al., 2007]{WaveScalar}
Swanson, S., Schwerin, A., Mercaldi, M., Petersen, A., Putnam, A., Michelson,
  K., Oskin, M., and Eggers, S.~J. (2007).
\newblock The wavescalar architecture.
\newblock {\em ACM Trans. Comput. Syst}, 25:4.

\bibitem[Szymanski and Woern, 2007]{szymanski07}
Szymanski, M. and Woern, H. (2007).
\newblock {JaMOS - A MDL}2$\epsilon$ based operating system for swarm micro
  robotics.
\newblock In {\em Proceedings of the 2007 {IEEE} Swarm Intelligence Symposium}.

\bibitem[Taylor et~al., 2002]{RAWarchitecture}
Taylor, M.~B., Kim, J., Miller, J., Wentzlaff, D., Ghodrat, F., Greenwald, B.,
  Hoffmann, H., Johnson, P., Lee, J.-W., Lee, W., Ma, A., Saraf, A., Seneski,
  M., Shnidman, N., Strumpen, V., Frank, M., Amarasinghe, S., and Agarwal, A.
  (2002).
\newblock The raw microprocessor: A computational fabric for software circuits
  and general purpose programs.
\newblock {\em IEEE Micro}, 22:25--35.

\bibitem[{Telecom Italia Lab}, 2011]{jade}
{Telecom Italia Lab} (2011).
\newblock {JADE} --- {Java Agent DEvelopment} framework.
\newblock http://jade.tilab.com/.

\bibitem[{The Klavins Lab}, 2012]{Gro}
{The Klavins Lab} (2012).
\newblock {\em Gro: The cell programming language}.
\newblock University of Washington.

\bibitem[Thies et~al., 2001]{StreamIT}
Thies, W., Karczmarek, M., Gordon, M., Maze, D., Wong, J., Hoffmann, H., and
  Brown, M. (2001).
\newblock Streamit: A compiler for streaming applications.
\newblock Technical Report MIT-LCS Technical Memo TM-622, Massachusetts
  Institute of Technology.

\bibitem[Tobis, 2005]{PyNSol05}
Tobis, M. (2005).
\newblock Pynsol: Objects as scaffolding.
\newblock {\em Computing in Science and Engineering}, 7(4):84--91.

\bibitem[Toffoli and Margolus, 1987]{CAMs}
Toffoli, T. and Margolus, N. (1987).
\newblock {\em Cellular Automata Machines: A new environment for modeling}.
\newblock MIT Press.

\bibitem[Trencansky and Cervenka, 2005]{trencansky2005agent}
Trencansky, I. and Cervenka, R. (2005).
\newblock Agent modeling language ({{AML}}): A comprehensive approach to
  modeling {{MAS}}.
\newblock {\em Informatica Ljubljana}, 29(4):391.

\bibitem[{University of Michigan Center for the Study of Complex Systems},
  2011]{swarmwebsite}
{University of Michigan Center for the Study of Complex Systems} (2011).
\newblock Swarm development wiki.
\newblock http://www.swarm.org/index.php/Main\_Page.

\bibitem[Usbeck and Beal, 2011]{usbeck2011agent}
Usbeck, K. and Beal, J. (2011).
\newblock An agent framework for agent societies.
\newblock In {\em Systems, Programming, Languages and Applications: Software
  for Humanity}.

\bibitem[Viroli et~al., 2011a]{VCMZ-TAAS2011}
Viroli, M., Casadei, M., Montagna, S., and Zambonelli, F. (2011a).
\newblock Spatial coordination of pervasive services through chemical-inspired
  tuple spaces.
\newblock {\em ACM Transactions on Autonomous and Adaptive Systems}, 6(2):14:1
  -- 14:24.

\bibitem[Viroli et~al., 2011b]{VNCMZ-WOA2011}
Viroli, M., Nardini, E., Castelli, G., Mamei, M., and Zambonelli, F. (2011b).
\newblock A coordination approach to spatially-situated pervasive service
  ecosystems.
\newblock In Fortino, G., Garro, A., Palopoli, L., Russo, W., and Spezzano, G.,
  editors, {\em WOA 2011 -- XII Workshop Nazionale "Dagli Oggetti agli
  Agenti"}, volume 741 of {\em CEUR Workshop Proceedings}, pages 19--27, Rende,
  Italy. Sun SITE Central Europe, RWTH Aachen University.

\bibitem[Walter et~al., 2004]{walter04}
Walter, J., Welch, J., and Amato, N. (2004).
\newblock Distributed reconï¬guration of metamorphic robot chains.
\newblock {\em Distributed Computing}, 17:171--189.

\bibitem[Weiss, 2001]{weissMCL}
Weiss, R. (2001).
\newblock {\em Cellular Computation and Communications using Engineered Genetic
  Regular Networks}.
\newblock PhD thesis, MIT.

\bibitem[Welsh and Mainland, 2004]{welsh2004regions}
Welsh, M. and Mainland, G. (2004).
\newblock Programming sensor networks using abstract regions.
\newblock In {\em Proceedings of the First USENIX/ACM Symposium on Networked
  Systems Design and Implementation (NSDI '04)}.

\bibitem[Werfel, 2006]{werfelphd}
Werfel, J. (2006).
\newblock {\em Anthills built to order: Automating construction with artificial
  swarms}.
\newblock PhD thesis, MIT, Cambridge, MA, USA.

\bibitem[Werfel et~al., 2005]{werfeletal}
Werfel, J., Bar-Yam, Y., and Nagpal, R. (2005).
\newblock Building patterned structures with robot swarms.
\newblock In {\em IJCAI}.

\bibitem[Werfel and Nagpal, 2007]{werfelAdaptiveConstruction}
Werfel, J. and Nagpal, R. (2007).
\newblock Collective construction of environmentally-adaptive structures.
\newblock In {\em 2007 IEEE/RSJ International Conference on Intelligent Robots
  and Systems (IROS 2007)}, Piscataway, NJ, USA. IEEE.

\bibitem[Whitehouse et~al., 2004]{hood}
Whitehouse, K., Sharp, C., Brewer, E., and Culler, D. (2004).
\newblock Hood: a neighborhood abstraction for sensor networks.
\newblock In {\em Proceedings of the 2nd international conference on Mobile
  systems, applications, and services}. ACM Press.

\bibitem[Wilensky, 2011]{netlogowebsite}
Wilensky, U. (2011).
\newblock Netlogo.
\newblock http://ccl.northwestern.edu/netlogo/.

\bibitem[Yamins, 2007]{yamins}
Yamins, D. (2007).
\newblock {\em A Theory of Local-to-Global Algorithms for One-Dimensional
  Spatial Mu lti-Agent Systems}.
\newblock PhD thesis, Harvard, Cambridge, MA, USA.

\bibitem[Yang et~al., 2011]{YangPNAS2011}
Yang, C., Wu, H., Huang, Q., Li, Z., and Li, J. (2011).
\newblock Using spatial principles to optimize distributed computing for
  enabling the physical science discoveries.
\newblock {\em Proc. Nat'l Academy of Sciences (PNAS)}, 108(14):5498--5503.

\bibitem[Yao and Gehrke, 2002]{Yao02thecougar}
Yao, Y. and Gehrke, J. (2002).
\newblock The cougar approach to in-network query processing in sensor
  networks.
\newblock {\em SIGMOD Record}, 31:2002.

\bibitem[Yim et~al., 2007]{yim07}
Yim, M., Shen, W., Salemi, B., Rus, D., Moll, M., Lipson, H., Klavins, E., and
  Chirikjian, G. (2007).
\newblock Modular self-reconfigurable robot systems: grand challenges of
  robotics.
\newblock {\em Robotics \& Automation Magazine, IEEE}, 14(1):43--52.

\bibitem[Zambonelli and Viroli, 2011]{ZV-JPCC2011}
Zambonelli, F. and Viroli, M. (2011).
\newblock A survey on nature-inspired metaphors for pervasive service
  ecosystems.
\newblock {\em International Journal of Pervasive Computing and
  Communications}, 7(3):186--204.

\bibitem[Zimmermann, 1980]{zimmermann1980osi}
Zimmermann, H. (1980).
\newblock Osi reference model--the iso model of architecture for open systems
  interconnection.
\newblock {\em Communications, IEEE Transactions on}, 28(4):425--432.

\end{thebibliography}
